\documentclass[twocolumn]{aastex631}

\usepackage{latexsym}		% to get LASY symbols
\usepackage{graphicx}		% to insert PostScript figures
\usepackage{rotating}		% for sideways tables/figures
\usepackage{natbib}  % Requires natbib.sty, available from http://ads.harvard.edu/pubs/bibtex/astronat/
\usepackage{savesym}
\usepackage{amssymb}
\usepackage{amsmath}
\usepackage{morefloats}
\usepackage{nicefrac}
\savesymbol{doublespace}
\usepackage{xspace}
\usepackage{color}
\usepackage{mdframed}
\usepackage{url}
\usepackage{grffile}
\usepackage{import}
\usepackage[utf8]{inputenc}
\usepackage{array}
\usepackage{fancyhdr}

\usepackage[hang,flushmargin]{footmisc}
\usepackage{ifpdf}

\newcommand{\msun}{\ensuremath{M_{\odot}}\xspace}			%  Msun

\newcommand{\methanol}{\ensuremath{\textrm{CH}_3\textrm{OH}}\xspace}

\newcommand{\water}{H$_{2}$O\xspace}		%  H2O
\newcommand{\kms}{\textrm{km~s}\ensuremath{^{-1}}\xspace}	%  km s-1

\def\eqref#1{Equation \ref{#1}}

\usepackage[version=4]{mhchem}

\newcommand{\ndisks}{9\xspace}
\newcommand{\ntentative}{3\xspace}
\newcommand{\ndiskstext}{nine\xspace}
\newcommand{\nnewdisks}{5\xspace}
\newcommand{\nregions}{6\xspace}

\shorttitle{ Salts on HMYSOs: Brinaries }
\shortauthors{ Ginsburg et al}

\begin{document}

\title{Salt-bearing disk candidates around high-mass young stellar objects}

\author[0000-0001-6431-9633]{Adam Ginsburg}
\affiliation{Department of Astronomy, University of Florida, P.O. Box 112055, Gainesville, FL, USA}

\author[0000-0003-1254-4817]{Brett A. McGuire}
\affiliation{Department of Chemistry, Massachusetts Institute of Technology, Cambridge, MA 02139}
\affiliation{National Radio Astronomy Observatory, Charlottesville, VA 22903} % chemist, SrcI
\author[0000-0002-7125-7685]{Patricio Sanhueza} % provided DIHCA
\affiliation{National Astronomical Observatory of Japan, National Institutes of Natural Sciences, 2-21-1 Osawa, Mitaka, Tokyo 181-8588, Japan}
\affil{Department of Astronomical Science, SOKENDAI (The Graduate University for Advanced Studies), 2-21-1 Osawa, Mitaka, Tokyo 181-8588, Japan}
\author[0000-0002-8250-6827]{Fernando Olguin} % provided DIHCA
\affil{Institute of Astronomy and Department of Physics, National Tsing Hua University, Hsinchu 30013, Taiwan} % :wave: Hi Luke! haha hey Adam AG: I'm leaving this here because it's funny.  Remember "Google Wave"?  That was a silly thing that was fun for about 5 minutes.
\author[0000-0002-7675-3565]{Luke T. Maud}% G17
\affiliation{ESO Headquarters, 
        Karl-Schwarzchild-Str 2 85748 Garching, Germany}
\author[0000-0002-6907-0926]{Kei E. I. Tanaka} % provided I16457
\affiliation{Center for Astrophysics and Space Astronomy, University of Colorado Boulder, Boulder, CO 80309, USA}
\affiliation{National Astronomical Observatory of Japan, National Institutes of Natural Sciences, 2-21-1 Osawa, Mitaka, Tokyo 181-8588, Japan}
\author[0000-0001-7511-0034]{Yichen Zhang} % provided I16457
\affiliation{Department of Astronomy, University of Virginia, Charlottesville, VA 22904, USA}
\affiliation{RIKEN Cluster for Pioneering Research, Wako, Saitama 351-0198, Japan}
\author[0000-0002-1700-090X]{Henrik Beuther} % provided G351
\affiliation{Max Planck Institute for Astronomy, Königstuhl 17, 69117, Heidelberg, Germany}
\author[0000-0001-8533-6440]{Nick Indriolo} % provided high-frequency G17 data
\affiliation{AURA for the European Space Agency (ESA), Space Telescope Science Institute, 3700 San Martin Drive, Baltimore, MD 21218, USA}

\begin{abstract}
Molecular lines tracing the orbital motion of gas in a well-defined disk are valuable tools for inferring both the properties of the disk and the star it surrounds.
Lines that arise only from a disk, and not also from the surrounding molecular cloud core that birthed the star or from the outflow it drives, are rare.
Several such emission lines have recently been discovered in one example case, those from NaCl and KCl salt molecules.
We studied a sample of 23 candidate high-mass young stellar objects (HMYSOs) in 17 high-mass star-forming regions to determine how frequently emission from these species is detected.
We present \nnewdisks new detections of water, NaCl, KCl, PN, and SiS from the innermost regions around the objects, bringing the total number of known briny disk candidates to \ndisks.
Their kinematic structure is generally disk-like, {though we are unable to determine whether they arise from a disk or outflow in the  sources with new detections}.
We demonstrate that these species are spatially coincident in a few resolved cases and show that they are generally detected together, suggesting a common origin or excitation mechanism.
We also show that several disks around HMYSOs clearly do not exhibit emission in these species.
Salty disks are therefore neither particularly rare in high-mass disks, nor are they ubiquitous.

\end{abstract}


\section{Introduction} \label{sec:intro}

Circumstellar accretion disks develop around forming new stars.
While the presence of disks around low-mass stars has been clear for decades, we have only definitively demonstrated that accretion disks exist around high-mass young stellar objects (HYMSOs) in the last decade \citep[e.g.,][]{Beltran2016,Maud2017,Ilee2018,Motogi2019,Johnston2020,Sanna2021,Moscadelli2021}.

One limiting factor in the detection and subsequent characterization of HYMSOs
has been the lack of molecular emission lines that arise from the disk, but are
not confused with or absorbed by the surrounding molecular cloud.  A select few
lines have recently been discovered that are uniquely produced in the disks of
some HMYSOs.  Emission from salt molecules has been detected in the
{surroundings} of four stars in three star-forming regions: Orion Source I
\citep{Ginsburg2019}, G17.64+0.16 \citep{Maud2019}, and a pair in IRAS
16547-4247 \citep{Tanaka2020} (hereafter, SrcI, G17, and I16547, respectively).
{Of these, SrcI and G17 are confirmed disks, while the I16547 pair remain 
candidates, and in all cases the salt emission comes from zones within
$\lesssim100$ au of the central source.}
Each of these sources also exhibits
H$_2$O emission from both the {(candidate)} disk and a slightly more
extended region, and thus we dub these objects `brinaries.'

Salts are detected in the atmospheres of AGB and post-AGB stars, giving some clues as to the physical conditions needed to produce them.
Salts have been detected in 
CRL2688 \citep{Highberger2003},
IRAS+10216 \citep{Cernicharo1987},
IK Tauri, VY Canis Majoris \citep{Milam2007, Decin2016},
VX Sgr \citep[mentioned in passing in][]{Danilovich2021},
and OH231.8+44.2 \citep{SanchezContreras2022}.
\citet{SanchezContreras2022} discovered a brinary disk surrounding the post-AGB mass transfer system OH231.8+4.2, highlighting the similarity between birth and death among moderately massive stars.
However, there are also some non-detections in well-surveyed sources, including the S-type AGB stars W Aql and $\pi^1$ Gru \citep{Homan2020,Danilovich2021}.

The development of line lists in the infrared, and more complete lists in the radio, has been driven in part by interest in salts as constituents of the atmospheres of hot planets, in which these species are predicted to be important components of upper cloud layers \citep{Barton2014}.
{There is therefore some motivation to understand where salts occur in disks and in or on dust grains.}

Studies of (post)-AGB stars and models of planetary atmospheres provide some clues about the physical conditions required to produce gas-phase salts.
NaCl is expected to change states (from solid to gas or vice-versa) around $\sim$500--600\,K at planetary atmospheric pressure  \citep[i.e., 1 bar;][]{Woitke2018}.
\citet{Decin2016} suggest that it comes off of grains at 100--300\,K based on the detection locations in supergiant stars.
These studies provide a first hint about where NaCl may come from if it precipitates out of cooling gas, though it remains unclear if the same mechanisms apply in {and around YSO} disks.

Motivated by the detection of salts in a few HMYSOs with ALMA in recent years, we present a first search for salt-bearing disks in the ALMA archives.
In Section \ref{sec:data},  we describe the ALMA observations we analyze.
In Section \ref{sec:analysis}, we describe the analysis approach (\ref{sec:wateranalysis}) and detections (\ref{sec:saltdetections}).
We discuss the chemical correlations observed in the sample and possible reasons for (non)detections in Section \ref{sec:discussion}, then conclude.

\section{Data \& Sample Selection}
\label{sec:data}

We utilize archival and new data from several projects.
We select data sets that have high angular resolution ($\lesssim0.1$ \arcsec) targeting high-mass star-forming regions.
Most of the sources in our sample come from the Digging into the Interior of Hot Cores with ALMA (DIHCA; PI: Sanhueza) program, which is surveying $\sim30$ {candidate} disks.
The main aims of the DIHCA survey are to study the interior of massive hot cores to determine whether they form high-mass stars collapsing monolithically or by fragmenting into binary (multiple) systems and to search for accretion disks around high-mass stars.
The DIHCA targets were selected from the literature as regions with previous interferometric (e.g., SMA) observations and having an expected flux of $>$0.1 Jy at 230 GHz.
All clumps follow the empirical threshold for high-mass star formation suggested by \cite{Kauffmann2010}. 
We only examine a small subset of the DIHCA sample here because not all data were available as of March 2022.
Because our sample is not uniformly selected, we can say little about completeness; we can only search for general trends.
Nevertheless, the trends we find are interesting and suggest that observing a more uniform sample in the future would be productive.

DIHCA observations of G335, G333.23, NGC6334, IRAS16562, G34.43mm1, G29.96, and G351.77 were obtained during July 2019, and observations of G5.89, G11.92, IRAS18089, and W33A were taken both in September 2017 and July 2019.
The observations were reduced using CASA (v5.4.0-70; \citealp{2007ASPC..376..127M}).
The data were then phase self-calibrated in three steps with decreasing solution intervals and the continuum subtracted following the procedure of \citet{Olguin2021}.
Dirty cubes were produced with a Briggs weighting robust parameter of 0.5 using the CASA \texttt{tclean} task.
We use dirty image cubes for expediency, so it is possible significant improved images of these objects could be obtained, though we note that the targeted lines are generally faint and would not be affected by cleaning with typical clean parameters.

\begin{table*}[htp]

\centering
\caption{Observation Summary}
\label{tab:obsstats}

\begin{tabular}{cccccccccccc}
\hline \hline
Field & Source Name & $\theta_{\rm maj }$ & $\theta_{\rm maj}$ & $\theta_{\rm min}$ & PA & $\sigma$ & $f(>5\sigma)$ & $N_{\rm beams}$ & $\sigma_{\rm avg}$ & $v$ ref. line & Distance \\
 &  & $\mathrm{AU}$ & $\mathrm{{}^{\prime\prime}}$ & $\mathrm{{}^{\prime\prime}}$ & $\mathrm{{}^{\circ}}$ & $\mathrm{K}$ &  &  & $\mathrm{K}$ &  & $\mathrm{kpc}$ \\
\hline
Orion & SrcI & 20 & 0.050 & 0.039 & 72.7 & 19.3 & 0.05 & 43 & 2.9 & NaCl J=18-17 v=0 & 0.4 \\
S255IR & SMA1 & 40 & 0.036 & 0.027 & 5.6 & 46.9 & 0.04 & 19 & 10.6 & no clear disk & 1.6$^k$ \\
G351.77 & mm12 & 50 & 0.027 & 0.022 & -89.3 & 6.0 & 0.08 & 24 & 1.2 & \water & 2.2$^c$ \\
G351.77 & mm2 & 50 & 0.027 & 0.022 & -89.3 & 6.0 & 0.05 & 21 & 1.3 & \water & 2.2$^c$ \\
G351.77 & mm1 & 50 & 0.027 & 0.022 & -89.3 & 6.0 & 0.14 & 41 & 0.9 & \water & 2.2$^c$ \\
G17 & G17 & 50 & 0.038 & 0.022 & 44.4 & 10.0 & 0.01 & 75 & 1.2 & \water & 2.2$^c$ \\
NGC6334I & mm1d & 50 & 0.074 & 0.041 & 62.8 & 19.6 & 0.10 & 20 & 4.3 & \water & 1.3$^d$ \\
NGC6334I & mm1b & 50 & 0.074 & 0.041 & 62.8 & 19.6 & 0.23 & 27 & 3.7 & \water & 1.3$^d$ \\
NGC6334I & mm2b & 50 & 0.074 & 0.041 & 62.8 & 19.6 & 0.07 & 7 & 7.0 & \water & 1.3$^d$ \\
NGC6334IN & SMA1b/d & 50 & 0.074 & 0.042 & 63.4 & 17.5 & 0.16 & 278 & 1.0 & \water & 1.3$^d$ \\
NGC6334IN & SMA6 & 50 & 0.074 & 0.042 & 63.4 & 17.5 & 0.06 & 22 & 3.7 & \water & 1.3$^d$ \\
IRAS18162 & GGD27 & 90 & 0.099 & 0.066 & -89.0 & 6.9 & 0.01 & 36 & 1.1 & SO $6_5-5_4$ & 1.3$^a$ \\
IRAS18089 & I18089-1732 & 100 & 0.064 & 0.045 & 65.8 & 18.4 & 0.33 & 57 & 2.4 & CH$_3$OH & 2.3$^f$ \\
G34.43 & mm1 & 110 & 0.105 & 0.069 & 59.5 & 9.2 & 0.35 & 53 & 1.3 & no clear disk & 1.6$^l$ \\
IRAS16562 & G345.4938+01.4677 & 120 & 0.106 & 0.053 & 81.6 & 10.5 & 0.04 & 25 & 2.1 & H30$\alpha$ & 2.3$^g$ \\
I16547 & A & 130 & 0.065 & 0.043 & 35.7 & 23.2 & 0.19 & 7 & 8.4 & \water & 2.9$^c$ \\
I16547 & B & 130 & 0.065 & 0.043 & 35.7 & 23.2 & 0.18 & 5 & 9.5 & \water & 2.9$^c$ \\
G5.89 & mm15 & 130 & 0.063 & 0.043 & 66.3 & 19.2 & 0.00 & 85 & 2.1 & \water & 3.0$^b$ \\
G335 & ALMA1 & 140 & 0.066 & 0.041 & 47.3 & 20.4 & 0.29 & 118 & 1.9 & no clear disk & 3.3$^i$ \\
W33A & mm1-main & 170 & 0.103 & 0.067 & -86.4 & 6.5 & 0.17 & 51 & 0.9 & \water & 2.6$^k$ \\
G333.23 & mm1 & 220 & 0.069 & 0.041 & 54.1 & 18.4 & 0.02 & 70 & 2.2 & SO $6_5-5_4$ & 5.3$^h$ \\
G333.23 & mm2 & 220 & 0.069 & 0.041 & 54.1 & 18.4 & 0.08 & 45 & 2.7 & SO $6_5-5_4$ & 5.3$^h$ \\
G11.92 & mm1 & 220 & 0.101 & 0.066 & -86.9 & 6.6 & 0.16 & 103 & 0.7 & SO $6_5-5_4$ & 3.3$^e$ \\
G29.96 & submm1 & 530 & 0.099 & 0.071 & 65.0 & 9.6 & 0.22 & 369 & 0.5 & no clear disk & 7.4$^j$ \\
\hline
\end{tabular}

\par
Observation properties.  The `Field' name indicates the region of the ALMA pointing.  The `Source Name' is the identifier of the disk candidate examined.
$\theta$ gives the beam parameters, with $\theta_{maj}$ [au] providing the physical size using the adopted distance.
$\sigma$ is the average noise level of the field, which is averaged down by $N_{beams}^{1/2}$ to give $\sigma_{\rm avg}$, the noise level in the stacked spectrum.
$f(>5\sigma)$ is the fraction of the stacked spectrum that is above five times $\sigma_{avg}$; it is used as a diagnostic of the line crowding in the spectrum covering 219.2--220.8 GHz, which is high for complex-molecule-rich regions.
The $v$ reference line is the line used to create a velocity map to produce stacked spectra.
Section \ref{sec:COMS} provides additional details.
  Distances come from the following sources:
      $^a$ \citet{Anez-Lopez2020},
      $^b$ \citet{Sato2014,Fernandez-Lopez2021},
      $^c$ \citet{Beuther2017},
      $^d$ \citet{Chibueze2014},
      $^e$ \citet{Sato2014},
      $^f$ \citet{Xu2011},
      $^g$ \citet{Guzman2020},
      $^h$ \citet{Whitaker2017},
      %$^i$ \citet{Olguin2022},
      $^i$ \citet{Peretto2013},
      $^j$ \citet{Kalcheva2018},
      $^k$ \citet{Reid2014},
      $^l$ \citet{Kurayama2011},
      $^m$ \citet{Ilee2018},
      %$^n$ \citet{Ginsburg2018}

\end{table*}

We use the SrcI data from \citet{Ginsburg2018,Ginsburg2019}. 
We use G17 data from \citet{Maud2019}.
We use G351.77 data from both DIHCA and \citet{Beuther2019}.
We use I16547 data from both DIHCA and \citet{Tanaka2020}.
For each of these data sets, we refer the reader to the cited papers for the data reduction description.  We include summary statistics of these observations in Table \ref{tab:obsstats}.
We give additional details about the physical resolution, distance to the targets, and the line used as a velocity guide (see \S \ref{sec:linestacking}) in  Table \ref{tab:obsstats}.

Finally, we use proprietary data toward Sh 255-IR SMA1 (hereafter S255IR) from project 2019.1.00492.S (PI Ginsburg).  We use the archive-produced data products, which were cleaned with the ALMA pipeline, and imaged them with CASA 6.4.3.4.
The images were cleaned to a depth of 10 mJy using Briggs robust=0 weighting.

\section{Results and Analysis}
\label{sec:analysis}
We search for lines of NaCl, KCl, \water, SiS, and H30$\alpha$ in each of the target pointings (see Table \ref{tab:lines}).
Since each target was selected for having a high luminosity or a strong HMYSO disk candidate beforehand, we used the literature identification of existing sources as our starting point.
We cut out cubes centered on the brightest continuum source in the ALMA images (except G5.89; see \S \ref{sec:g5.89}).
We also searched fainter continuum sources, selecting small sub-regions around each of the compact continuum peaks that could plausibly contain disks.
Because the source selection is based on a by-eye examination of the data over a limited field of view (in most cases, only the inner 5--10 arcseconds of the ALMA field of view was imaged), the sample presented here has unknown completeness - the conclusions we draw will therefore be only suggestive, not conclusive.

Most of our line detections come from stacked spectra of resolved disk-like objects.
We describe in Section \ref{sec:linestacking} the line stacking approach used to obtain higher signal-to-noise ratio spectra that represent average values over the {candidate} disk.
Section \ref{sec:saltdetections} describes the detections in individual sources and shows some of the extracted images and spectra.
Additional images and spectra are displayed in the appendices.

\begin{figure*}
    \centering
    \includegraphics[width=0.49\textwidth]{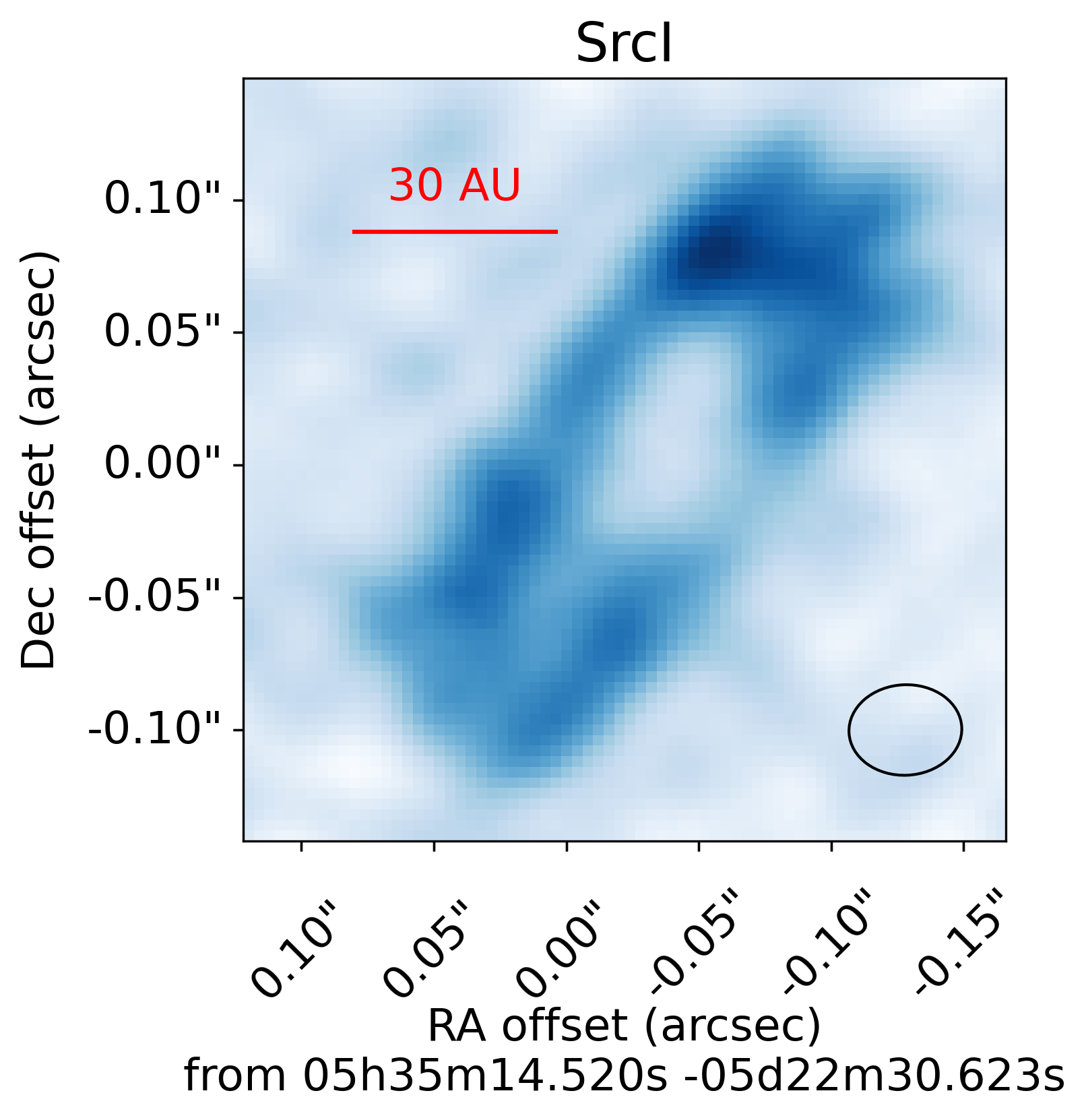}
    \includegraphics[width=0.49\textwidth]{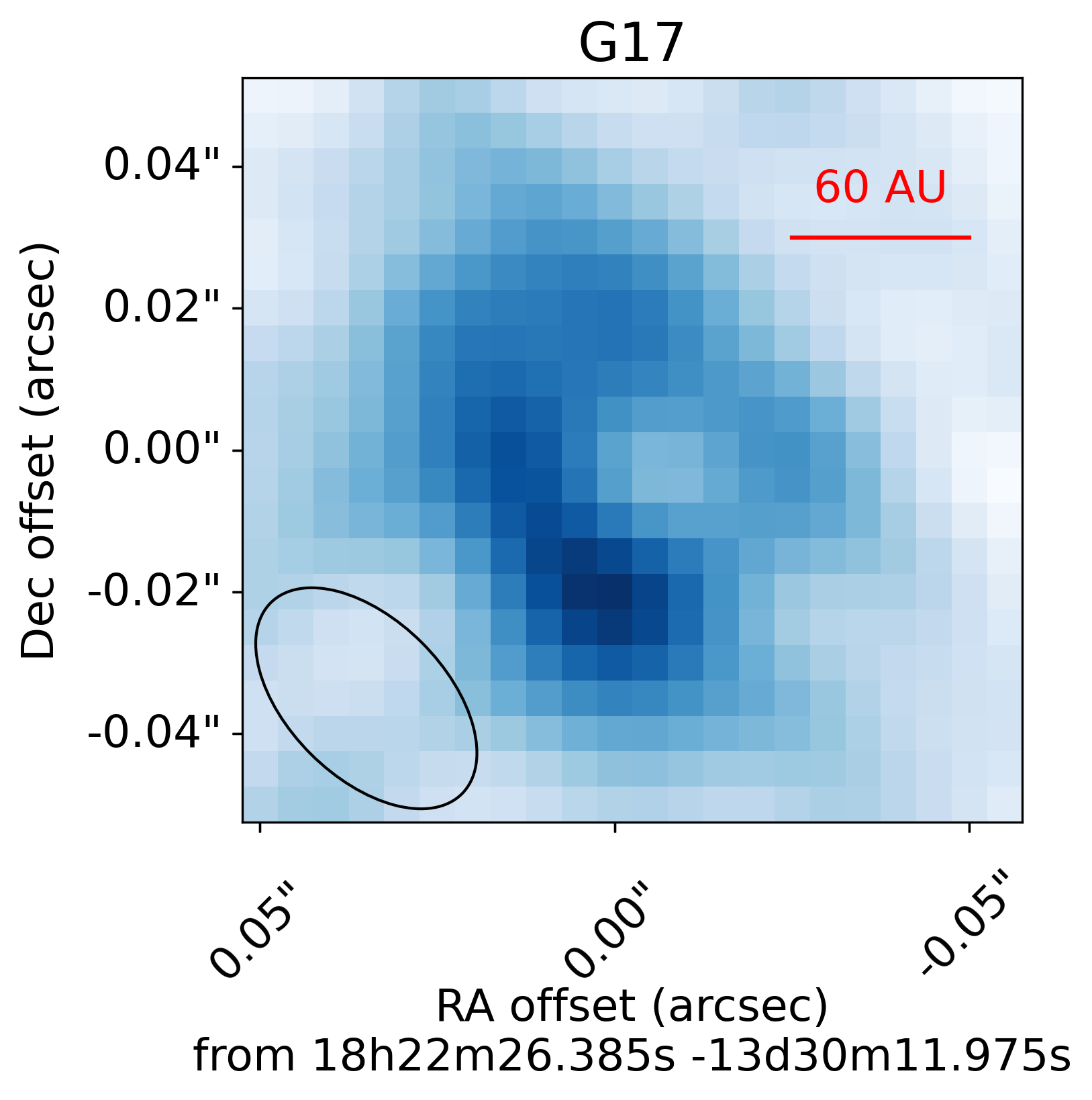}
    \includegraphics[width=0.49\textwidth]{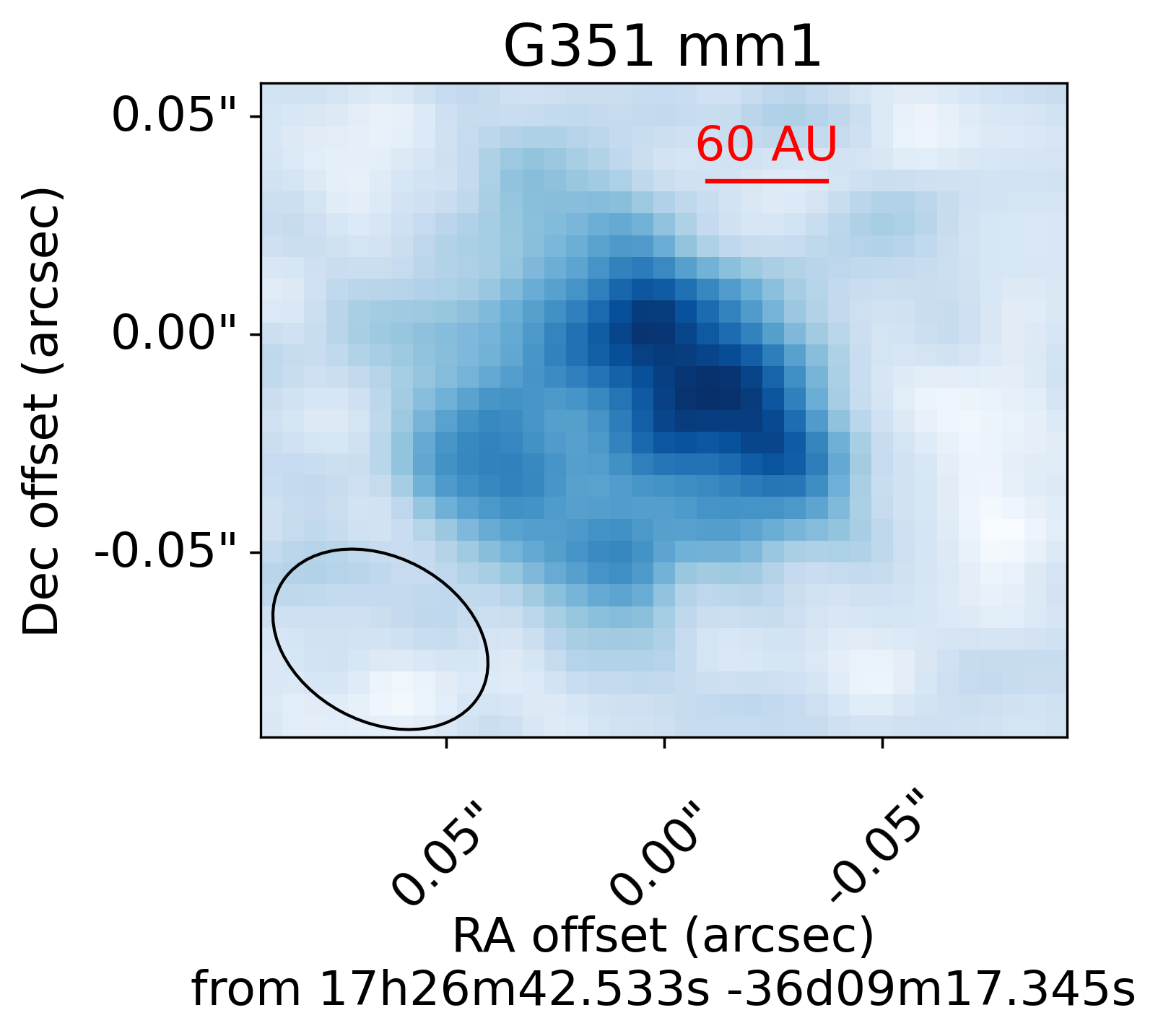}
    \includegraphics[width=0.49\textwidth]{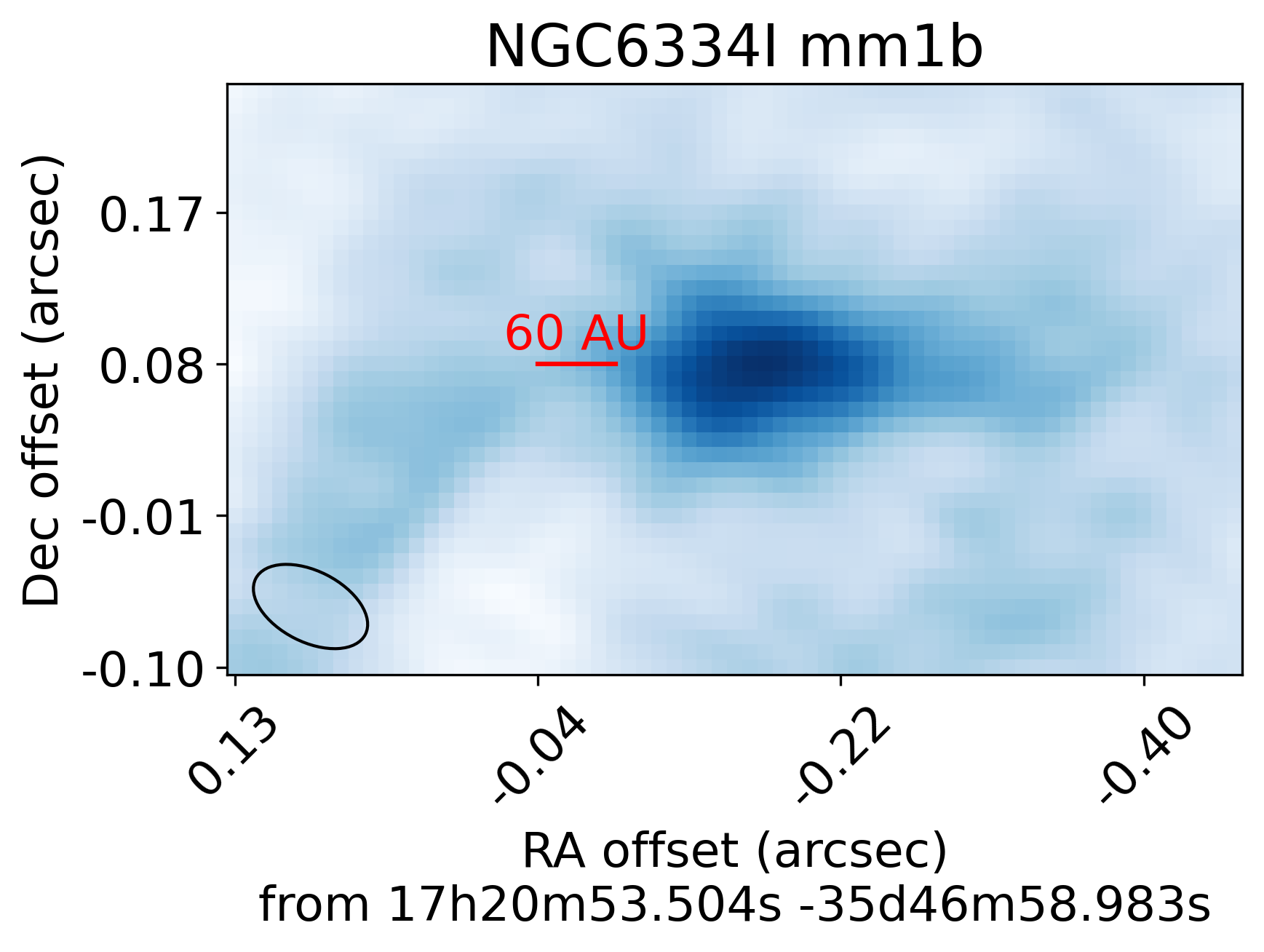}
    \caption{Moment-0 (integrated intensity) images of the resolved sources in NaCl lines.
     For SrcI (top-left), this is the integrated intensity of the NaCl $J$=18-17 $v$=0
     line.  For the remainder, G17 (top right), G351 mm1 (bottom left), and NGC 6334I mm1b (bottom right),
     these are the average of the $J$=18-17 and $J$=17-16 transitions
     of both the $v$=0 and $v$=1 states.  The coordinates are given in RA/Dec offset from the
     central position specified under the abscissa.
     The scalebars show physical sizes as labeled.
     The ellipses in the corners show the full-width half-maximum beam ellipse.}
    \label{fig:resolveddisks}
\end{figure*}

\begin{figure*}
    \centering
    \includegraphics[width=0.32\textwidth]{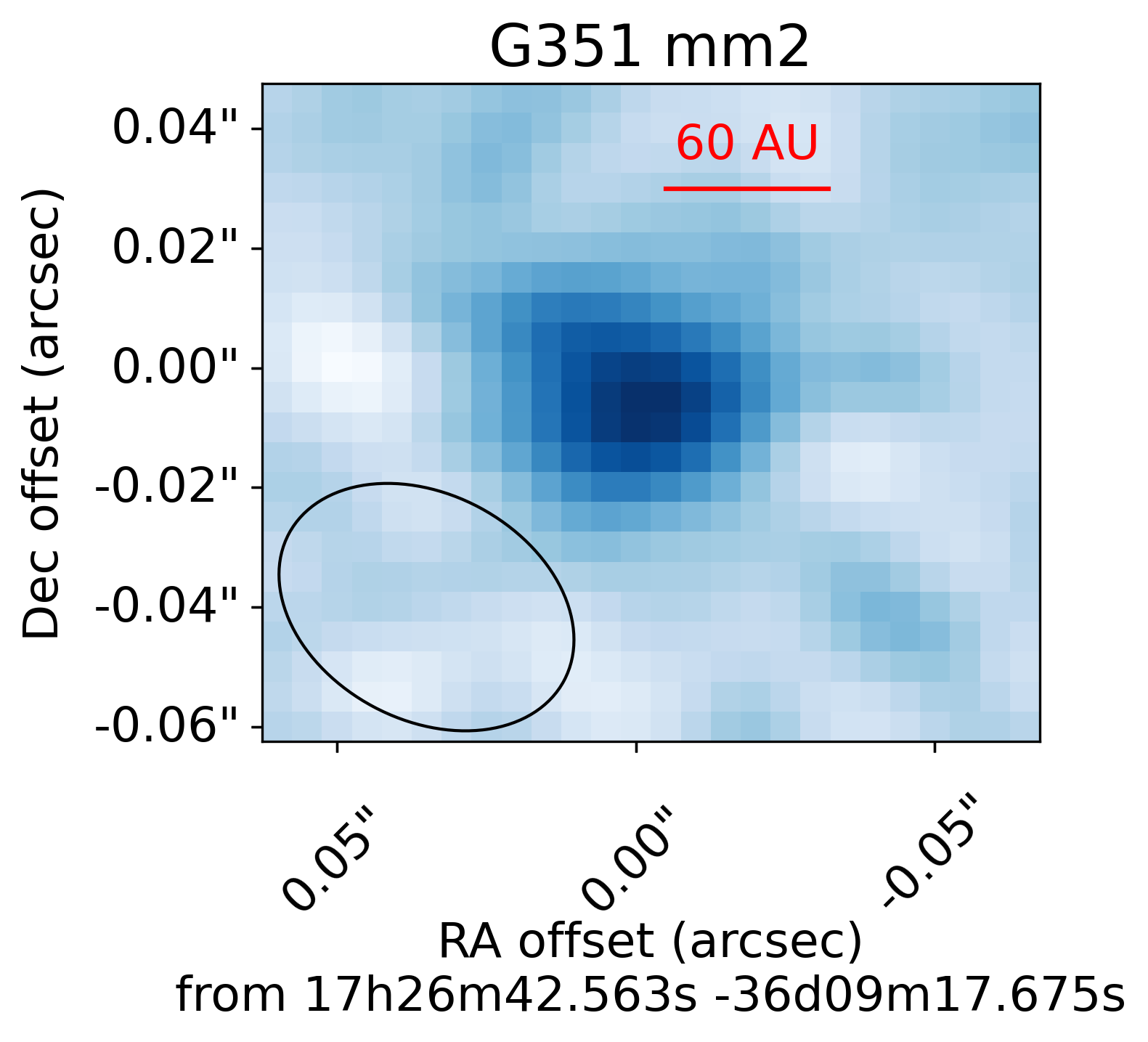}
    \includegraphics[width=0.32\textwidth]{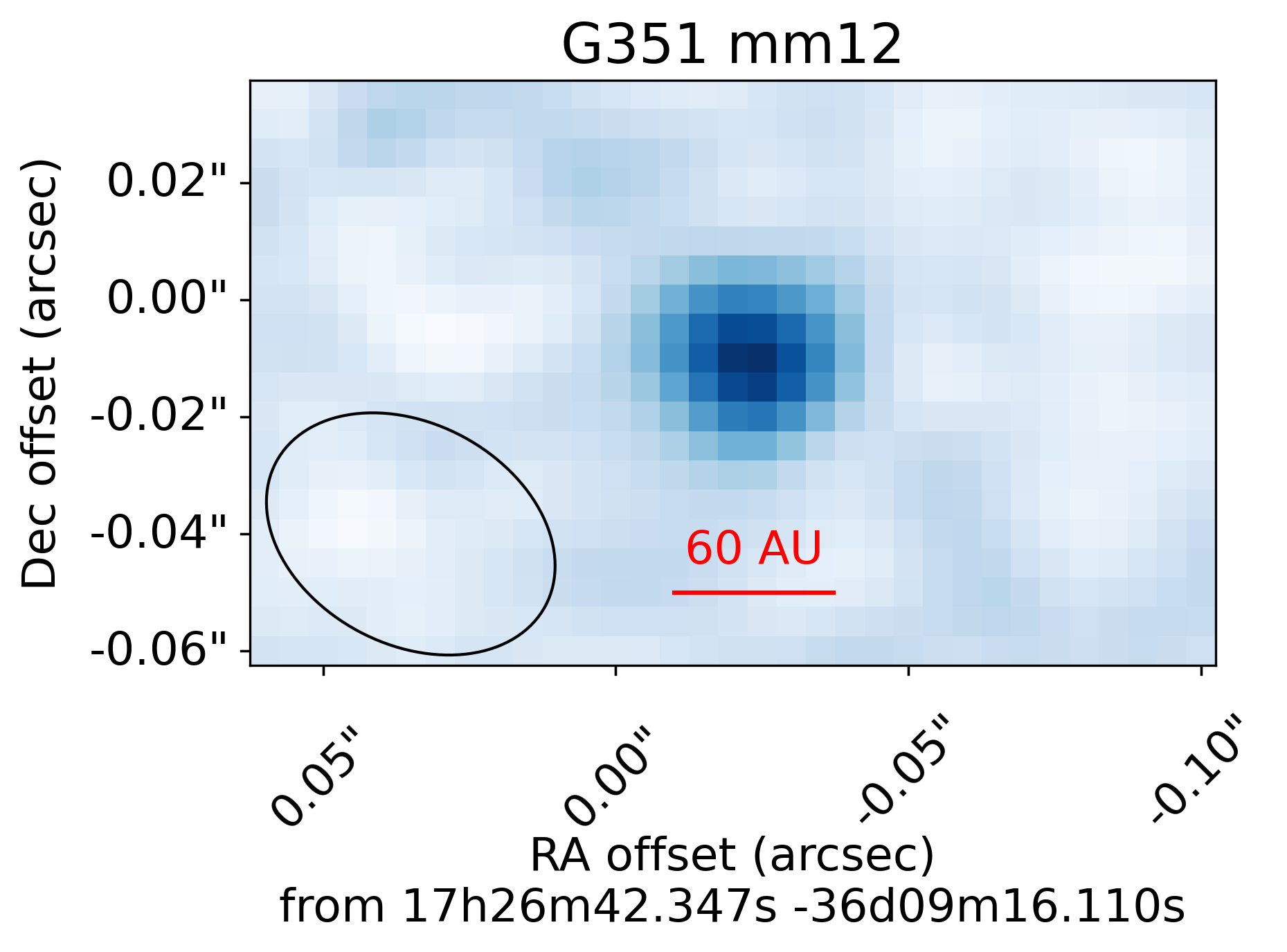}
    \includegraphics[width=0.32\textwidth]{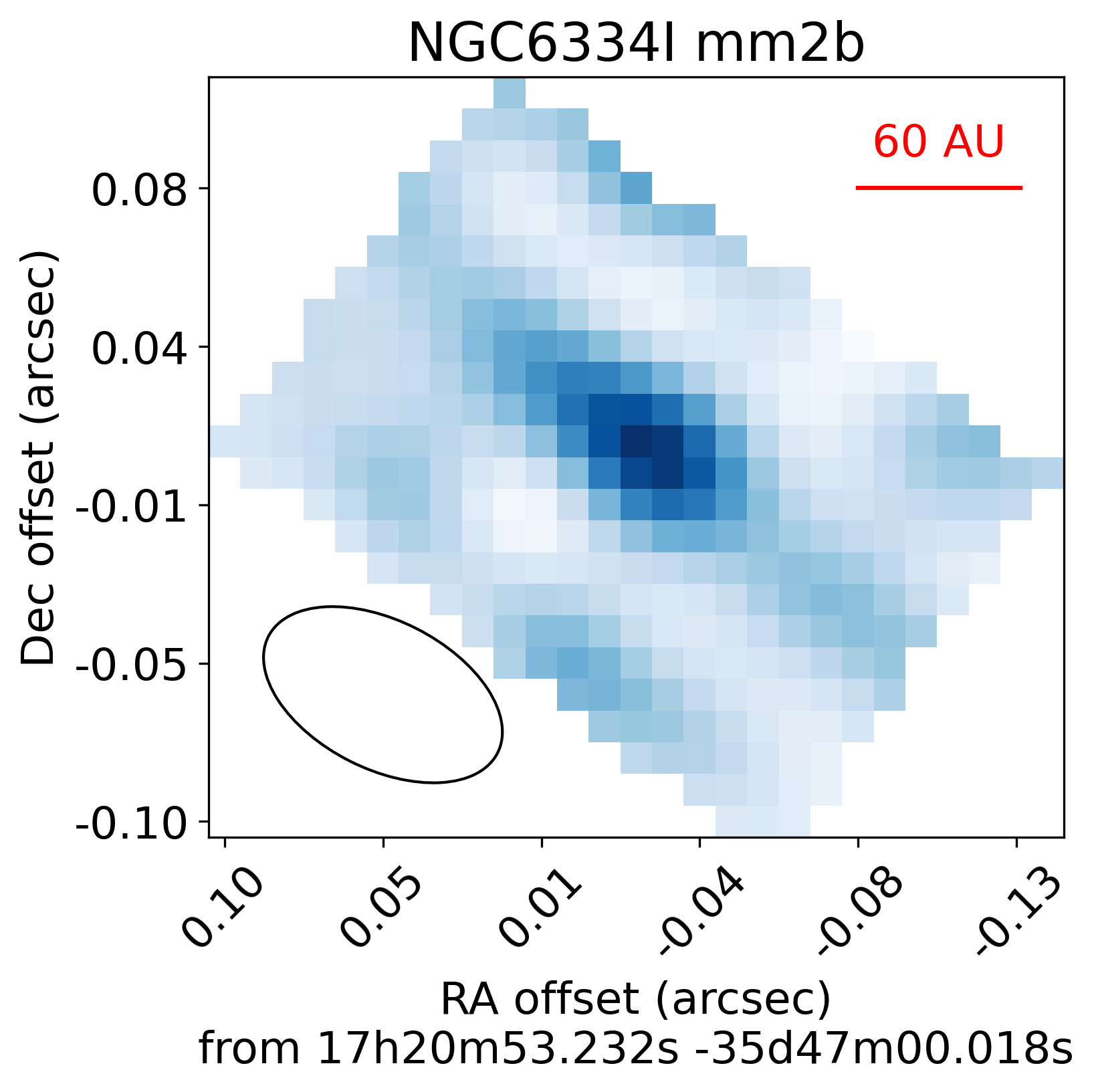}
    \includegraphics[width=0.32\textwidth]{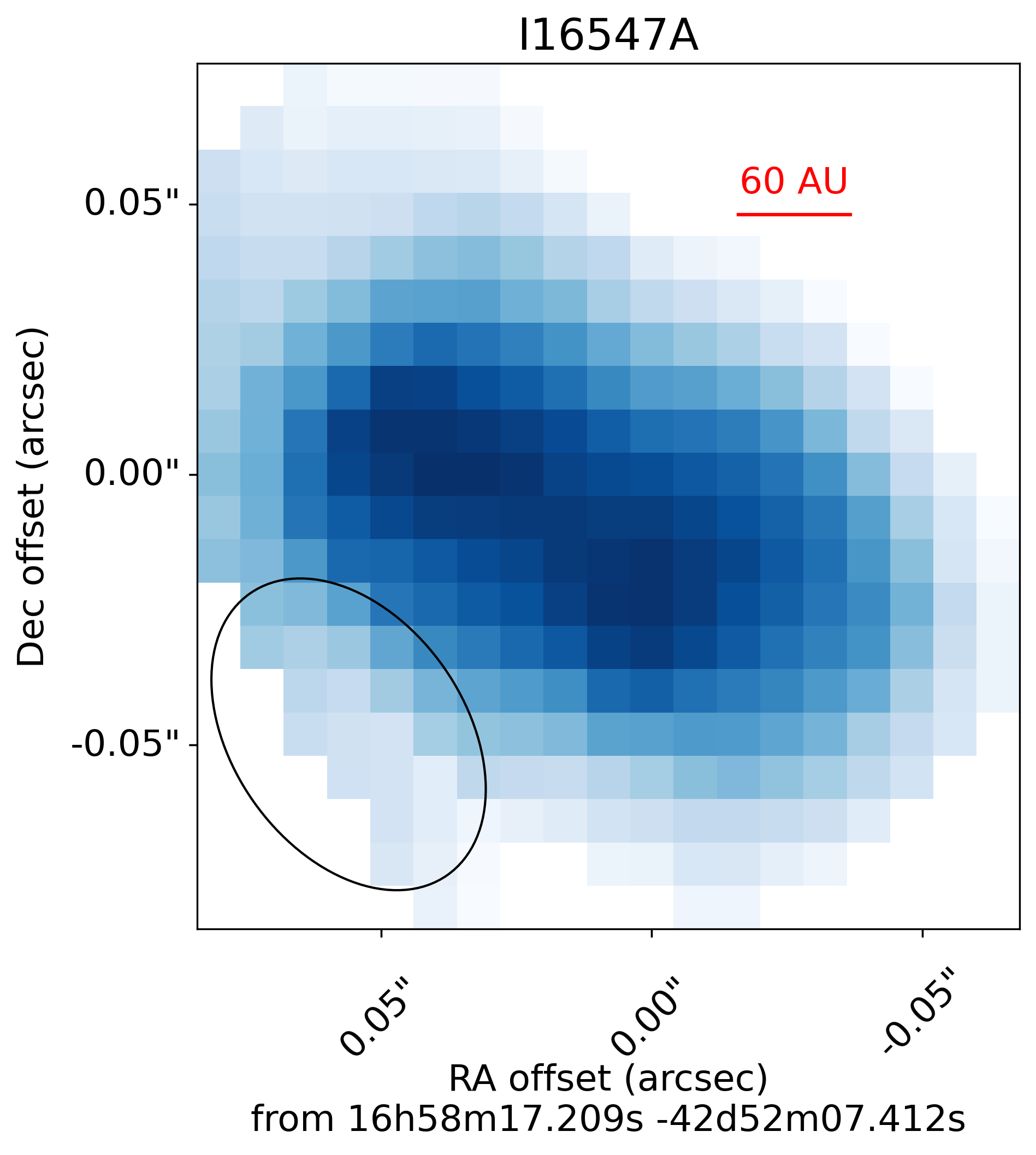}
    \includegraphics[width=0.32\textwidth]{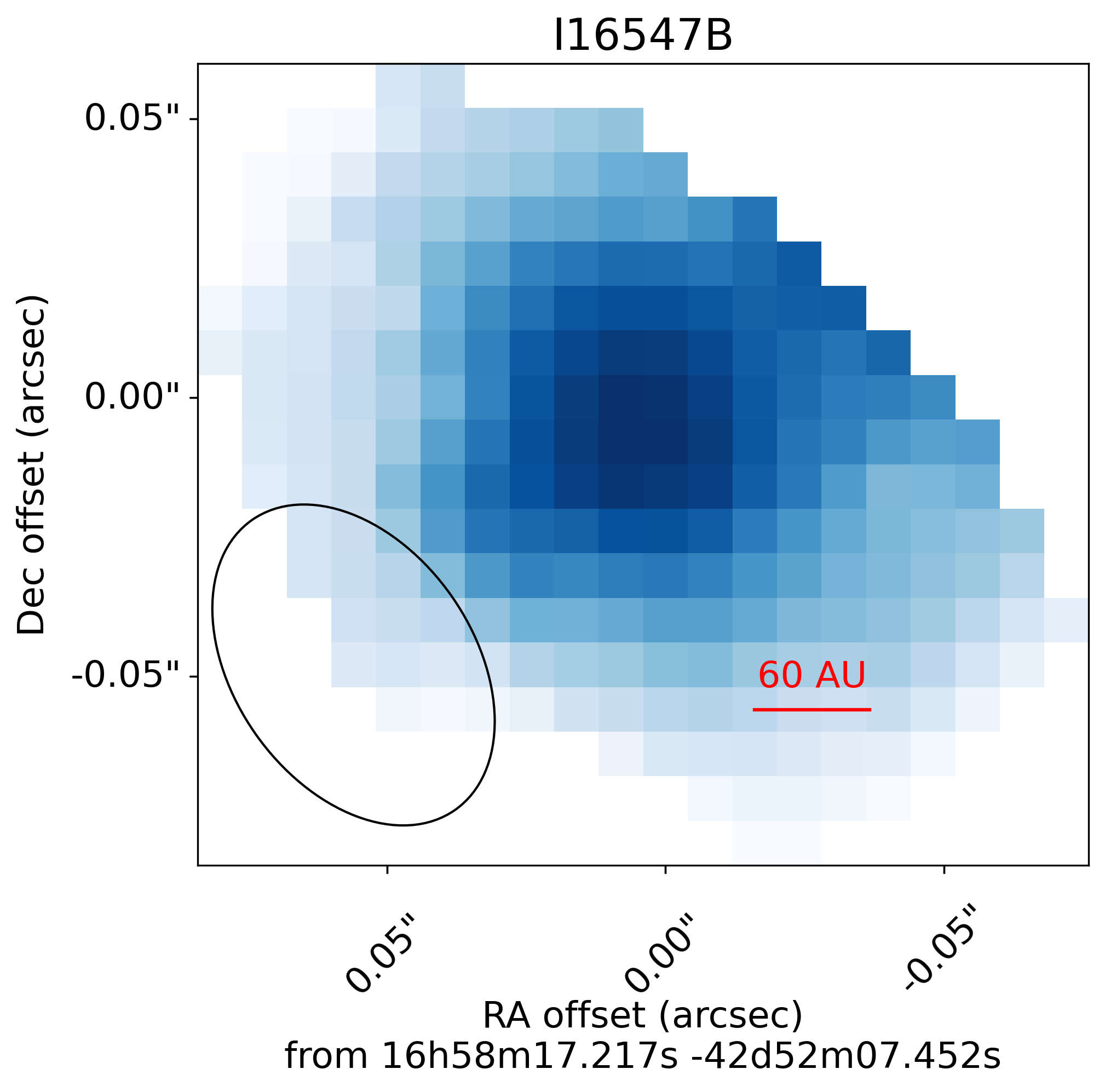}
    \includegraphics[width=0.32\textwidth]{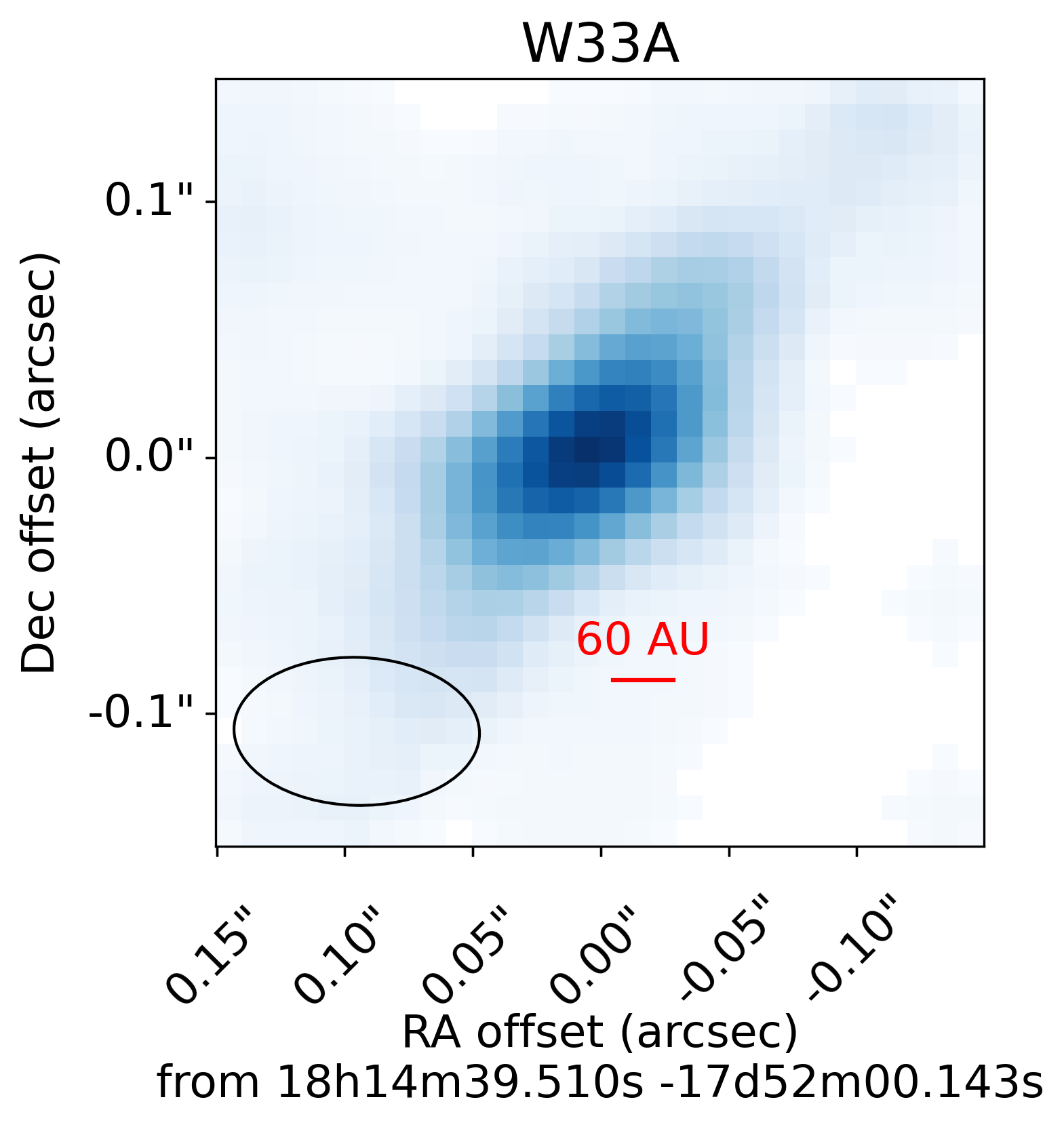}
    \caption{Integrated intensity (moment-0) images of the NaCl stacked lines toward G351mm2 (top left), G351mm12  (top middle), NGC 6334I mm2b - {which is only a tentative detection} (top right), I16547A (bottom left) and B (bottom center), and W33A (bottom right).
    These disk {candidates} are either unresolved or marginally resolved.}
    \label{fig:i16547}
    \label{fig:ngc6334i_m0}
    \label{fig:g351mm12m0}
    \label{fig:unresolveddisks}
\end{figure*}

{The main result is the detection of the `brinary' lines toward the \ndisks sources (of which \ntentative are tentative detections) shown in Figures \ref{fig:resolveddisks} and \ref{fig:i16547}.
These initial figures show moment maps of the NaCl lines as described in Section \ref{sec:saltstack}.
}


\subsection{Line stacking extraction}
\label{sec:linestacking}

From source candidate identification from the continuum data, the analysis forks down two different paths.
We start by searching by eye for emission associated with the \water line, which is quite bright in SrcI, in \S \ref{sec:wateranalysis}, which we then use as a kinematic reference.
If water is not detected, we use a different line as our kinematic reference as described in \S \ref{sec:nowateranalysis}.

\subsubsection{Water-driven analysis}
\label{sec:wateranalysis}
If the water line is detected, we use it to create a `velocity map' following this procedure:
\begin{enumerate}
    \item Cut out a cube containing the region around the estimated central $v_{\rm LSR} \pm 20\:\kms$ in velocity and encompassing only the candidate disk region in space.
    \item Create a peak intensity map of the line in that region.
    \item Create a threshold mask including only the (by-eye) estimated significant emission in the peak intensity map.  Use one iteration of binary erosion and three to seven iterations of binary dilation to remove isolated bright noise pixels and fill back in the mask.\footnote{Binary dilation refers to mathematical morphology operations in which any pixel having value False and a neighbor with the value True is set to True.  Binary erosion is the inverse operation.}
    \item Create a volumetric threshold mask including only pixels above the estimated noise level.  Then, use one iteration of binary erosion followed by one to three iterations of binary dilation to fill in the mask.  As in the previous step, this step is to eliminate isolated bright pixels, but because it is in 3D, a different threshold can be adopted.
    \item Create a moment-1 (intensity-weighted velocity) map of the spatially and volumetrically masked data cube.
\end{enumerate}

The erosion and dilation steps are performed to exclude isolated bright pixels and to maximally include all pixels associated with source emission.

We then use the velocity map to stack the spectra obtained across the disk candidate.
Each spectrum {from each spatial pixel} in the cube that has a measured velocity is shifted such that the line peak is moved to 0 \kms.
The spectra are then averaged to produce the stacked spectrum.

This stacking process assumes that all spectra through the {candidate} disk will have similar peak intensity and width but different central velocities; this assumption held well in the SrcI spectrum \citep{Ginsburg2018}.
We assume that the kinematics of our selected line are the same as the lines of interest; this assumption is justified by position-velocity diagrams in \S \ref{sec:g351}.

\begin{figure}
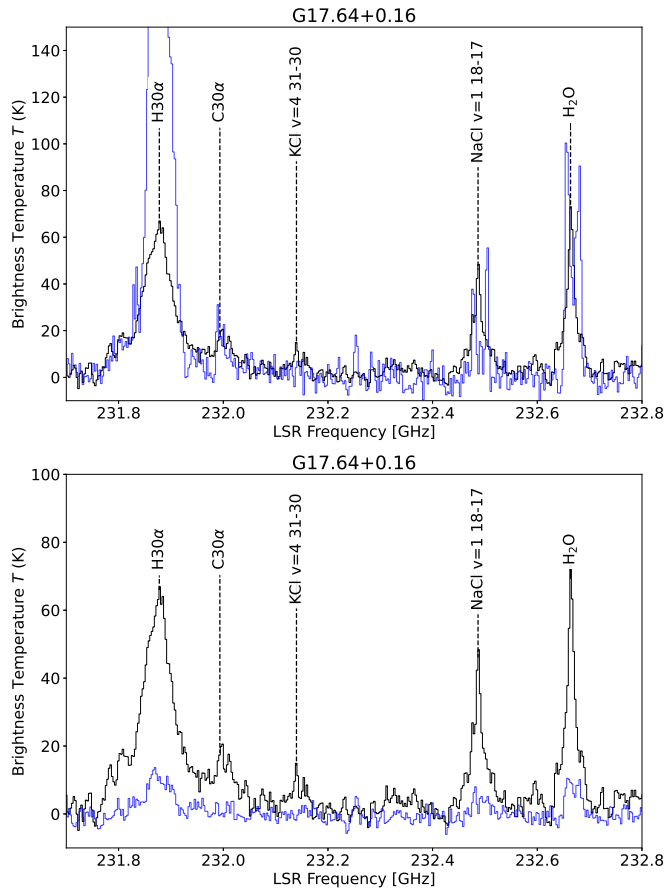

    \centering
    \includegraphics[width=0.49\textwidth]{f11}
    \includegraphics[width=0.49\textwidth]{f12}
    \caption{Demonstration of the utility of stacking analysis toward G17 in one spectral window.
    The black spectrum is stacked by shifting each individual spectrum to match the location of the peak intensity of the \water line; it is the same in both panels.
    The top panel shows the peak intensity spectrum taken over the same area as the stacked spectrum, and the bottom shows the mean spectrum in blue.
    It is clear that the stacked spectrum has clearer features and better signal-to-noise ratios than either the peak or average spectrum.
    }
    \label{fig:G17stacks}
\end{figure}

We demonstrate the advantage of this line stacking approach in Figure \ref{fig:G17stacks}.
The signal-to-noise in targeted lines is substantially increased, and faint lines appear that are otherwise missed.
In the example figure, the most obvious case is KCl $v$=4 $J$=31-30, which is not apparent in either the peak intensity or average spectrum, but is strongly evident in the stacked spectrum.
Those sources that are best-resolved - in our sample, G351 and G17 - show the most improvement.

We label the resulting averaged spectra with known NaCl and KCl transitions and selected other lines, including those of SiO, SiS, \water, H30$\alpha$, and several prominent other molecules (see Table \ref{tab:lines}).
We then use these plots to populate the detection table, Table \ref{tab:similarities}.
We regard the lines as firm detections only if they:
\begin{itemize}
    \item Are prominent, bright, and relatively isolated (e.g., \water, as in Figure \ref{fig:g17diskspectrum}) 
    \item Exhibit an appropriately broad linewidth (H30$\alpha$ is expected to have $\sigma_{v}\gtrsim5$ \kms because it comes from hot plasma ($T\sim10^4$ K), while molecular species are expected to be narrower, coming from gas at $T<1000$ K). 
\end{itemize}
For species for which we label the vibrationally excited transitions, we consider only the $v$=0, $v$=1, and $v$=2 levels, as higher levels are expected to be weaker if present.

There are several cases where a compelling detection of one line of a species (NaCl or KCl) is detected, but an adjacent state is not detected.  For example, the $J$=18-17 $v$=1 line is detected, but the $J$=16-17 $v$=1 line is not.  In these cases, we regard the detection as tentative and note it in Table \ref{tab:similarities} with an asterisk.  These were cases in which the line may still be present, but may be hidden by either blending with neighboring species or absorption by a line in the surrounding core.

We report detections qualitatively rather than quantitatively because the detections are generally obvious and high signal-to-noise.
When a detection is ambiguous, it is not because of high noise but because of confusion with neighboring lines (e.g., $^{41}$KCl $J$=31-30 {is confused with NaCl $v=1$ $J=18-17$}) or absorption by non-disk material (e.g., KCl $v$=4 $J$=31-30). %  in Fig. \ref{fig:g351mm1}

\subsubsection{Water nondetection-driven analysis}
\label{sec:nowateranalysis}
The path for water non-detections is less linear.
If we are unable to clearly identify emission in the \water line, we search for other lines to use as the basis for stacking.
We first create a simple averaged spectrum over the selected disk candidate's emitting region by examining several different lines in the cube.
We then search for other plausible guiding lines, focusing on those that exhibit gradients in the direction expected given known outflows (i.e., we look at lines rotating perpendicular to outflows).
We look at SiS, SO, and, when truly desperate, \methanol lines.
If we are able to identify a reasonably disk-like line from among these lines, we use it to produce a velocity map as above (\S \ref{sec:wateranalysis}).
If, after this search, we are still unable to find a line that traces disk-like kinematics, we
{remove the source from further consideration}.

For both water- and non-water-driven stacking, we report the achieved noise level in Table \ref{tab:obsstats}.

\subsubsection{Cube stacking}
\label{sec:saltstack}
When NaCl is detected, it is generally seen in multiple transitions.
The NaCl $v$=0 and $v$=1 $J$=18-17 and $v$=1 and $v$=2 $J$=17-16 transitions are present in {the DIHCA} observational setups and are not too badly contaminated by neighboring lines.
We therefore create `stacked' NaCl cubes by cutting out cubes centered on each of those transitions, smoothing them to a common beam, regridding them to a common spectral resolution, and then averaging the cubes.
We use these stacked cubes for further analysis of the NaCl lines, producing both moment-0 images and position-velocity diagrams {(except for SrcI, for which the signal-to-noise ratio in individual lines was high enough to produce moment maps without stacking)}.
The stacked cubes have higher signal-to-noise than the individual cubes and de-emphasize contaminant lines that are adjacent to the target lines, since the contaminants arise at different relative velocities for each {NaCl line (for example, while the $^{41}$KCl 31-30 line is 13 \kms to the red of the NaCl $v=1$ $J=18-17$ line, there is no line 13 \kms to the red of the three other NaCl lines included here).}
This stacking approach is not strictly necessary for use or analysis of the NaCl lines, but it aesthetically improves the resulting images and makes visual inspection and comparison more straightforward.

{We created these cubes whether or not we first noted an NaCl detection in the spectrum.
In cases where no detection was apparent in the stacked spetrum, we nevertheless checked the stacked NaCl cubes to see if extended emission at the expected velocity was apparent.
No additional detections were obtained through this approach.}

\subsubsection{COMs}
\label{sec:COMS}
We identify the presence of complex organic molecules (COMs) in the spectrum in a very broad-strokes manner.
We do not identify specific species, though we note that \methanol and CH$_3$CN are commonly detected, but instead characterize the spectra by the richness of the `line forest.'
For each observation, in the spectral range 219.2-220.8 GHz (which is COM-rich and includes the CH$_3$CN {$J=12-11$ ladder}), we measure the per-pixel noise ($\sigma_{\rm cube}$) by obtaining the standard deviation over the full field of view over all pixels; this approach slightly overestimates the noise because it includes signal in the noise estimate.
For image cubes that were not already continuum-subtracted, we estimate, and then subtract, the continuum by performing pixel-by-pixel sigma clipping to 3-$\sigma$, then taking the median across the spectral axis \citep[i.e., as in][]{Sanchez-Monge2018}.
Then, for each extracted region around a candidate disk, we average the spectra within that region, then determine what fraction of the spectrum exceeds five times the expected noise level, where the expected noise level is $\sigma_{\rm cube} n_{\rm beams}^{-1/2}$, where $n_{\rm beams}$ is the number of beams included in the averaging area.
Note that we only search for COMs in emission; absorption by C- and O-bearing species is seen toward most sources, but is not directly associated with the {candidate} disk, instead, it likely comes from the surrounding envelope or molecular cloud.

Table \ref{tab:obsstats} gives a summary of these statistics in addition to general properties of the data.

\subsection{Line Identification}

{
We briefly discuss the key lines used for identification of NaCl, KCl, PN, SiS, and SiO in this section.
The summary of lines considered is in Table \ref{tab:lines}.
} %% was tab:lines that is the correct reference.
{
The NaCl $v$=0 and $v$=1 $J$=18-17 and $v$=1 and $v$=2 $J$=17-16 transitions are present in {the DIHCA} observational setups.
The $v=1$, $J=18-17$ line is very close to the $^{41}$KCl 31-30 line, but the latter can be ruled out as a contaminant because the $^{41}$KCl 29-28 is also included in the observations.
In Orion SrcI, the peak intensity ratio was NaCl $v=1$ $J=18-17$ $\approx5\times$ $^{41}$KCl 29-28.
}

\begin{table}[htp]
\centering
\caption{Summary of spectroscopic lines used in this analysis}
\begin{tabular}{ccc}
\hline \hline
Line Name & Frequency & E$_U$ \\
 & $\mathrm{GHz}$ & $\mathrm{K}$ \\
\hline
SiS v=1 12-11 & 216.757603 & 1138.75 \\
SiO v=0 5-4 & 217.104980 & 31.26 \\
KCl v=4 29-28 & 217.228912 & 1733.21 \\
$^{41}$KCl 29-28 & 217.543178 & 156.71 \\
SiS v=0 12-11 & 217.817644 & 67.95 \\
NaCl v=2 17-16 & 217.979967 & 1128.38 \\
H$_2$CO $3_{0,3}-2_{0,2}$ & 218.222192 & 20.96 \\
HC$_3$N 24-23 & 218.324788 & 1084.99 \\
CH$_3$OH $4_{2,2}-3_{1,2}$ & 218.440063 & 1422.49 \\
KCl v=3 29-28 & 218.579708 & 1345.06 \\
C$^{18}$O 2-1 & 219.560354 & 15.81 \\
NaCl v=1 17-16 & 219.614936 & 614.51 \\
HNCO $10_{10}-9_9$ & 219.798274 & 58.02 \\
KCl v=2 29-28 & 219.936113 & 671.38 \\
SO $6_5-5_4$ & 219.949440 & 34.98 \\
K$^{37}$Cl v=2 30-29 & 221.078543 & 948.25 \\
$^{41}$KCl v=1 31-30 & 231.088150 & 572.25 \\
K$^{37}$Cl 31-30 & 231.218839 & 177.68 \\
H30$\alpha$ & 231.900928 & - \\
C30$\alpha$ & 232.016632 & - \\
KCl v=4 31-30 & 232.163002 & 1755.14 \\
$^{41}$KCl 31-30 & 232.499840 & 178.67 \\
NaCl v=1 18-17 & 232.509950 & 625.67 \\
Si$^{33}$S 13-12 & 232.628545 & 78.16 \\
H$_2$O v$_2$=1 $5_{5,0}-6_{4,3}$ & 232.686700 & 3461.91 \\
K$^{37}$Cl v=4 32-31 & 232.907553 & 1739.18 \\
PN v=1 J=5-4 & 233.271800 & 1937.30 \\
KCl v=3 31-30 & 233.605698 & 1367.12 \\
NaCl v=0 18-17 & 234.251912 & 106.85 \\
SiS v=1 13-12 & 234.812968 & 865.40 \\
PN J=5-4 & 234.935663 & 33.83 \\
KCl v=2 31-30 & 235.055578 & 687.16 \\
K$^{37}$Cl v=2 31-30 & 235.768235 & 970.53 \\
SiS v=0 13-12 & 235.961363 & 79.28 \\
\hline
\end{tabular}

\par
Lines covered by one or more observations in this work.
The frequency and upper state energy levels are pulled from Splatalogue and refer
either to CDMS, SLAIM, or JPL values.
For KCl v=3 and v=4 values, E$_U$ is drawn from the modified 
version of the \citet{Barton2014} line list used in \citet{Ginsburg2019}.
\label{tab:lines}

\end{table}

\subsection{Salt detections}
\label{sec:saltdetections}

We detect salts in \ndisks sources {(of which \ntentative are tentative)} in \nregions regions.
Figure \ref{fig:resolveddisks} gives an overview of those objects that are spatially resolved.
In the following sections, we describe the salt-bearing sources in more detail:
G17 (\S \ref{sec:g17}),
G351 mm1, mm2, {and mm12} (\S \ref{sec:g351}),
W33A (\S \ref{sec:w33a}),
NGC6334I mm1b and mm2b (\S \ref{sec:ngc6334i}),
and I16547 A and B (\S \ref{sec:i16547}).
The sixth region and disk, Orion SrcI, is not discussed in detail here because the same analysis was already done in \citet{Ginsburg2019}, but it is included in the discussion section below.

\begin{table*}[htp]
    \centering
    \caption{Summary of detections, tentative detections, and non-detections of target species in the source sample}
    \label{tab:similarities}
    \begin{tabular}{ccccccccccc}
    Source         & disk  & H$_2$O &  NaCl & KCl & SiO & RRL & COMs & SiS & SO  & PN  \\
    \hline   
    Orion SrcI     & yes   & yes    &  yes  & yes & yes & no  & no   & yes & yes & ?   \\
    G17            & yes   & yes    &  yes  & yes & yes & yes & no   & yes & yes*& no  \\
    I16547A        & yes-c & yes    &  yes* & no  & yes & no  & yes  & yes & yes & yes \\
    I16547B        & yes-c & yes    &  yes* & no  & yes & no  & yes  & yes & yes & yes \\
    G351.77mm1     & yes-c & yes    &  yes  & no  & yes & no* & no   & yes & no  & yes \\
    G351.77mm2     & unres & yes    &  yes* & no  & no  & yes & no   & no  & no  & no* \\
    G351.77mm12    & unres & yes    &  yes* & no  & yes & no  & no   & yes & no  & yes \\
    W33A mm1-main  & unres & yes*   &  yes* & no  & yes & yes*& yes  & yes*& yes & yes \\
    NGC6334Imm1b   & yes-c & yes    &  yes  & yes & yes & no  & yes* & yes & no* & yes*\\
    \hline  
    G5.89 mm15     & cont  & no     &  no   & no  & ?   & no  & no   & no  & no  & no  \\
    IRAS18162 GGD27& yes   & no     &  no   & no  & ?   & no  & yes  & no  & yes & no  \\
    NGC6334IN SMA6 & no    & yes*   &  no   & no  & no  & no  & yes  & no  & no  & yes*\\
 NGC6334IN SMA1b/d & no    & no*    &  no   & no  & no  & no  & yes  & no  & no  & no* \\
    G11.92mm1      & yes   & no     &  no   & no  & yes & no  & yes  & yes*& yes & yes*\\
IRAS18089  I18089-1732 & yes-c & no     &  no   & no  & yes & no  & yes  & yes*& yes & no* \\
IRAS16562 G345.4938+01.4677      & no    & no     &  no   & no  & yes & yes & ext  & no  & ext & no  \\
    G333.23mm1     & no    & no     &  no   & no  & no* & no  & yes  & no  & yes & yes*\\
    G333.23mm2     & yes*  & no     &  no   & no  & no* & no  & yes  & no  & yes & yes*\\
    G335 ALMA1     & no    & no     &  no   & no  & yes & no  & yes  & no  & no* & no  \\
    G29.96 submm1  & no    & no     &  no   & no  & yes & no  & yes  & no* & no* & no* \\
    G34.43mm1      & no    & no     &  no   & no  & no* & no* & yes  & no  & yes & no* \\
    S255IR SMA1    & no*   & no*    &  no   & no  & yes & yes & yes  & no  & yes & no  \\   
    NGC6334Imm1d   & yes*  & yes*   &  no   & no  & no  & no  & no   & no  & no  & yes*\\
    NGC6334Imm2b   & unres & yes    &  no*  & no  & yes & no  & yes  & yes & yes & yes*\\
    
    \end{tabular}
    \par
    We use `?' to indicate ``not observed'', `yes' for a definitive detection,
    `no' for definitive non-detections, `yes*' for tentative detections, and `no*' for tentative nondetections, 
    where this uncertainty can either be from line confusion or low signal-to-noise,
    and `ext' for those associated with the envelope but not the disk.
    These all refer to detections in emission; COMs are seen in absorption toward many sources, but we
    do not consider these.
    For SiO, we do not attempt to distinguish between SiO from the outflow and from the disk; in Orion SrcI, we know that SiO is present in both.
    For the `disk' column, we either answer `yes' for a clear disk detection {from literature kinematic characterization}, {`yes-c' for a candidate disk toward which a kinematic signature consisten with rotation has been observed, but for which the kinematics have not yet been confirmed to be Keplerian,} `continuum' if a disk-like (linear, $\lesssim300$ au long) feature is seen in the continuum, `unresolved' if the targeted source is too small for us to say, and `no' if neither of the above hold; `no' does not indicate that no disk is present, merely that we did not identify one.  In many cases, we suspect a disk must be present because there is an outflow, but we say `no' if we can't see it.
\end{table*}

\subsubsection{G17}
\label{sec:g17}
\begin{figure*}
    \centering
    \includegraphics[width=0.7\textwidth]{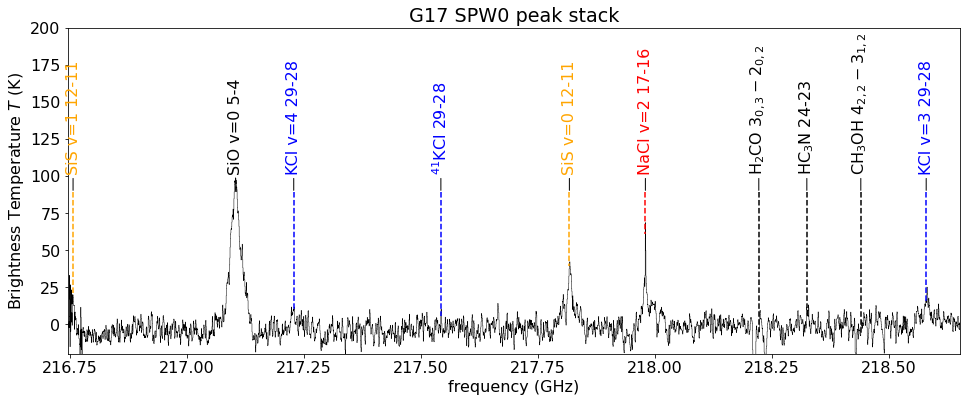}
    \includegraphics[width=0.7\textwidth]{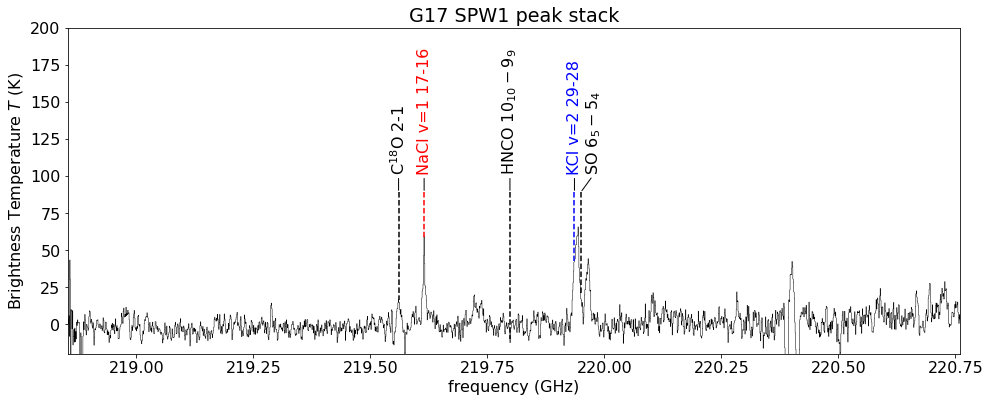}
    \includegraphics[width=0.7\textwidth]{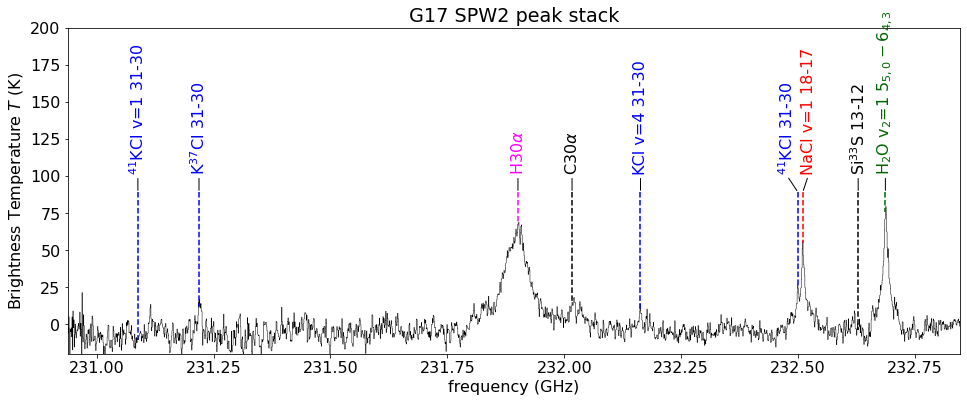}
    \includegraphics[width=0.7\textwidth]{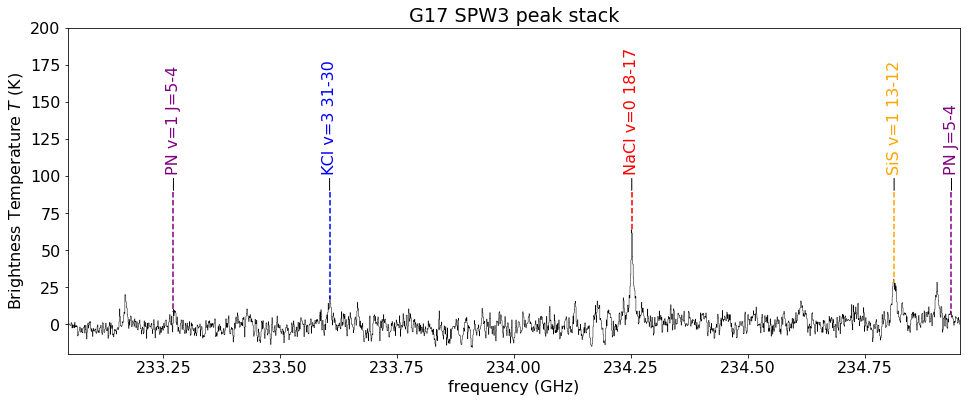}
    \caption{Stacked spectra toward G17 from the \citet{Maud2019} data set.
    The stacking was based on the H$_2$O line.
    Line IDs are shown.
    Different colors are used for targeted species with multiple transitions in-band: orange for SiS, blue for KCl, red for NaCl, {magenta for H30$\alpha$, purple for PN, and green for \water}.
    The remaining species, with only one transition marked, are shown in black.
    }
    \label{fig:g17diskspectrum}
\end{figure*}

The G17 disk is the best resolved source in our sample after SrcI \citep[Fig \ref{fig:resolveddisks};][]{Maud2019}.
{It is a confirmed Keplerian disk \citep{Maud2019}.}
{Because of its high signal-to-noise and well-resolved structure, we investigate it in somewhat more detail than the other sources in the sample.}
{Figure \ref{fig:g17diskspectrum} shows the stacked spectrum based on the water line as described in Section \ref{sec:wateranalysis}.}

\begin{figure*}
    \centering
    \includegraphics[width=0.49\textwidth]{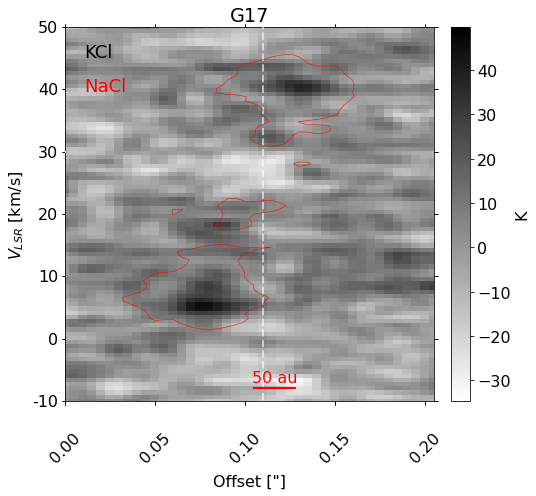}
    \includegraphics[width=0.49\textwidth]{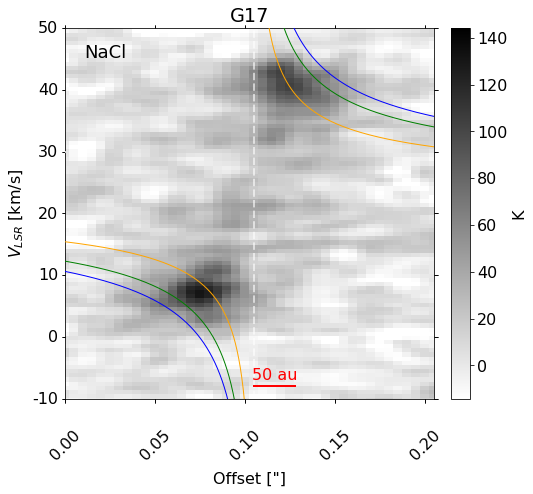}
    \includegraphics[width=0.49\textwidth]{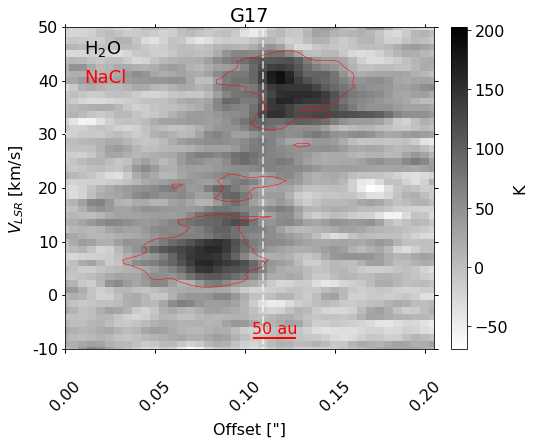}
    \includegraphics[width=0.49\textwidth]{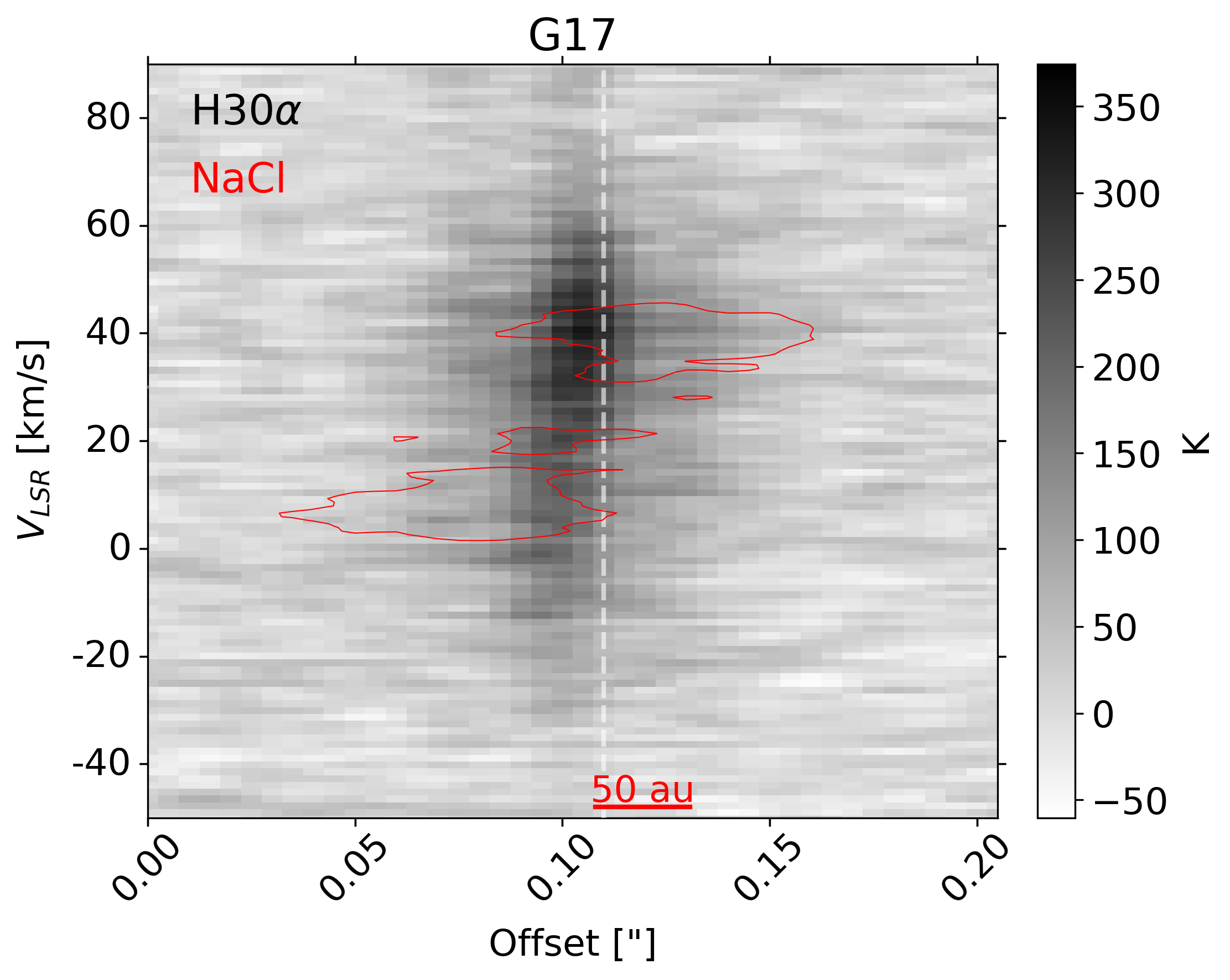}
    \caption{Position-velocity diagrams extracted across the G17 disk.  (top left) KCl grayscale with NaCl contours
    (top right) NaCl grayscale with Keplerian rotation curves drawn for a 15, 30, 40 \msun star with a $i=40^\circ$ disk \citep{Maud2019}
    in orange, green, and blue, respectively.
    (bottom left) Water (H$_2$O) in grayscale with thin NaCl contours.
    {In all cases, the velocity gradient matches that of the disk identified in \citet{Maud2019}.}
    (bottom right) Position-velocity diagram of the H30$\alpha$ line toward G17, with NaCl contours overlaid.
    The H30$\alpha$ emission is clearly confined to within the NaCl disk.
    While there is a hint of a velocity gradient {in the H30$\alpha$ line, it is unclear whether
    this gradient traces rotation.}
    }
    \label{fig:G17rotationcurves}
\end{figure*}

\begin{figure*}
    \centering
    \includegraphics[width=\textwidth]{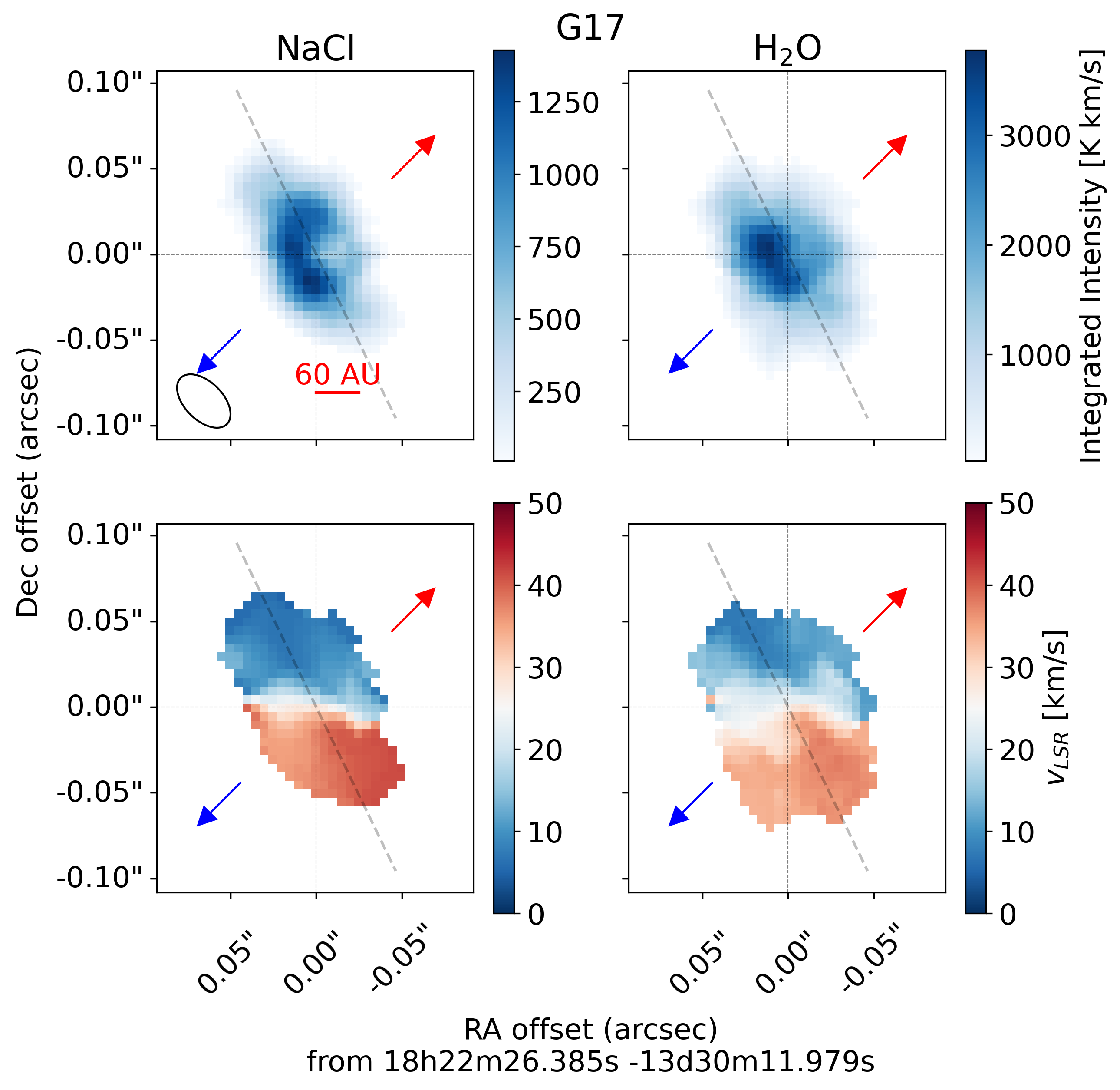}
    \caption{Moment-0 (integrated intensity) and moment-1 (intensity-weighted velocity) images of {stacked} NaCl (left) and \water (right) for G17.
    {The red and blue arrows indicate the outflow direction from \citet{Maud2017}.  The dashed gray line shows the direction the position-velocity diagram was extracted from (Figure \ref{fig:G17rotationcurves}).
    }
    }
    \label{fig:g17moments}
\end{figure*}

The salt detections toward G17 resemble that toward SrcI, with both NaCl and KCl tracing the same rotational structure.
Figure \ref{fig:G17rotationcurves} shows position-velocity diagrams perpendicular to the outflow axis measured by \citealt{Maud2018} {at position angle $\theta\approx135^\circ$ based on the large-scale CO outflow.
\citet{Maud2019} measured the disk position angle to be 25.9$^\circ$, which traces the direction of maximum gradient in the H$_2$O disk and through the emission peak of the continuum structure.
This angle is nearly perpendicular to the large-scale outflow; we adopt this angle as the disk PA.
\citet{Maud2019} measured the disk inclination to be $i=40\pm4^\circ$ from the axis ratio of the continuum image, so we adopt that inclination when overplotting Keplerian curves.
}
As with SrcI, both the water and salt lines trace out orbits consistent with Keplerian rotation around a high-mass star.
{Figure \ref{fig:g17moments} shows integrated intensity (moment-0) and intensity-weighted velocity (moment-1) maps of the G17 disk.}

G17 chemically resembles SrcI in several ways: the only lines seen directly toward the source are \water, NaCl, KCl, SiS, and SiO.  There is little `contamination' in the spectrum from COMs.
As in SrcI, KCl lines are detected at about half the peak brightness of NaCl lines with similar $E_{\rm U}$; no $v$=0, $v$=1, or $v$=2 transitions of KCl are covered by the observations (except KCl v=2 J=29-28, which is blended with SO $6_5-5_4$ and cannot be clearly identified).

\begin{figure*}
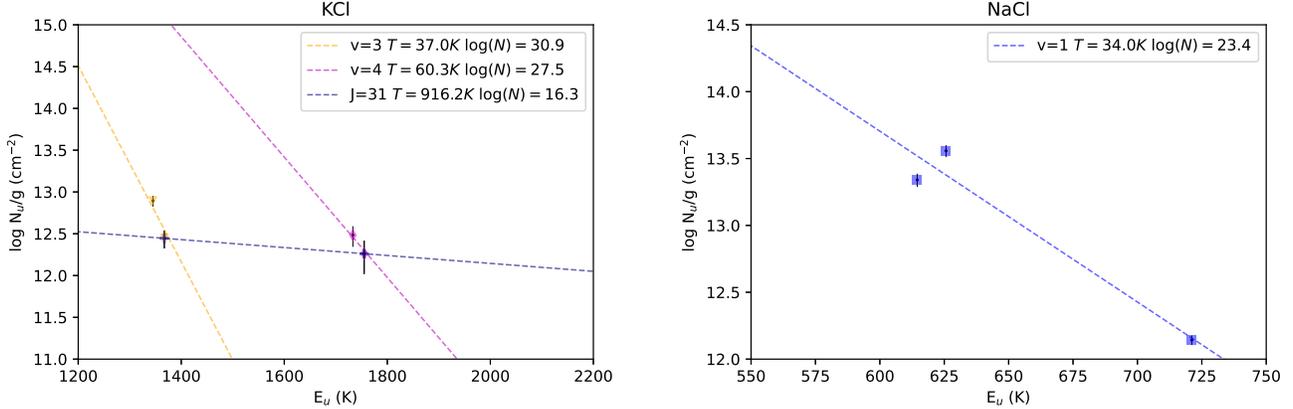

    \centering
    \includegraphics[width=0.49\textwidth]{f22}
    \includegraphics[width=0.49\textwidth]{f23}
    \caption{Rotation diagrams from the fitted lines toward G17.
     (left) KCl rotation diagram
     (right) NaCl rotation diagram}
    \label{fig:G17NaClRot}
\end{figure*}

We measure enough lines to produce both rotational and vibrational diagrams for KCl, but only a rotational diagram for NaCl (Figure \ref{fig:G17NaClRot}).
These plots provide temperature and column density measurements of the target molecules if the observed transitions are in local thermodynamic equilibrium (LTE).
As in SrcI, the rotational temperatures are much cooler ($T_{rot}\sim35-60$ K) than the vibrational temperatures ($T_{vib}\sim900$ K), indicating that non-LTE effects are important.
We have yet to determine the underlying mechanism, but several possibilities are discussed in \citet{Ginsburg2019}.
We defer further discussion of the excitation to a future work in which we will integrate additional transitions.

A key difference between SrcI and G17 is that G17 exhibits radio recombination line (RRL) emission, while SrcI does not.
Figure \ref{fig:G17rotationcurves} shows the H30$\alpha$ RRL in grayscale with NaCl contours on top, showing that the RRL comes from a smaller region contained within the NaCl-bearing disk.

Both SrcI and G17 have central `holes' in which there is no NaCl emission.
In SrcI, \citet{Ginsburg2019} inferred the presence of this hole from the gas kinematics, as the \emph{apparent} hole seen in Figure \ref{fig:resolveddisks}a is an observational effect caused by high optical depth in the edge-on dust disk.
In G17, which is less inclined, the hole is directly observed {(Figure \ref{fig:resolveddisks}b)}. 
Since the radiation field is more energetic in G17 than in SrcI, it is possible that the salt hole is related to the gas temperature.

\subsubsection{G351.77-0.54}
\label{sec:g351}
\citet{Beuther2017} and \citet{Beuther2019} published high-resolution observations of the G351.77-0.54 region,
{which we use here over the lower-resolution DIHCA data}.
We focus on the two brightest mm sources, mm1 and mm2, which both exhibit H$_2$O and salt line emission.
{G351mm2 is a substantially fainter source, but similarly exhibits brine lines.}
{Spectra of the G351 sources and subsequent sources are presented in the Appendices.}

We identify several lines in mm1, including NaCl $v$=1 and $v$=2 $J$=18-17 and SiS $v$=0 and $v$=1 $J$=13-12 (these latter were clearly detected, and used for disk kinematic study, in \citet{Beuther2019}, but were listed as `unidentified').
Because of the slightly different spectral configuration adopted in these observations as compared to the DIHCA observations, both the $J$=12-11 and $J$=13-12 lines of SiS $v$=0 are covered and detected.

The morphology and range covered by NaCl and SiS are very similar (Appendix \ref{appendix:g351}), though the gap in the center for SiS is more pronounced than for NaCl.
These molecules arise in similar, but not identical, regions.

{The velocity gradients in NaCl, H$_2$O, and SiS in mm1 appear to trace a bipolar outflow.}
\citet{Beuther2019} discussed the SiS emission lines in detail, comparing the velocity gradient observed in this line to that seen in SiO.
The SiO $J$=5-4 and CO $J$=6-5 lines both appear in an extended redshifted lobe to the northwest of the source.
The outflow is asymmetric and truncates at the position of mm1, suggesting that mm1 is the source \citep{Beuther2017}.  
Since the redshifted lobe occurs on the redshifted side of the observed SiS velocity gradient, \citet{Beuther2019} interpreted the lines as part of an outflow rather than a disk.
{While there is a blueshifted component opposite the redshifted flow, it is detected in only two channels and is only seen in the beam adjacent to the central source, so its direction cannot be determined independent from the elongated red lobe.}
They also noted the presence of a velocity gradient perpendicular to the outflow direction in CH$_3$CN, suggesting that {CH$_3$CN traces the disk} in a disk-outflow system.
{The observed velocity gradient in briny lines is perpendicular to the CH$_3$CN disk, indicating either that the emission comes from outflow or that the disk changes orientation with scale.}

\begin{figure*}
    \centering
    \includegraphics[width=\textwidth]{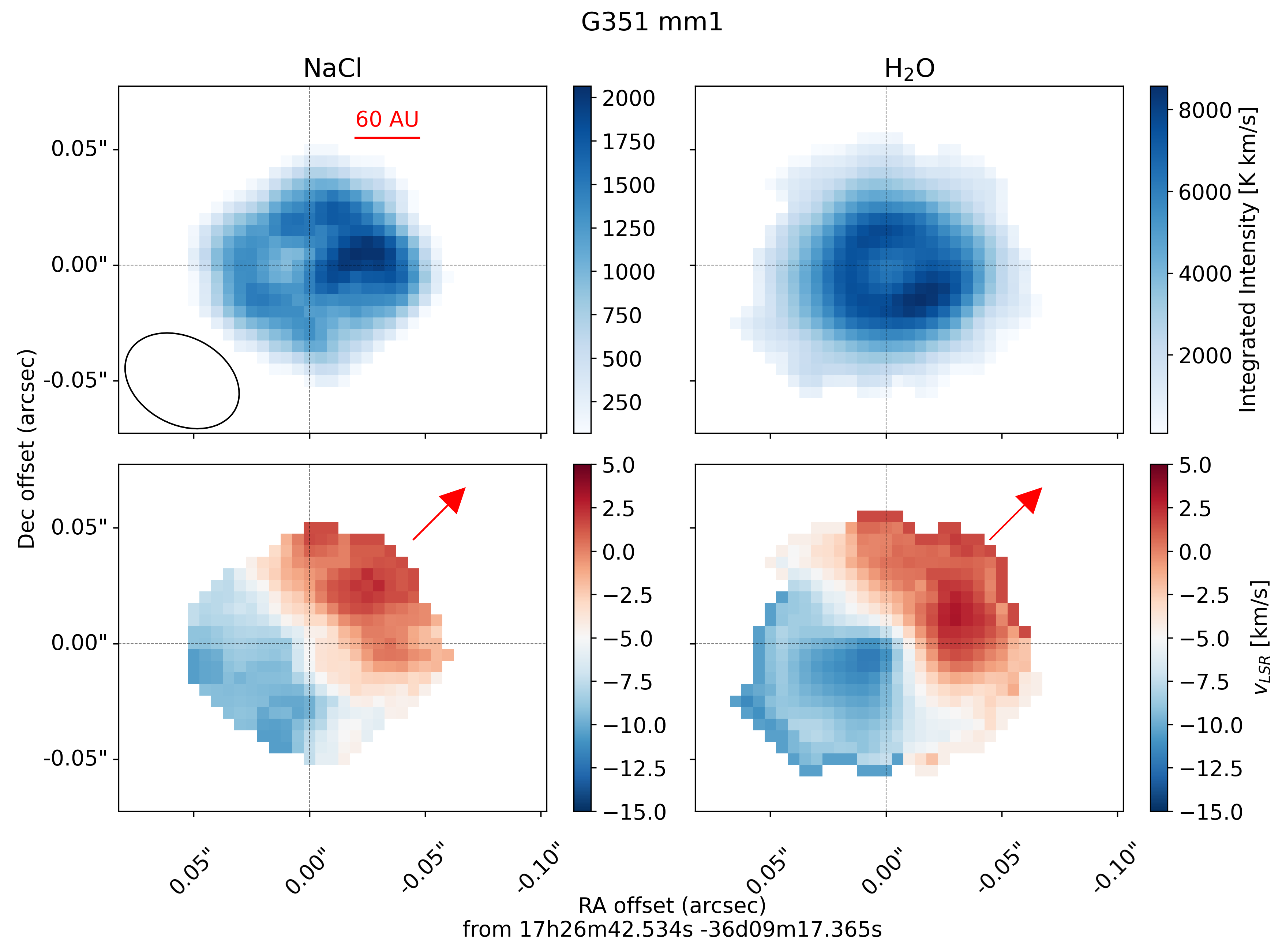}
    \caption{Moment-0 (integrated intensity) and moment-1 (intensity-weighted velocity) images of {stacked} NaCl (left) and \water (right) for G351 mm1.
    The position-velocity diagrams in Figure \ref{fig:g351momentpv} are
    taken along position angle $130^\circ$, parallel to the velocity gradient seen in the lower panels here.
    {The red arrow shows the direction of the outflow from \citet{Beuther2019} {that extends $\gtrsim0.2$ pc to the northwest}.
    A compact blueshifted {feature} was also weakly detected b{in SiO} opposite the redshifted flow, but it is unresolved.
    {The blue side of the NaCl, \water, and (in Appendix \ref{appendix:g351}) SiS and PN corresponds with the SiO blueshifted lobe seen in Fig. 7 of \citet{Beuther2019}, but since its extent is limited to the $\lesssim100$ AU scale shown in this figure, we cannot confirm whether it is comprised of outflowing material.
    }}
    }
    \label{fig:g351brinemoments}
\end{figure*}

We re-examine the kinematics of the now-identified lines here.
Figure \ref{fig:g351brinemoments} shows moment-0 and moment-1 maps of both NaCl (stacked) and \water.
The kinematics resemble a disk, with kinematics consistent with orbital motion around a $\sim20/\sin i$ \msun central potential, 
However, the extended SiO / CO $J$=6-5 outflow feature is parallel, rather than perpendicular, to the axis of the velocity gradient.
Assuming the SiO structure is tracing an outflow, the velocity structure seen in Figure \ref{fig:g351brinemoments} is the base of the outflow.
{The lack of any gradient perpendicular to the outflow direction suggests that the briny lines are not tracing a disk wind in G351mm1, as they are in SrcI \citep{Hirota2017}, since they would retain that rotation signature for at least some distance above the disk.}
{However, the lines are marginally resolved and extended in the direction perpendicular to the outflow, suggesting that the emission arises from an area at least comparable to be beam size ($\sim50$ AU).
The extended launching region is difficult to reconcile with the lack of disklike kinematics.}
We are not able to definitively conclude on the nature of the velocity gradient in mm1 and suggest that it should be studied further at high resolution, particularly to trace the kinematics of the disk-outflow system(s).
{Nevertheless, we note that the briny emission is limited to a region $<100$ AU across.}

By contrast to mm1, mm2 has H$_2$O emission (Appendix \ref{appendix:g351mm2}), but only tentative NaCl emission (the $v$=1 line is detected at $\sim{\mathrm{few}}-\sigma$, but the $v$=2 line is below the noise).
No SiS emission is detected.
However, H30$\alpha$ emission is fairly clearly detected.
{There is a velocity gradient along the NW-SE axis in the \water line {(Appendix \ref{appendix:g351mm2})}.
Curiously, this is {perpendicular to} the direction of the gradient shown in Figure 9 of \citet{Beuther2019}, which shows the unidentified 231.986 GHz line, suggesting that there may be perpendicular gradients here tracing outflow and disk.
However, no outflow is known toward this source, and we do not see clear signs of outflow in any of the lines studied here, including SiO.}
While SiO is detected in emission and absorption toward mm1, it is detected only in absorption toward mm2.

Despite the presence of many KCl transitions in band, there are no detections - but our limits on these lines are relatively weak, as all of the $v$=0 and $v$=1 transitions tend to land in confused regions or come from doubly-rare isotopologues.

A third source, G351mm12, also has salt detected.
Appendix \ref{sec:g351mm12} shows the standard suite of figures for this detection.
{This source is barely resolved.
Its moment maps indicate a hint of velocity gradient, but the gradient is at the limit of our sensitivity and may be spurious.
}

\subsubsection{NGC6334I}
\label{sec:ngc6334i}
\begin{figure}
    \centering
    \includegraphics[width=0.5\textwidth]{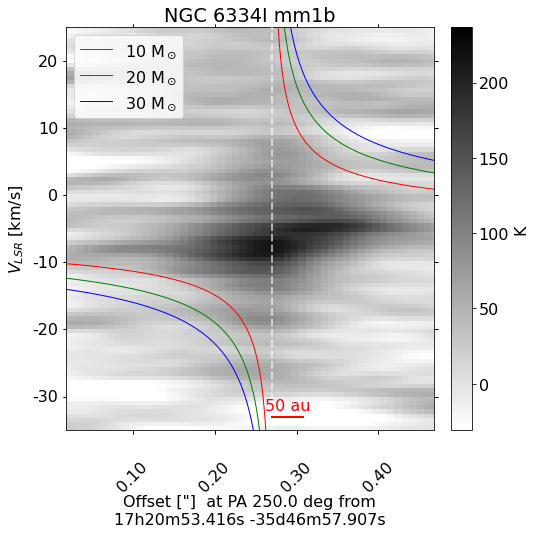}
    \caption{Position-velocity diagram of {NaCl in} the NGC 6334I mm1b disk.
    Keplerian velocity curves for edge-on 10, 20, and 30 \msun central potentials
    are overplotted; the central potential here appears to have $M \sin i<10$ \msun.}
    \label{fig:ngc6334i_mm1_pv}
\end{figure}

There are three sources within the NGC 6334I region that may exhibit brine emission.
Sources mm1b and mm2b are both notable for being faint continuum sources adjacent to bright sources identified in lower-resolution data (the `b' designation indicates that there are sources mm1a and mm2a that are brighter).

mm1b exhibits clear signatures of \water, multiple NaCl lines, SiS, and several likely KCl detections (Appendix \ref{appendix:ngc6334imm1b}). %Figure \ref{fig:ngc6334imm1b}).
The \water line shows {hints of} rotation perpendicular to the outflow axis  (Fig. 8 of \cite{Brogan2018} shows the outflow, which is aligned to $\mathrm{PA}\approx0^{\circ}$ to $-5^{\circ}$, close to straight north-south), though the other lines do not as clearly exhibit this signature.
{\citet{Brogan2018} note several other less prominent outflows centered on this source, however, which hints that this object cannot be interpreted as a single disk-outflow system.}
In the PV diagram {of NaCl (Fig. \ref{fig:ngc6334i_mm1_pv})}, we show curves at 10, 20, and 30 \msun for an edge-on Keplerian disk.
{Since the emission is confined to lower velocities than the 10 \msun curve,} it appears that this source is $<10$ \msun.
{However,} it may also be significantly inclined to the line of sight, {in which case it may be more massive}.
KCl is tentatively detected toward mm1b, with reasonably strong peaks appearing in the KCl $v$=2 $J$=29-28 and \ce{K^37Cl} $v$=0 $J$=31-30 lines.  Some others that might be expected to be bright, e.g., KCl $v$=3 $J$=29-28 and $v$=3 $J$=31-30, are ambiguous or blended.

{The other sources are less clear detections and are therefore considered only candidate brinary sources.}
Similar to mm1b, mm2b has a reasonably clear \water detection and strong signs of SiS $v$=0 $J$=12-11 emission, but it has no clear detection of any salt line (Appendix \ref{appendix:ngc6334imm2b}).
{No clear outflow is present.}
The SiS line profile is broad and somewhat different from that of water, {so it is not obvious that they trace the same kinematics}.

mm1d has only has marginal SiS and \water detections.  
We do not include it in the figures or detection statistics.
It is a strong candidate for follow-up observations.

\subsubsection{W33A}
\label{sec:w33a}
We marginally detect \water, H30$\alpha$, and NaCl toward the bright source at the center of W33A (Appendix \ref{appendix:w33a}).
At the current resolution, the emitting region is unresolved.
The detection of any of these lines individually is tentative {because we have only one firm line detection for each of these molecules and they are potentially blended with other emission lines}.
{The PN 5-4, \water, and NaCl v=1 18-17 lines are the most prominently detected.
The NaCl v=1 and v=2 J=17-16 lines are too weak to confirm the NaCl detection.
While we detect only one PN line, it is isolated enough that confusion is unlikely to affect it, so it is a reasonably firm detection.
}
This source is a prime candidate for further followup.

\subsubsection{I16547}
\label{sec:i16547}
I16547A and B were reported to have salt, water, and SiS emission in \citet{Tanaka2020}.
We confirm their detections both with their original data and with coarser-resolution observations from the DIHCA program.
Despite the clear detection of those three briny species {(see spectra in Appendix \ref{appendix:i16547a} and \ref{appendix:i16547b})}, there is no sign of KCl in the data.
The non-detection is in part driven by confusion, in that many of the lower-$J$ transitions of KCl lay atop transitions from other molecular species that are {spatially extended and} not filtered out.  
Line stacking was not very helpful for these two targets because they are only marginally resolved (Fig. \ref{fig:i16547}).
Nevertheless, {velocity gradients consistent with rotation are} apparent in position-velocity diagrams extracted along the direction of the maximum gradient (Appendices \ref{appendix:i16547a} and \ref{appendix:i16547b}), {which is perpendicular to the outflow direction \citep{Tanaka2020}}, suggesting that these are both likely disks around high-mass YSOs.

\subsection{Unsalted sources}
\label{sec:unsalted}
The remainder of our targets do not have salt or water detections.
We describe them in slightly more detail in Appendix \ref{appendix:unsalted}.
We note here that several of these, i.e., G11.92, GGD27, and I18089, have clear extended disks that are detected in other molecules (e.g., CH$_3$CN) but not in brinary lines.

\section{Discussion}
\label{sec:discussion}

\subsection{Are briny lines disk-only tracers?}
{
Many of our targets are only candidate disk sources, in that no resolved Keplerian rotation curve has been observed.
We therefore consider the question: Could the briny emission be from outflows?
Under the assumption that outflows are driven by either a wind from a disk or from accretion from a disk onto a star, the presence of an outflow still indicates the presence of a disk, but that does not mean the lines we observe necessarily arise within that disk.
}

{
In SrcI, it is clear that the water is partly outflowing, but it arose from a disk wind and had a small observable scale height \citep[$h<40$ au; see Fig. 10 of ][]{Ginsburg2018}.
Even in that case, the water line was dominated by disk kinematics, not outflow.
}

{
In most of the observed sources, a velocity gradient across the lines of interest was observed.
Such a velocity gradient can be produced by either outflow ejection or disk rotation.
In the best-resolved cases, G17 and G351, the emission forms a complete ring, which is expected of a disk or disk wind.
The {circular extent of the} briny emission shows that it is not tracing a collimated jet feature.
{However, in G351mm1, the direction of the gradient is along the known outflow, suggesting that the briny lines do not trace a disk or a disk wind; this source remains difficult to interpret.
}
}

{While we cannot definitively determine the general origin of briny emission, we observe here that it is restricted to radii $<300$ au in our full sample, such that it is always consistent with arising in either the disk or the very inner portion of the outflow.}

\subsection{Chemical similarities between the salt disks}

We examine the general chemical properties of the brinary disks.
Each has several properties in common, but several differences.
Table \ref{tab:similarities} lists the detections and non-detections toward each examined source.
We observe that NaCl, \water, and SiS lines are often detected toward the same sources, and generally if one is absent, all are.
In other words, they exhibit comparable brightness when they are observed.
This correlation hints that they come from similar regions within disks {or outflows}, either because of excitation or chemical (formation/destruction) conditions.

 %We check whether each source exhibits H30$\alpha$ RRL,
 %complex organic molecule (COM), SO, and SiS emission.
 %For the COM check, we simply summarize whether there is or
 %is not a line forest; we do not identify the carrier species,
 %but note that the targeted sources generally exhibit CH$_3$CN
 %and CH$_3$OH emission.  

\begin{figure}[b]
    \centering
    \includegraphics[width=0.5\textwidth]{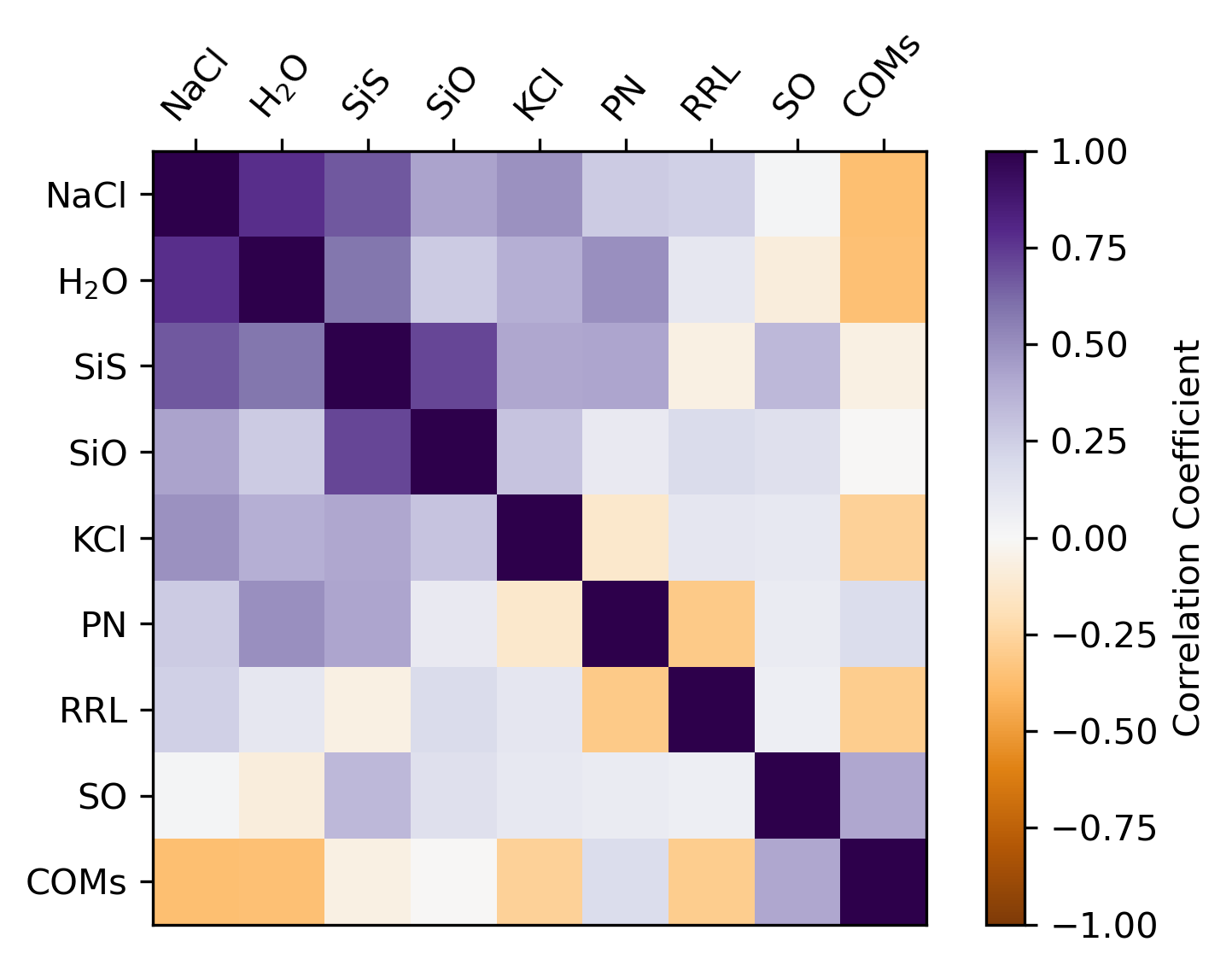}
    \caption{Cross-correlation plot made from Table \ref{tab:similarities}.
    We cross-correlate purely on the boolean value; those with a `yes' are marked
    True, and those without a `yes' are marked False.
    The rows are sorted in order of correlation with NaCl.
        }
    \label{fig:xcorrplot}
\end{figure}

Figure \ref{fig:xcorrplot} shows the cross-correlation of the boolean value (``yes'' or ``no'') encoded in table \ref{tab:similarities};
the tentative detections marked with `*'s are assigned to their corresponding `yes' or `no' prefix.
It shows that the presence of NaCl, SiS, and \water are strongly correlated.
The presence of COMs in the spectra is moderately anti-correlated with these `brinary' species.
RRLs are strongly anti-correlated with COMs.
The correlations among the other molecules (and RRLs) are less pronounced. 

We note that our data are extremely incomplete, as we explicitly select against COM-bearing targets by selecting \water-bearing disks.
There are many fainter disk candidates in the fields of view of our observations that we did not include in our sample or in this plot.
Most of these sources do exhibit COM emission and do not exhibit RRL emission, though, so adding them to this plot would generally strengthen the trends shown.

Because of the incompleteness of our sample, we caution against over-interpretation.
A repeat of this analysis with a consistently-selected sample will be needed to draw firm conclusions.
However, the correlations in the top-left of the plot, between \water, NaCl, and SiS appear firm - these species coexist in this sample of disk candidates.

\subsubsection{The low detection rate of KCl is an observational effect}
The KCl detections are not perfectly correlated with NaCl detections, but we argue here that this is an observational selection effect.
Most likely, KCl is better-correlated with NaCl than is apparent in Figure \ref{fig:xcorrplot}.

In the targeted band, there are no $v$=1 or $v$=0 transitions of the most common isotopologue ($^{39}$K-$^{35}$Cl) of KCl, such that all detections are of higher-excitation, lower-brightness transitions.
The targeted KCl lines are also partly hidden by confusion with other lines.
The KCl lines are strongly anti-correlated with the presence of COMs, which is primarily (or entirely) because of confusion: the KCl lines in this band are weaker than NaCl when they are detected, so they are more difficult to distinguish from the molecular line forest present in dense spectra.
They land inconveniently near bright COM lines in the 215-235 GHz range more often than the NaCl lines.

The comparison to SrcI illustrates some of these effects.
In SrcI, the peak brightness temperature of the NaCl and KCl lines with E$_U < 1000$ K were similar, with $T_{\rm B,max}\sim100-200$ K.
The NaCl $v$=0 lines were no more than twice as bright as the KCl $v$=0 lines.
The SrcI data set was targeted on outflow-tracing lines, including $^{12}$CO $J$=2-1 and SiO $v$=1 $J$=5-4, which both have nearby $v$=0 and $v$=1 KCl transitions, while the DIHCA observations and others presented here chose to target the CH$_3$CN ladder at 220.4 GHz {and therefore did not cover these low-$v$ transitions}.

\subsubsection{PN in the brine}
The PN J=5-4 line appears to be detected in several of our sources.
The PN v=1 J=5-4 transition, at 233.27182 GHz, is not detected.
For the line-poor brinaries, there is little confusion around this line, and its velocity lines up perfectly with that of salts, so this identification is reasonably certain.
For the rest of the sample, the case is less clear; while there are some tentative detections with clear lines at this velocity, the spectra are so rich that we cannot definitively identify PN as the carrier species.
The presence of PN in the same regions as the highly-excited salt lines may suggest that PN occupies a similar location within and binding energy to dust grains.
\citet{Rivilla2020} suggest that PN is present in the cavity walls of an outflow toward AFGL 5142, but that it is released to the gas phase as PH$_3$, and the PN molecules are subsequently formed under the influence of the star's UV radiation.
Given the lack of UV photons in these sources, as indicated by the lack of correlation between PN and RRLs in Figure \ref{fig:xcorrplot}, our data may indicate that PN is present in the grains and not formed in the gas phase.

\begin{figure*}
    \centering
    \includegraphics[width=\textwidth]{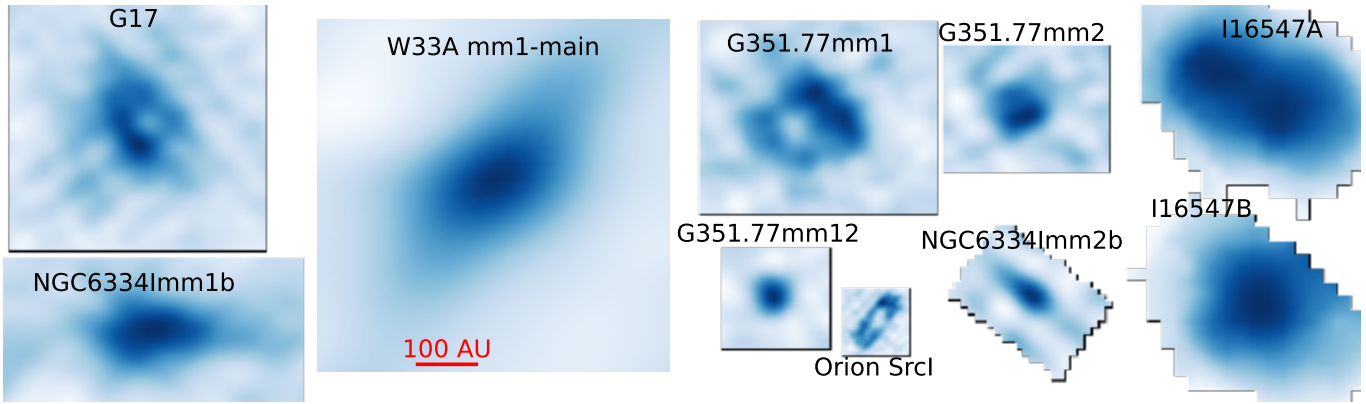}
    \caption{{The integrated intensity (moment-0) maps of NaCl shown in Fig. \ref{fig:resolveddisks} and \ref{fig:unresolveddisks},
    but now with each image scaled to the same physical resolution.  The 100 AU scalebar, shown on the W33A mm1-main panel, applies to all
    panels.  The intensity scales are arbitrary, as the intent is only to show the disk candidate structure.}
    }
    \label{fig:overviewsameres}
\end{figure*}

\subsubsection{RRLs}
{
Hydrogen recombination lines are likely to be produced in the ionized regions surrounding accreting high-mass young stars once they have contracted onto the main sequence.
In our sample, few RRLs are detected.
Their presence, or absence, is only weakly anti-correlated with the presence of COMs and PN, but has little correlation with other molecules.
While we might expect RRL emission to become detectable from accreting HMYSOs toward the end of their accretion phase, as the ionization rate is able to overtake the accretion rate of fresh neutral material, our data provide little evidence for this process. 
}

\subsection{Excitation}
\label{sec:excitation}
At least in the best-resolved cases, G17 and G351, vibrationally excited states of NaCl are detected ($v$=1, and 2).
This feature is in common with SrcI, where states up to $v$=6 were convincingly detected.
The $v$=2 detections in particular, with $E_{\rm U}\gtrsim1000$ K, suggest that an excitation pattern similar to that in
SrcI, in which $T_{\rm vib}>T_{\rm rot}$, is common.
This general feature is also common in evolved stars that exhibit salt emission, suggesting that these salt lines only appear in regions with strong radiative backgrounds in the infrared.
Figure \ref{fig:G17NaClRot} highlights the high excitation, though we defer deeper analysis of the excitation properties to a future work.

\subsection{Spatial Resolution}
\label{sec:spatialresolution}
Our observed sample has non-uniform spatial resolution (Table \ref{tab:obsstats}), which helps explain several of the non-detections.
{We show a version of Figure \ref{fig:resolveddisks} with all images resized to the same physical scale shown in Figure \ref{fig:overviewsameres}, highlighting that the detected NaCl disks are small.}
We did not detect NaCl toward any sources observed with beam size $>300$ au, which included four of our targeted fields.
Five of the fifteen targeted fields included detections.
We performed a logistic regression of the detection of salt against resolution and found that the likelihood of salt detection in our sample is $>50\%$ for resolution $<120$ au, peaking at $\sim80\%$ at infinite resolution.
While our haphazard sample selection prevents drawing strong conclusions, these results suggest that {salty regions} are limited to small scales ($\lesssim300$ au) and are challenging to observe if they are not well-resolved.
This observational limitation highlights the need for more extensive extremely high-resolution observations with ALMA's longest-baseline configurations to obtain dynamical mass measurements for high-mass YSOs.

It is possible that more sensitive observations with poor physical resolution may detect brinary lines now that we know to look for them.
However, the main difficulty in detecting these lines is not raw sensitivity, but confusion with other lines.
It will be beneficial to search other parts of the spectrum for more isolated brinary lines, which might be expected to be more common at lower frequencies.
It may be possible to determine the presence or absence of brinary lines by stacking the various transitions across species to average down the `contaminant' noise provided by other lines; we leave investigation of this possibility to future work.

\section{Conclusions}

\begin{itemize}
    \item We have substantially increased the number of high-mass protostellar objects with detection of salt, hot water, and SiS, increasing the number from three to \ndiskstext published detections across six regions. 
    \item Salt, water, and SiS tend to coexist.  PN may also coexist with these species.  When any of these molecules is detected in a HMYSO disk, all of them are likely to be.
    \item Brinary {sources} are line-poor compared to typical hot cores.  The lack of COMs in the briny regions suggests that the chemistry of these regions is different from hot cores, even when the {central objects} are surrounded by hot cores.
    \item These emission lines do not come from the same volume as ionized hydrogen.  While some of the HMYSO candidates targeted exhibit both RRL and brine emission, the presence of an RRL is a poor predictor of whether NaCl and \water are detected.   The resolved case of G17 shows that the ionized gas comes from a smaller radius than the NaCl.
    \item With the \ndiskstext brinaries presented here, we demonstrate that salt emission is not rare.
    The primary reason it is not often detected is resolution: the emission comes exclusively from small ($\lesssim100$ au) size scales,
    confirming that either chemistry or excitation restricts the millimeter lines described here to disk-sized regions.  Line confusion limits our ability to detect these lines even when they are present, though confusion can be alleviated with high spatial resolution.%(Maybe mention that line confusion hamper the identification of these lines even if they are present. Maybe it could be a new bullet point with technical difficulties: angular resolution, line forest, sensitivity)
    \item However, salt and water emission is also not ubiquitous in HMYSO disks.  Some of the most compelling Keplerian disks around HMYSOs in the literature, such as GGD27 and G11.92mm1, show no sign of salt emission despite their similarities to other disks and superior data quality.  
\end{itemize}

There is clearly substantial future work to do with these data and expanded samples of brinaries.
Some of the more obvious questions about brinary lines include:
\begin{itemize}
    \item Are they correlated with source luminosity or stellar mass?
    \item {What disks, or outflows, produce them?}
    \item Do they occur around low-mass YSOs?
    \item Why are highly vibrationally excited lines ($v>3$) detected?
\end{itemize}
Many of these questions will require establishing less biased, more systematic samples of YSO candidates.  Others will require more detailed, multi-line studies toward a limited sample.

\textit{Acknowledgements}
{We thank the anonymous referee for a thorough and constructive report that significantly improved the paper, particularly in terms of structure and readability.}
AG acknowledges support from NSF AAG 2008101 and NSF CAREER 2142300.
HB acknowledges support from the European Research Council under the European Community's Horizon 2020 frame-work program (2014-2020) via the ERC Consolidator Grant ‘From Cloud to Star Formation (CSF)' (project number 648505). HB also acknowledges support from the Deutsche Forschungsgemeinschaft (DFG, German Research Foundation) – Project-ID 138713538 – SFB 881 (“The Milky Way System”, subproject B01).
P.S. was partially supported by JSPS KAKENHI grant Nos. JP18H01259 and JP22H01271.
K.E.I.T. acknowledges support by JSPS KAKENHI grant Nos. JP19K14760, JP19H05080, JP21H00058, and JP21H01145.

\begin{sloppypar}
This paper makes use of the following ALMA data: \\
{ADS/JAO.ALMA\#2016.1.01036.S},
{ADS/JAO.ALMA\#2017.1.00237.S},
{ADS/JAO.ALMA\#2016.1.00550.S}, 
{ADS/JAO.ALMA\#2019.1.00492.S},
{ADS/JAO.ALMA\#2017.1.00098.S},
{ADS/JAO.ALMA\#2018.1.01656.S},
and
{ADS/JAO.ALMA\#2013.1.00260.S}. ALMA is a partnership of ESO (representing its member states), NSF (USA) and NINS (Japan), together with NRC (Canada), MOST and ASIAA (Taiwan), and KASI (Republic of Korea), in cooperation with the Republic of Chile. The Joint ALMA Observatory is operated by ESO, AUI/NRAO and NAOJ. In addition, publications from NA authors must include the standard NRAO acknowledgement: The National Radio Astronomy Observatory is a facility of the National Science Foundation operated under cooperative agreement by Associated Universities, Inc.
\end{sloppypar}

\software{
The source code underlying this work are available from github at \url{https://github.com/keflavich/saltmining/releases/tag/accepted-2022-11-04}.
This work used 
CARTA \citep{Comrie2021}, 
Jupyter  notebooks \citep{Kluyver2016}.
numpy \citep{vanderWalt2011,Harris2020},
scipy \citep{Virtanen2020},
astropy \citep{AstropyCollaboration2013,AstropyCollaboration2018,TheAstropyCollaboration2022},
spectral-cube \citep{Ginsburg2019b},
radio-beam \citep{Koch2021},
and CASA-6 (\url{https://casa.nrao.edu/casadocs/casa-5.6.0/introduction/casa6-installation-and-usage}).
matplotlib \citep{Hunter2007},
astroquery \citep{Ginsburg2019a},
and pyspeckit \citep{Ginsburg2022}.
}

\bibliographystyle{aasjournal}

\appendix

\section{Additional figures for salted sources}
For the salted sources, we present moment maps and spectra with lines identified in the following Appendices.

\subsection{G351.77mm1}
\label{appendix:g351}
\begin{figure*}
    \centering
    % Why are these in a weird order?
    \includegraphics[width=0.7\textwidth]{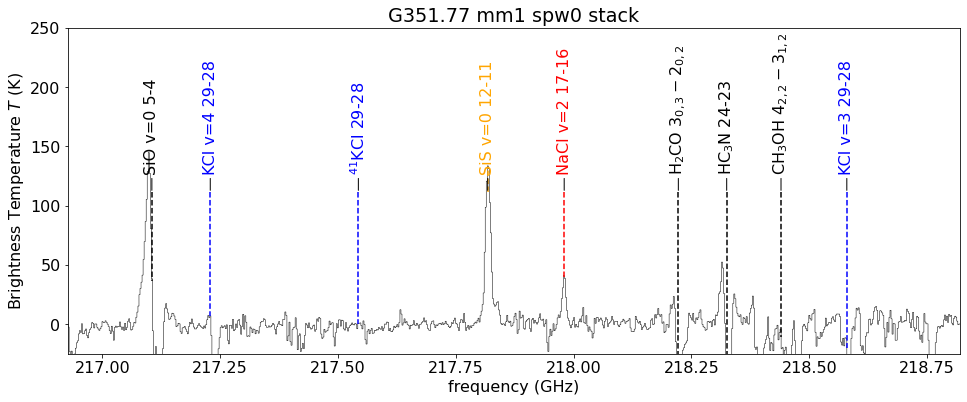}
    \includegraphics[width=0.7\textwidth]{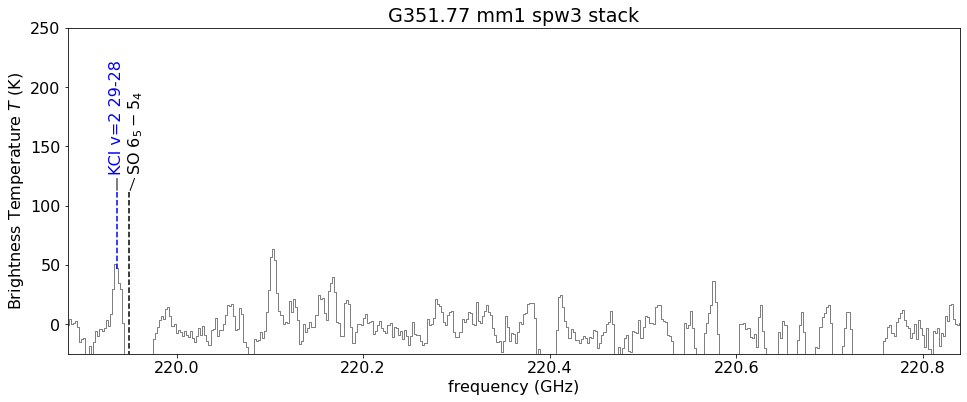}
    \includegraphics[width=0.7\textwidth]{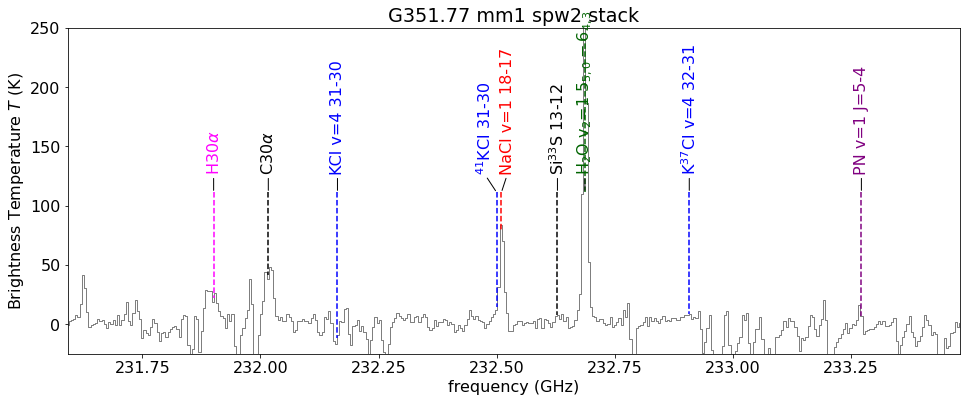}
    \includegraphics[width=0.7\textwidth]{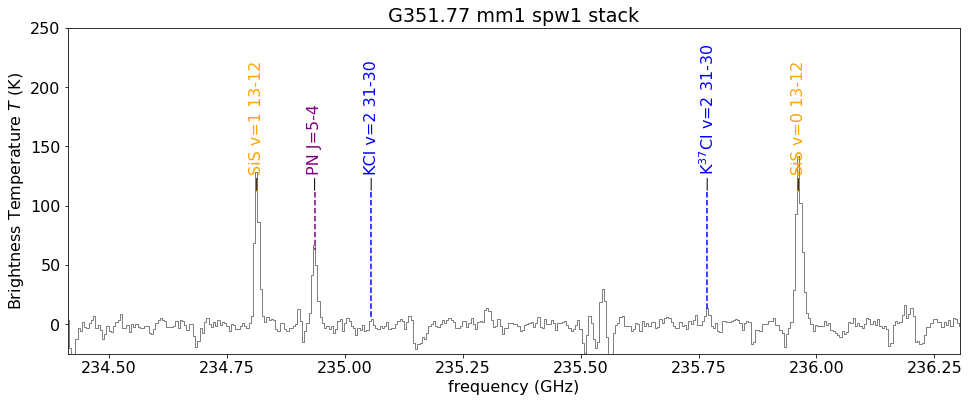}
    \caption{Stacked spectra from G351.77mm1 from the \citet{Beuther2019} data set.
    The stacking was based on the H$_2$O line.
    Line IDs are shown; no KCl detections are clear.
    Different colors are used for targeted species with multiple transitions in-band: orange for SiS, blue for KCl, red for NaCl, {magenta for H30$\alpha$, purple for PN, and green for \water}.
    The remaining species, with only one transition marked, are shown in black.
    The PN line identification should be taken with a grain of salt
    since it is the only transition we observe from PN.
    }
    \label{fig:g351mm1}
\end{figure*}

\begin{figure*}
    \centering
    \includegraphics[width=\textwidth]{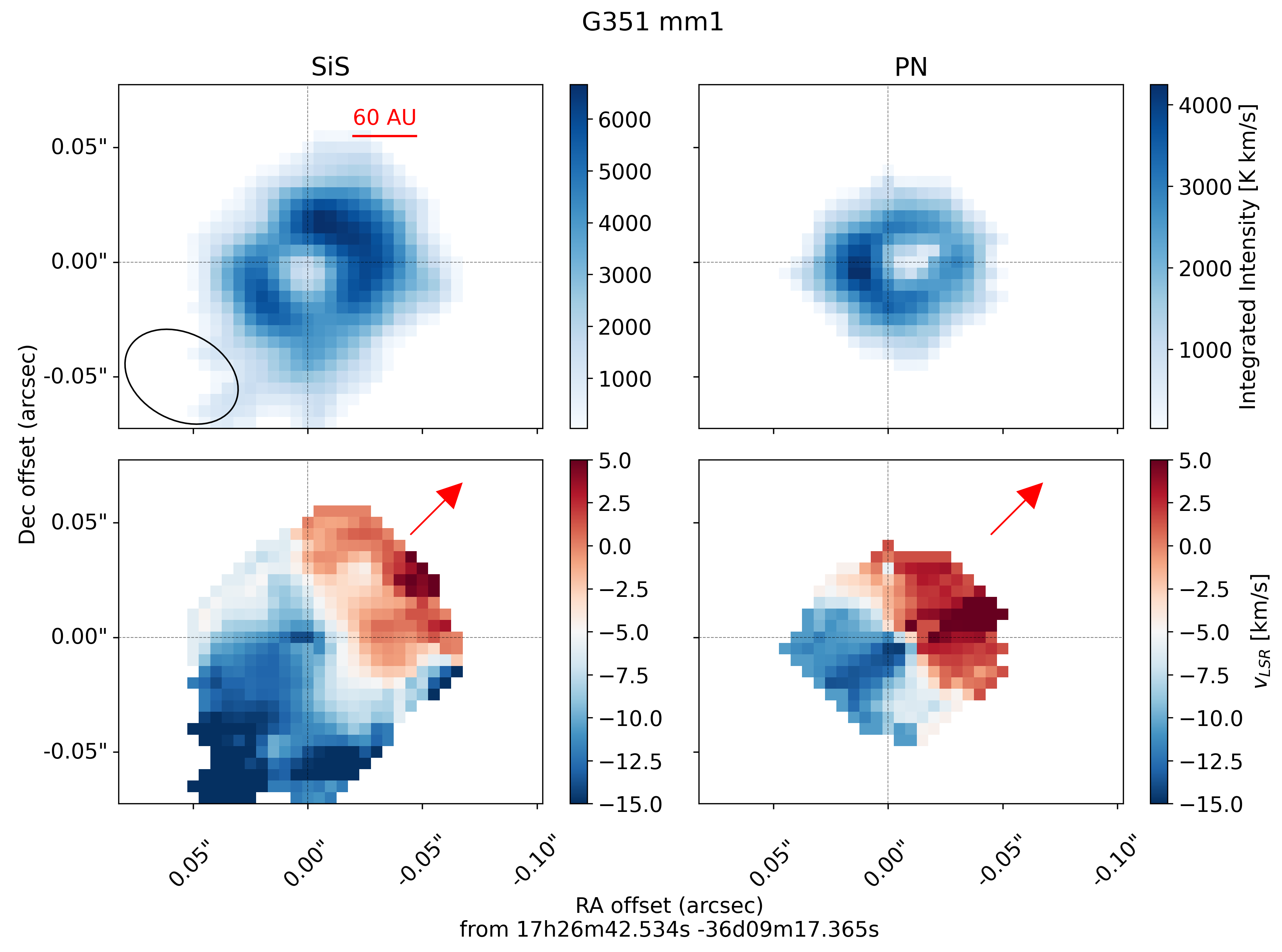}
    \caption{Moment-0 and 1 images as in Figure \ref{fig:g351brinemoments},
    but for the SiS v=0 J=13-12 and PN 5-4 lines of G351 mm1.}
    \label{fig:g351sispn}
\end{figure*}

\begin{figure*}
    \centering
    \includegraphics[width=0.49\textwidth]{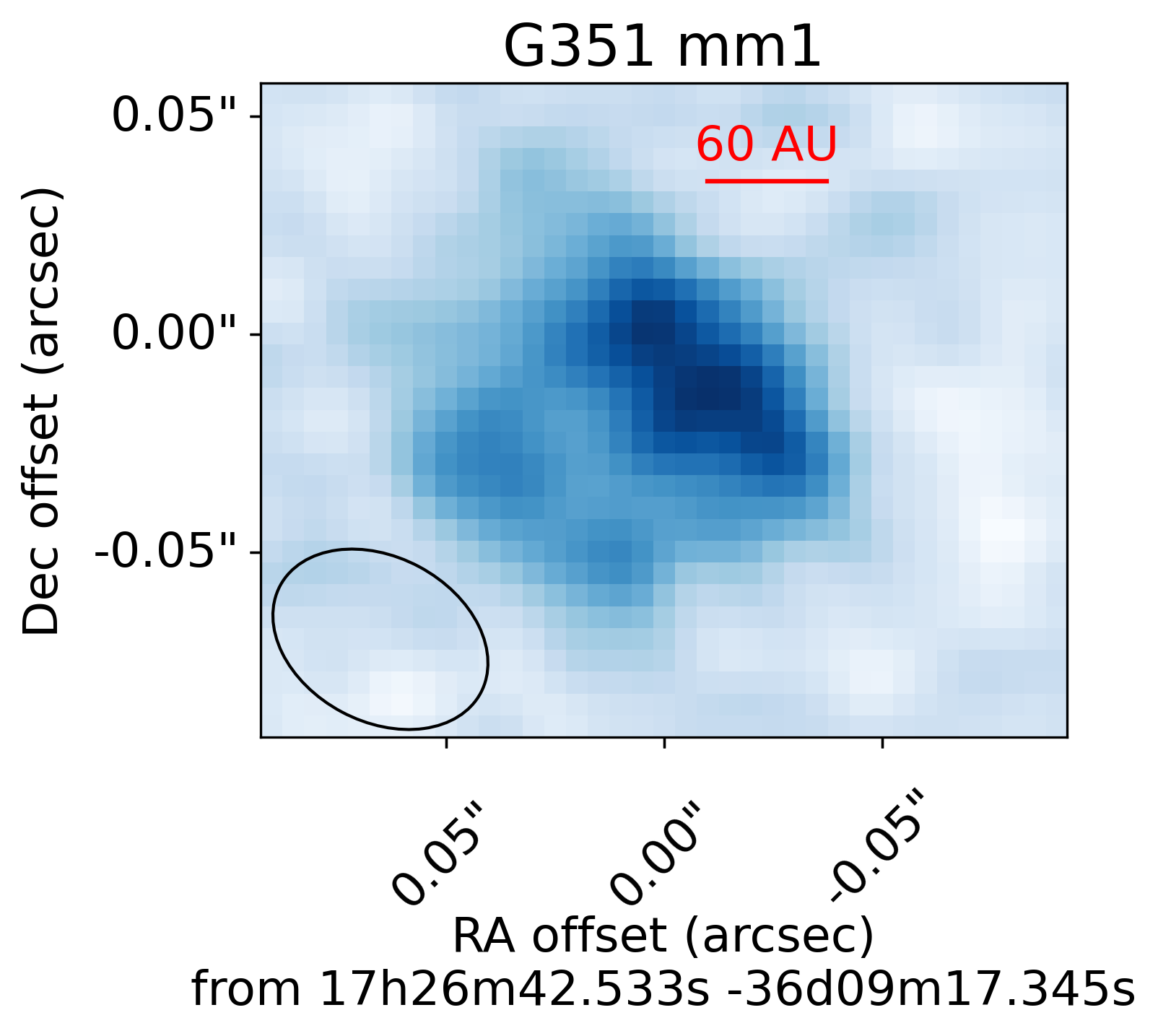}
    \includegraphics[width=0.49\textwidth]{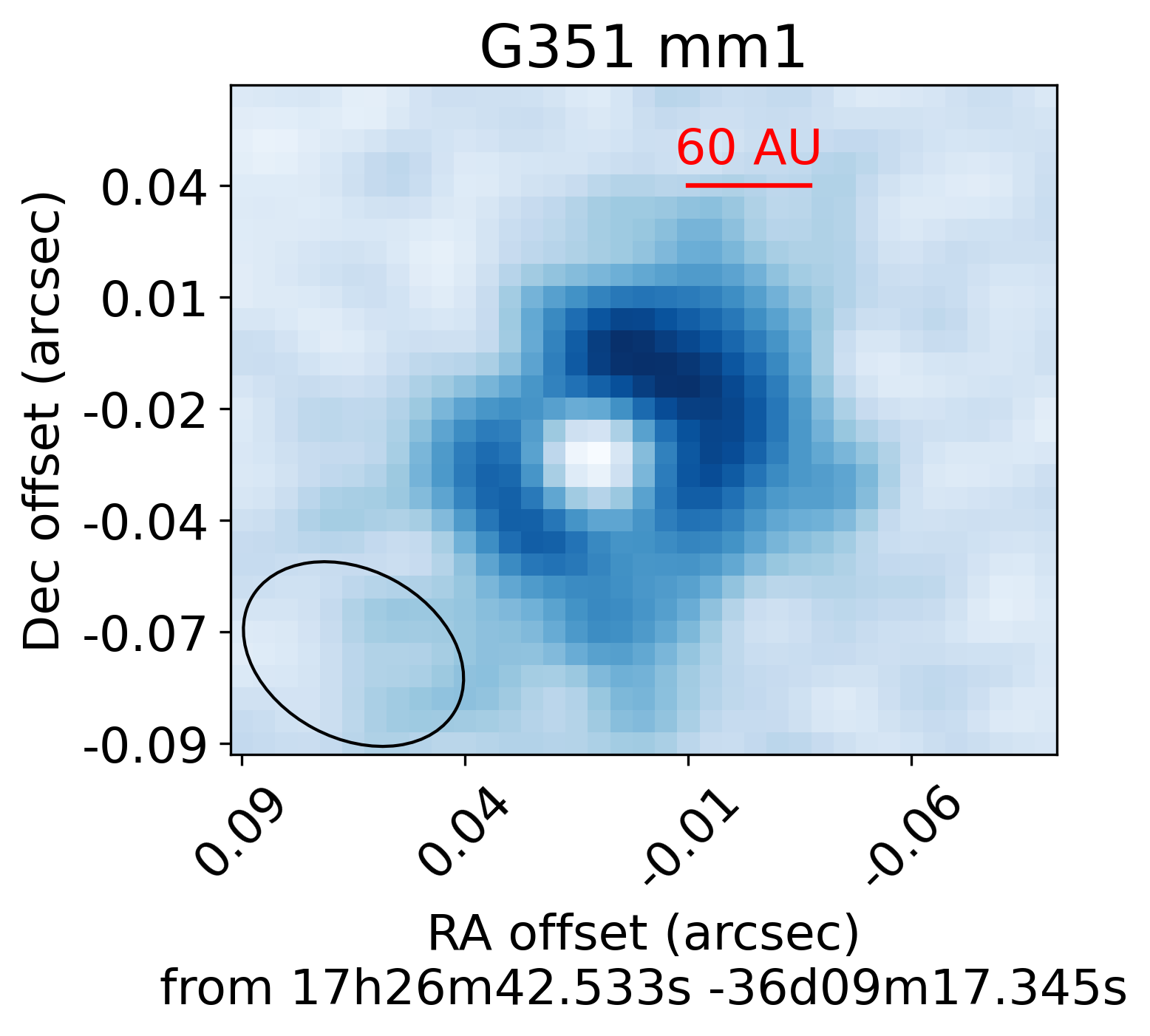}
    \includegraphics[width=0.46\textwidth]{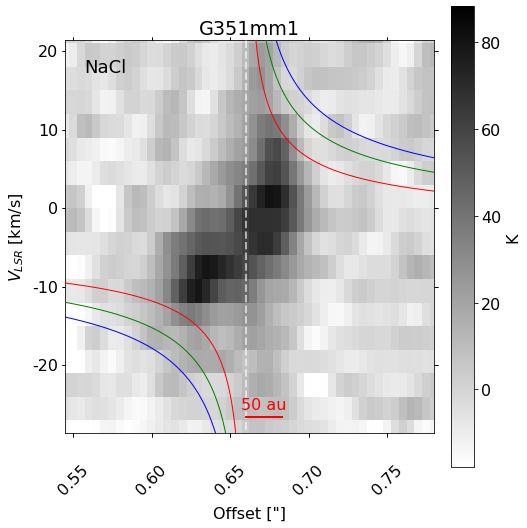}
    \includegraphics[width=0.52\textwidth]{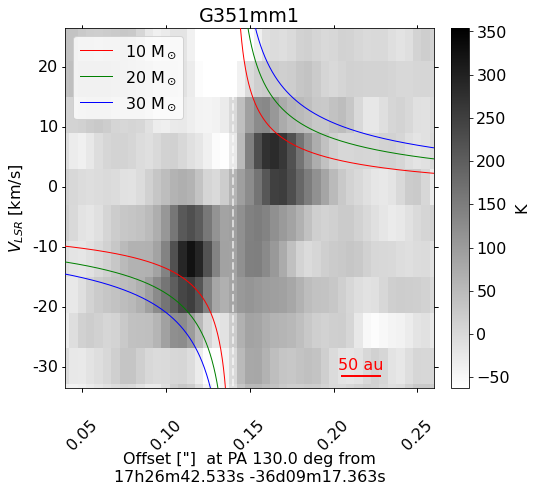}
    \caption{Comparison of NaCl and SiS moment-0 and position-velocity maps of G351.77mm1.
    (top left) NaCl stack moment 0 image, as seen in Figure \ref{fig:resolveddisks}.
    (top right) SiS 13-12 moment 0 image, showing similar morphology.
    (bottom left) NaCl stack cube position-velocity diagram extracted {along the direction of maximum gradient}.
    (bottom right) SiS 13-12 position-velocity diagram extracted  {along the direction of maximum gradient}.
    Keplerian rotation curves assuming an edge-on central source with the listed mass are shown in colored lines; these curves are not fits to the data and do not account for inclination, they are just provided to guide the eye.
    Furthermore, as discussed in Section \ref{sec:g351}, the velocity gradient shown here may trace an outflow rather than a disk.
    }
    \label{fig:g351momentpv}
\end{figure*}

\begin{figure}
    \centering
    \includegraphics[width=0.49\textwidth]{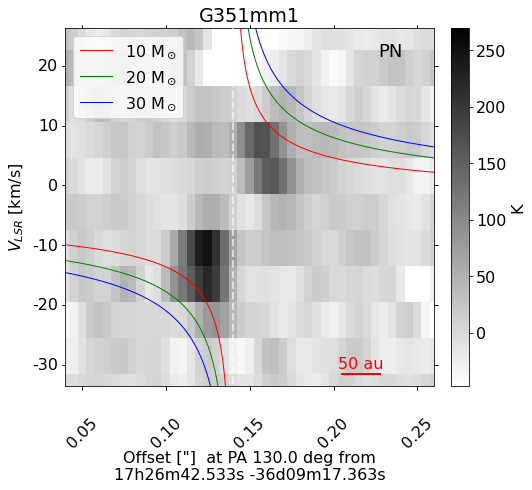}
    \includegraphics[width=0.49\textwidth]{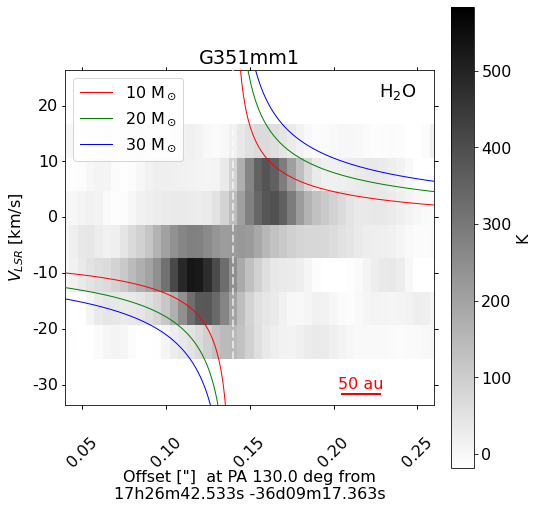}
    \caption{Two more position-velocity diagrams of G351 mm1, showing the 235 GHz PN 5-4 line (newly identified here) and the H$_2$O line.
    The common structure seen in these diagrams, and in Figure \ref{fig:g351momentpv}, justifies our assumption that these have common kinematics.
    Keplerian rotation curves assuming an edge-on central source with the listed mass are shown in colored lines; these curves are not fits to the data and do not account for inclination, they are just provided to guide the eye.
    {Furthermore, the direction of maximum gradient, along which these diagrams are extracted, points in the direction of the outflow.}
    }
    \label{fig:g351pv2}
\end{figure}

We show additional moment maps, spectra, and position-velocity diagrams from G351.77mm1 in this section.
This object is one of the most compelling, high signal-to-noise objects in the sample, yet its kinematics are perplexing, exhibiting disk-like rotation in the same direction as the larger-scale outflow rather than rotating perpendicular to the outflow, as we would expect.
{Figure \ref{fig:g351mm1} shows the stacked spectra.}
Figure \ref{fig:g351brinemoments} shows moment-0 and moment-1 maps of NaCl and \water.
Figure \ref{fig:g351sispn} shows the same for SiS and PN.  
All four molecules exhibit similar morphology and kinematics, including a prominent central hole, which is strongly suggestive of a disk.
Figure \ref{fig:g351momentpv} shows moment-0 maps and position-velocity diagrams, and Figure \ref{fig:g351pv2} more position-velocity diagrams, illustrating that rotation is a plausible explanation for the observed kinematics.

\subsection{G351.77mm2}
\label{appendix:g351mm2}

\begin{figure*}
    \centering
    \includegraphics[width=0.7\textwidth]{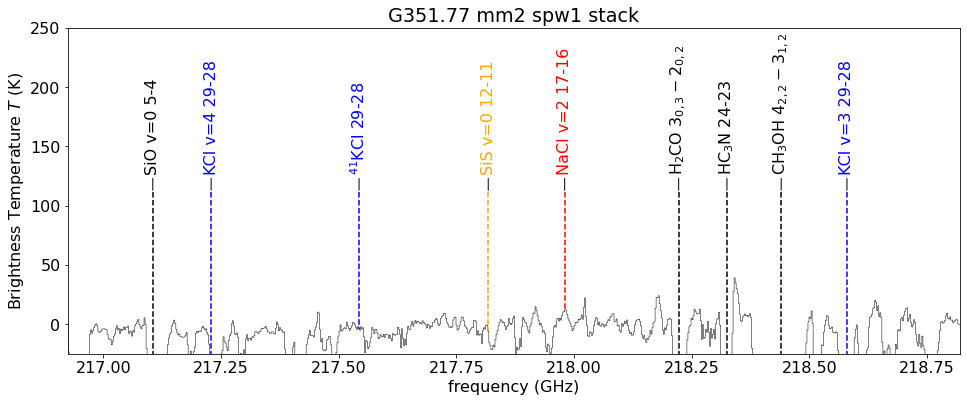}
    \includegraphics[width=0.7\textwidth]{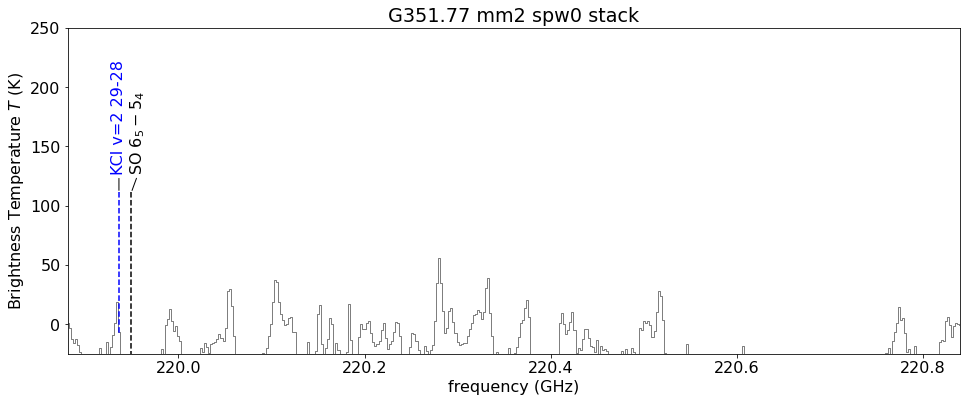}
    \includegraphics[width=0.7\textwidth]{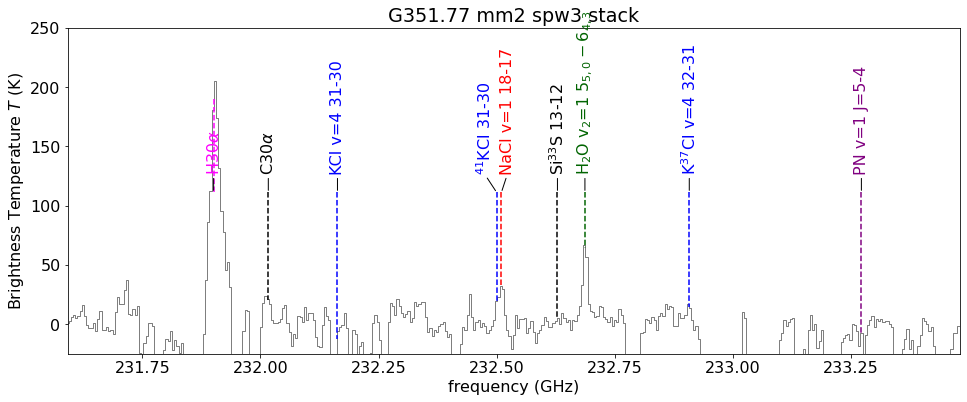}
    \includegraphics[width=0.7\textwidth]{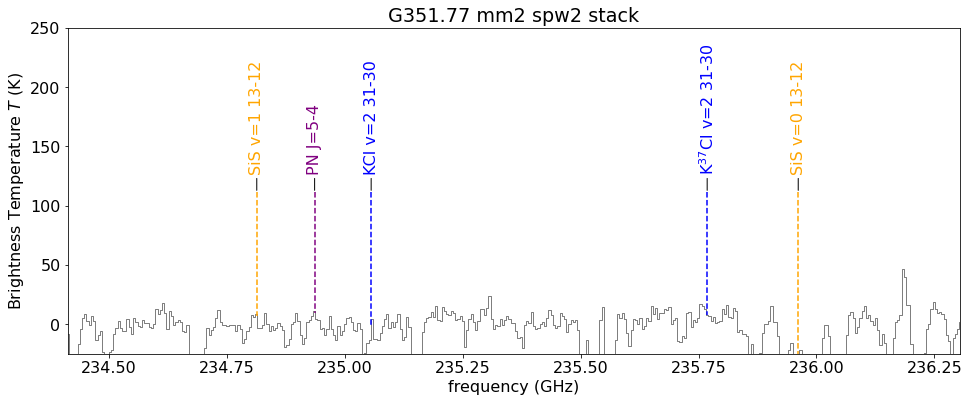}
    \caption{
    The G351 mm2 disk candidate stacked spectrum.  Like mm1 (Fig. \ref{fig:g351mm1}),
    the stacking was done on the \water line.  However, other lines are at most
    weakly detected; the NaCl v=1 J=18-17 line is evident, but no other clear detections
    are present in this or other bands.
    In all spectra, but particularly in spectral window 0, much of the spectrum is absorbed by material unassociated with the disk; we cut off the absorption features to emphasize emission features here.
    }
    \label{fig:g351mm2}
\end{figure*}

\begin{figure*}
    \centering
    \includegraphics[width=\textwidth]{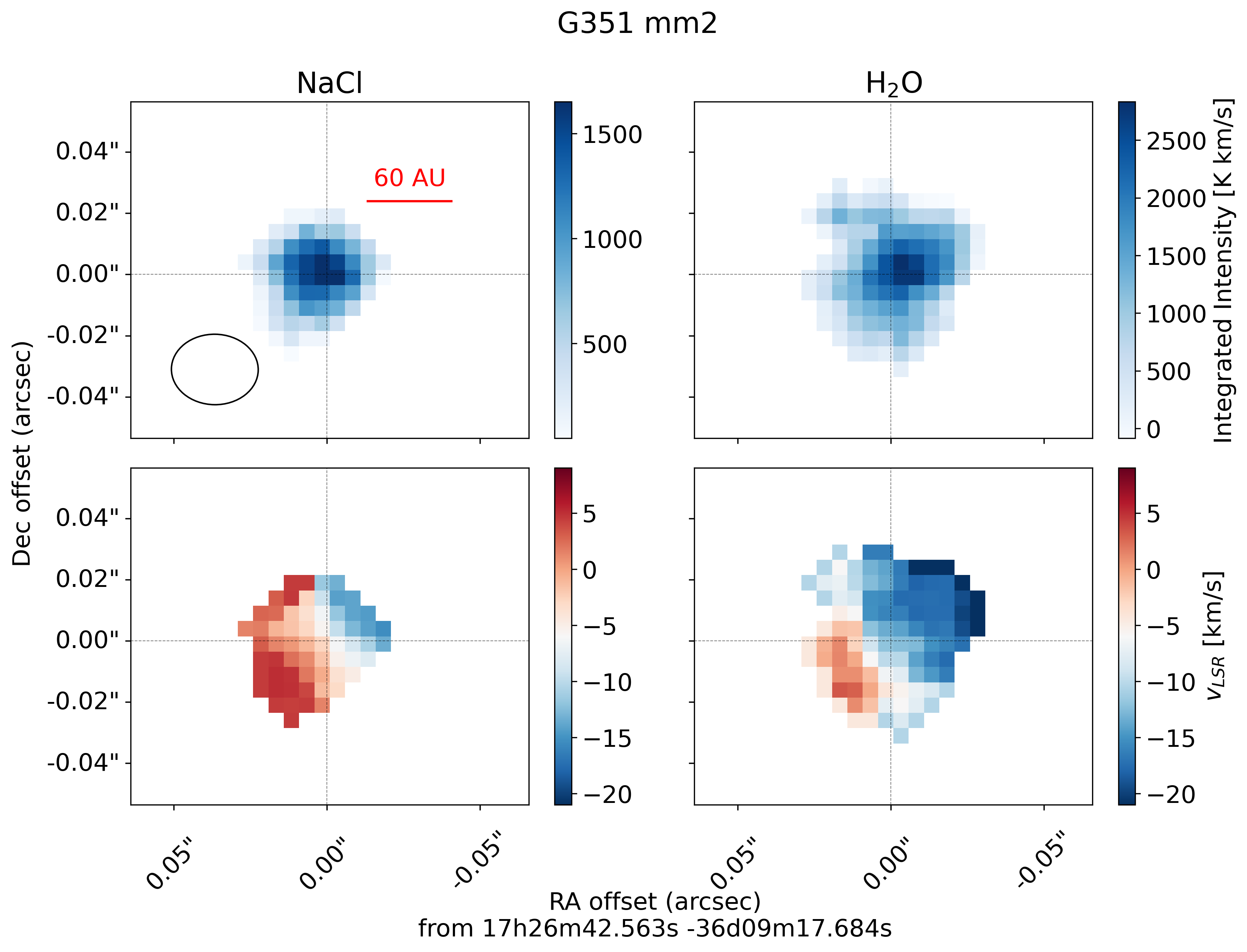}
    \caption{{Moment-0 (integrated intensity) and moment-1 (intensity-weighted velocity) images of {stacked} NaCl (left) and \water (right) for G351 mm2.}
    }
    \label{fig:g351mm2brinemoments}
\end{figure*}

Figure \ref{fig:g351mm2} shows the stacked spectra of G351 mm2.
{Figure \ref{fig:g351mm2brinemoments} shows the moment-0 and moment-1 maps of NaCl and H$_2$O, which are only marginally resolved.}

\subsection{G351.77mm12}
\label{sec:g351mm12}

\begin{figure*}
    \centering
    \includegraphics[width=0.7\textwidth]{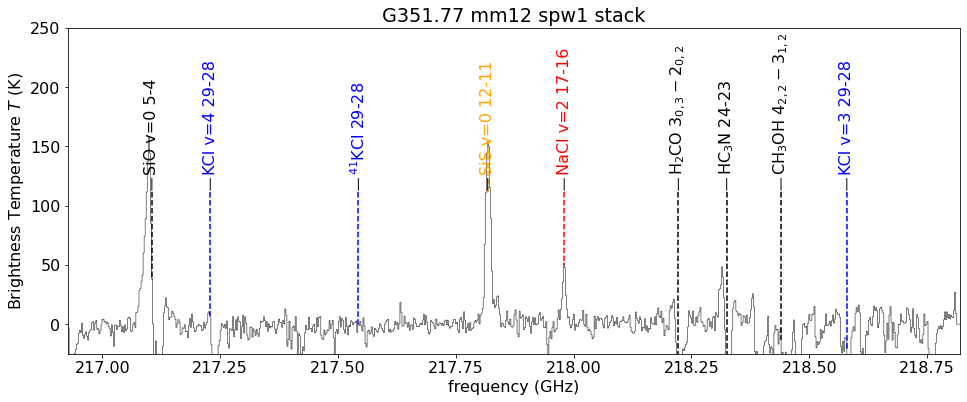}
    \includegraphics[width=0.7\textwidth]{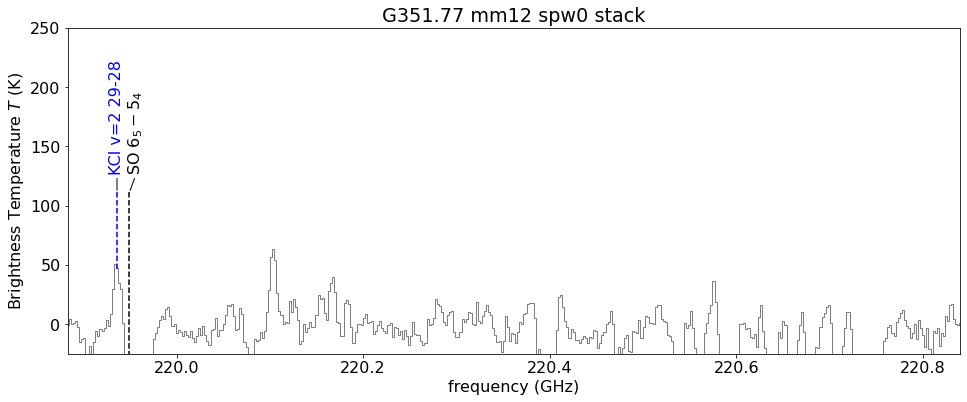}
    \includegraphics[width=0.7\textwidth]{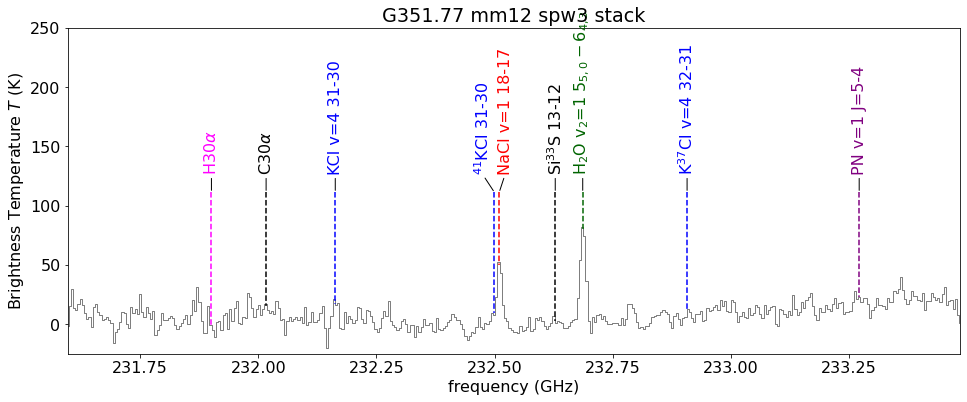}
    \includegraphics[width=0.7\textwidth]{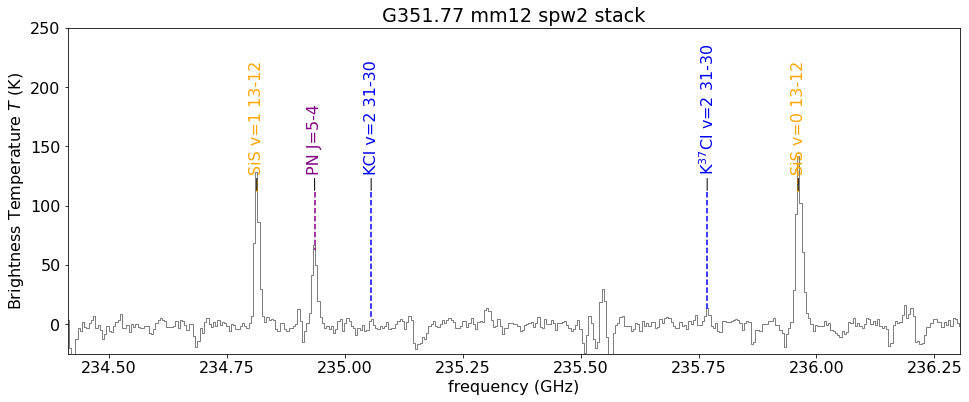}
    \caption{
    The G351 mm12 disk candidate stacked spectrum.
    % Like mm1 and mm2 (Fig. \ref{fig:g351mm1}, \ref{fig:g351mm2}),
    % the stacking was done on the \water line.  
    See Figure \ref{fig:g351mm1} for additional description.
    }
    \label{fig:g351mm12}
\end{figure*}

\begin{figure*}
    \centering
    \includegraphics[width=\textwidth]{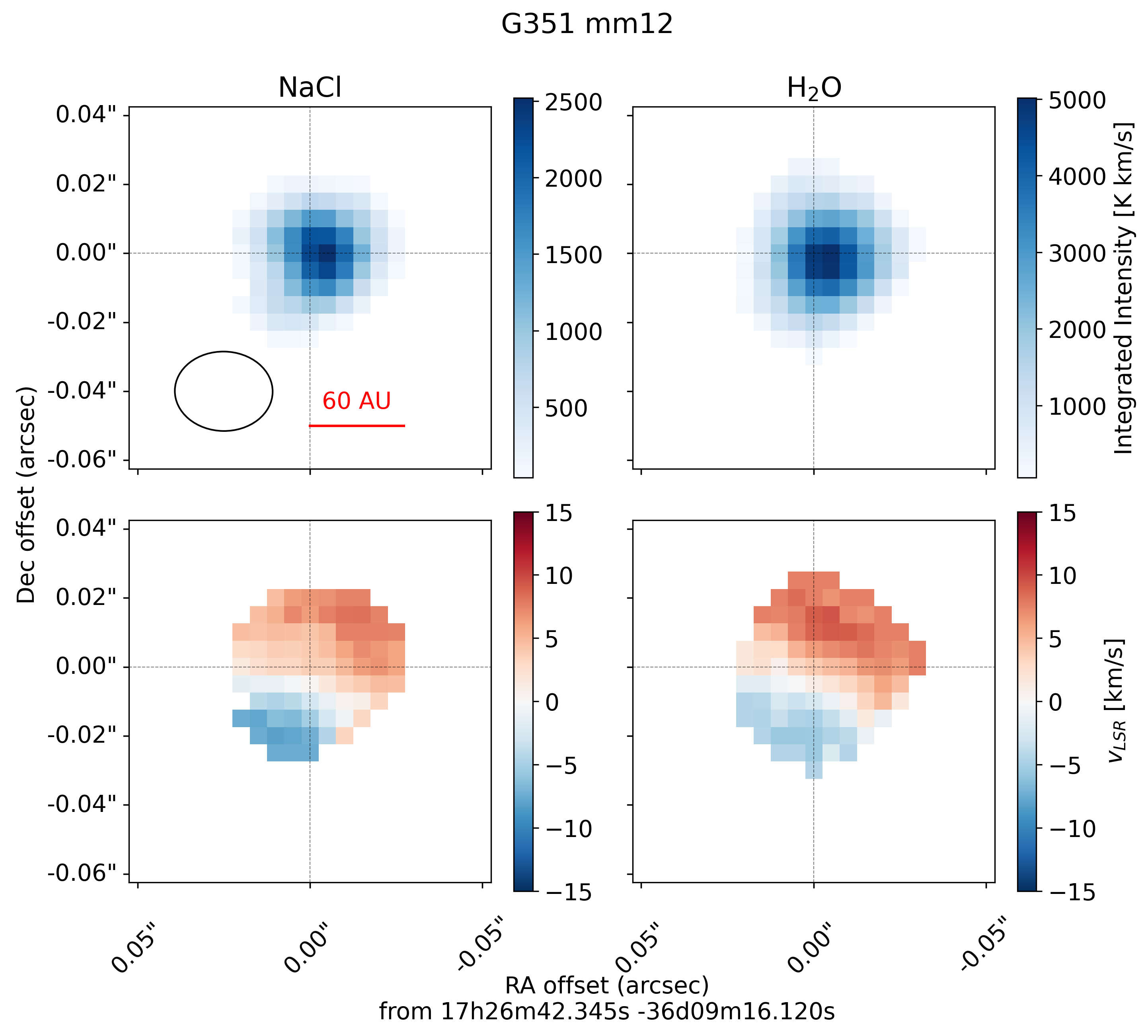}
    \caption{{Moment-0 (integrated intensity) and moment-1 (intensity-weighted velocity) images of {stacked} NaCl (left) and \water (right) for G351 mm12.}
    }
    \label{fig:g351mm12brinemoments}
\end{figure*}

Figure \ref{fig:g351mm12} shows the stacked spectra of G351 mm12.
This source is unresolved, as shown in Fig \ref{fig:g351mm12m0}.
{Figure \ref{fig:g351mm12brinemoments} shows the moment-0 and moment-1 maps of NaCl and H$_2$O, which are only marginally resolved.}

\subsection{NGC6334I}
\label{appendix:ngc6334imm1b}
\label{appendix:ngc6334imm2b}
\begin{figure*}
    \centering
    \includegraphics[width=0.7\textwidth]{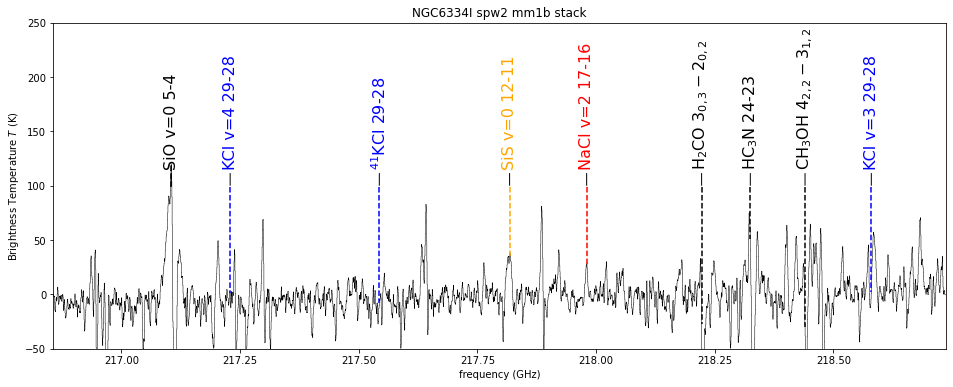}
    \includegraphics[width=0.7\textwidth]{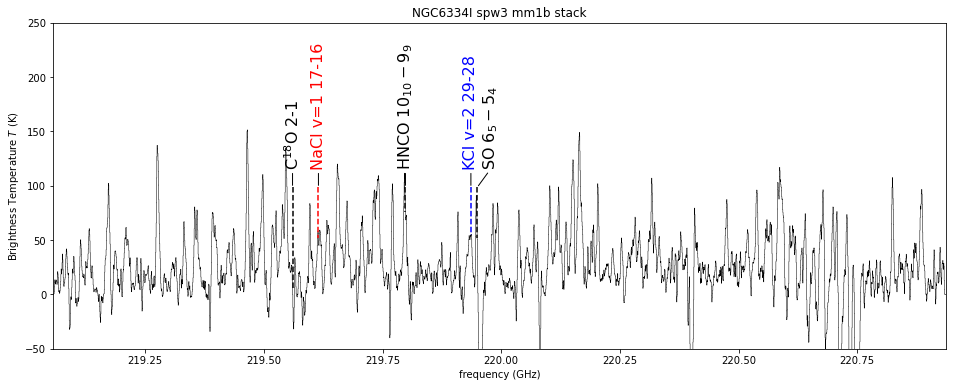}
    \includegraphics[width=0.7\textwidth]{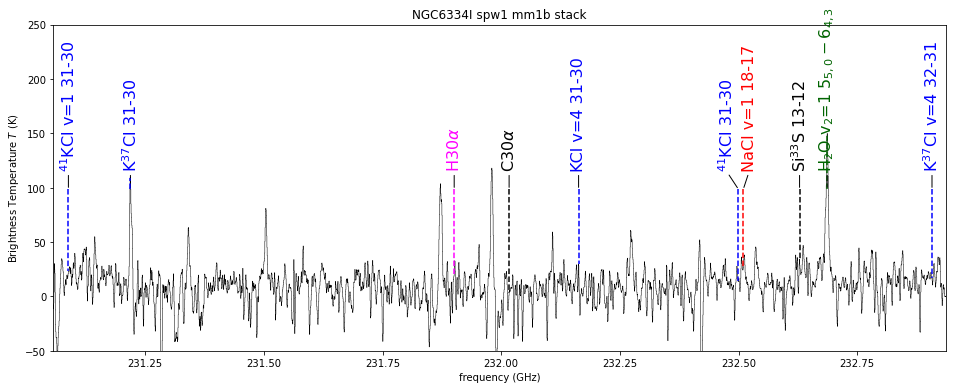}
    \includegraphics[width=0.7\textwidth]{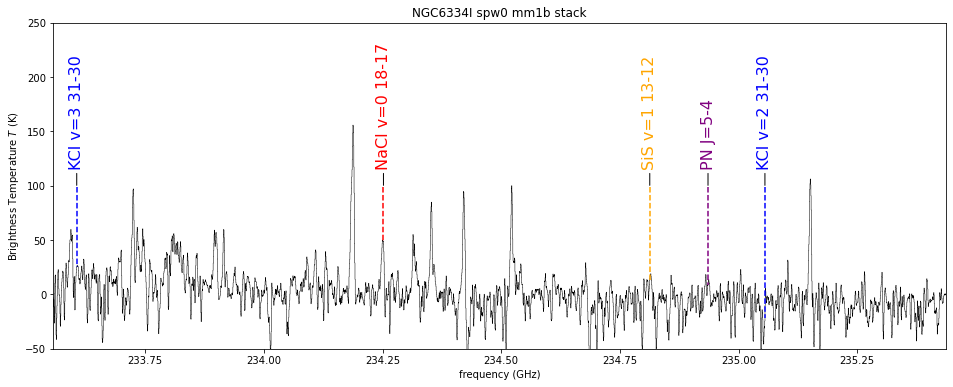}
    \caption{The NGC6334I mm1b disk stacked spectrum.  
    See Figure \ref{fig:g351mm1} for additional description.
    }
    \label{fig:ngc6334imm1b}
\end{figure*}

\begin{figure*}
    \centering
    \includegraphics[width=\textwidth]{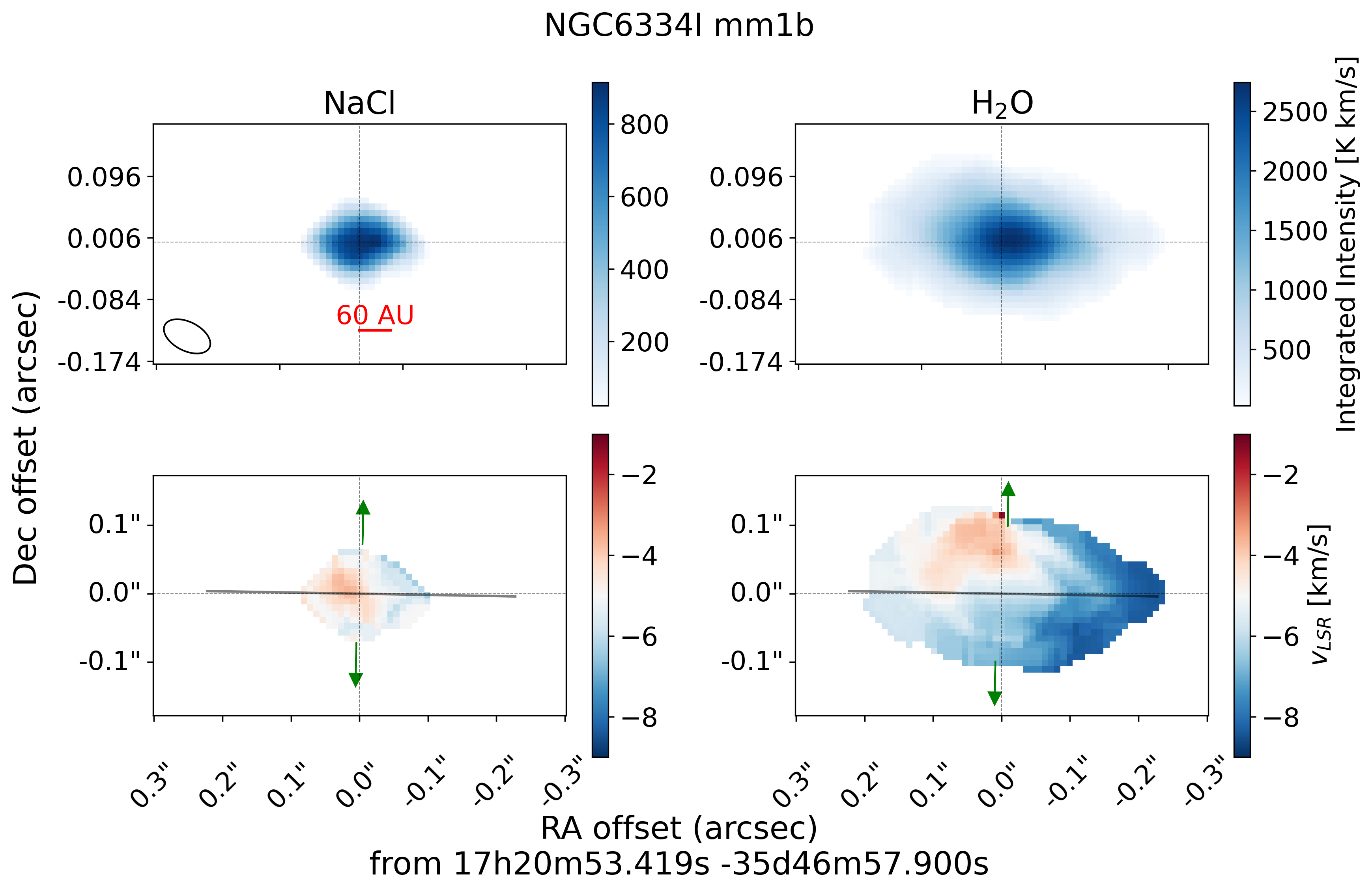}
    \caption{{Moment-0 and 1 images as in Figure \ref{fig:g351brinemoments},
    but for NGC 6334I mm1b.
    The outflow noted by \citet{Brogan2018} is shown at PA=-5$^\circ$ with green arrows,
    as it appears to be primarily in the plane of the sky.
    The solid gray line shows the orientation from which the position-velocity diagram
    (Figure \ref{fig:ngc6334i_mm1_pv}) is extracted.
    }
    }
    \label{fig:ngc6334imm1bmoments}
\end{figure*}

\begin{figure*}
    \centering
    \includegraphics[width=0.7\textwidth]{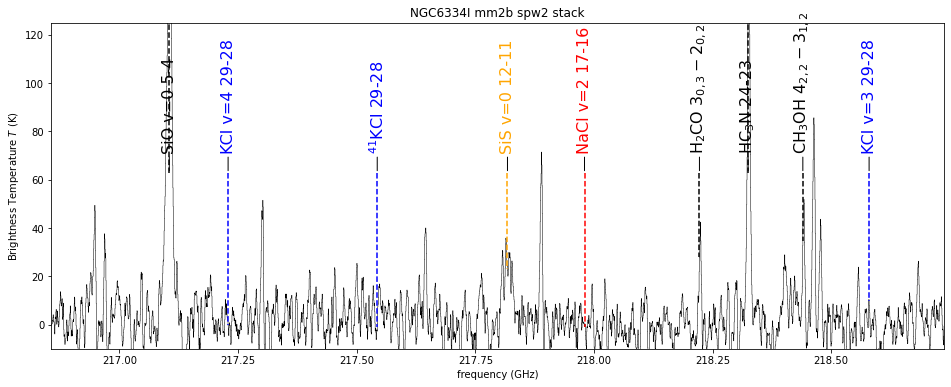}
    \includegraphics[width=0.7\textwidth]{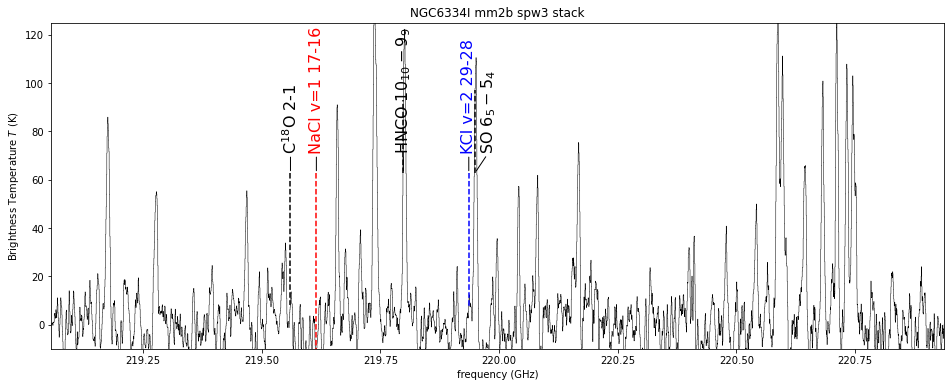}
    \includegraphics[width=0.7\textwidth]{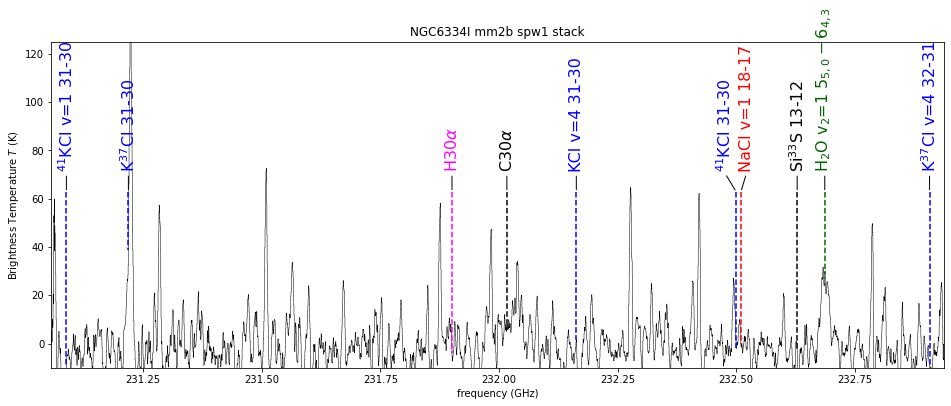}
    \includegraphics[width=0.7\textwidth]{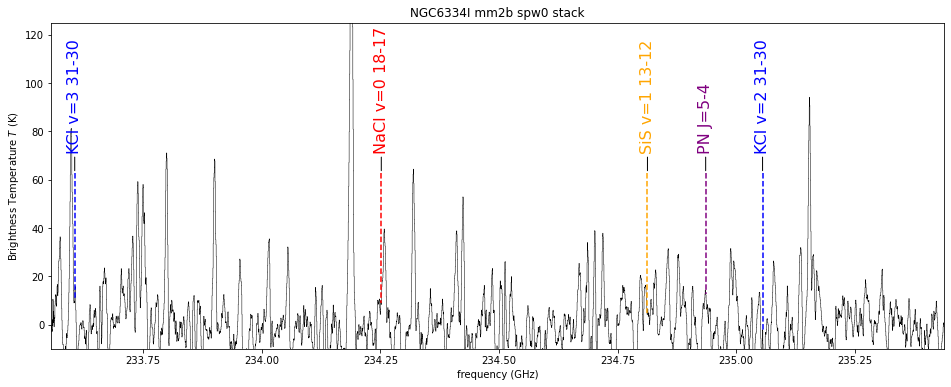}
    \caption{The NGC6334I mm2b disk stacked spectrum. 
    See Figure \ref{fig:g351mm1} for additional description.
    }
    \label{fig:ngc6334imm2b}
\end{figure*}

We show additional figures of NGC6334I, including
spectra (Fig \ref{fig:ngc6334imm1b} and
moment maps (Fig \ref{fig:ngc6334imm1bmoments}.
For mm2b, we show only the stacked spectrum (Fig \ref{fig:ngc6334imm2b}), since the source is unresolved

\subsection{W33A}
\label{appendix:w33a}
\begin{figure*}
    \centering
    \includegraphics[width=0.7\textwidth]{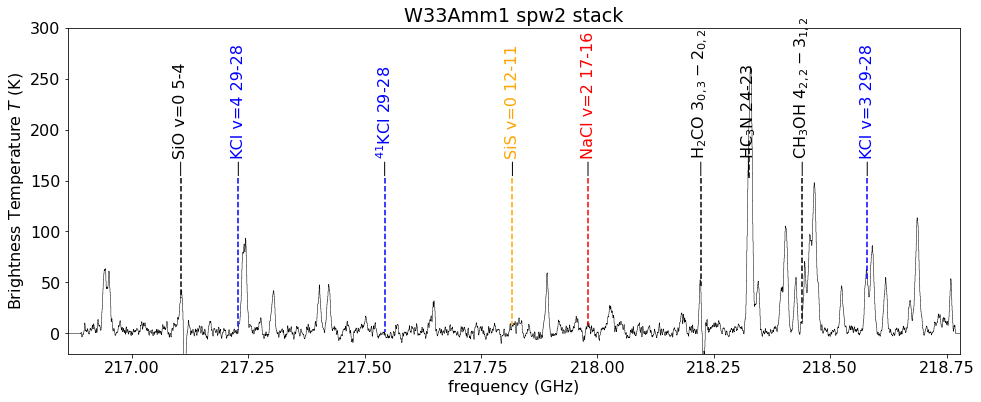}
    \includegraphics[width=0.7\textwidth]{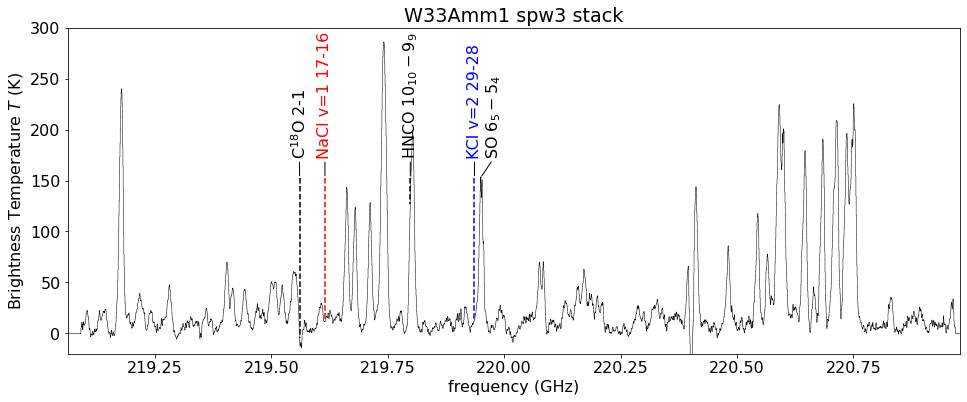}
    \includegraphics[width=0.7\textwidth]{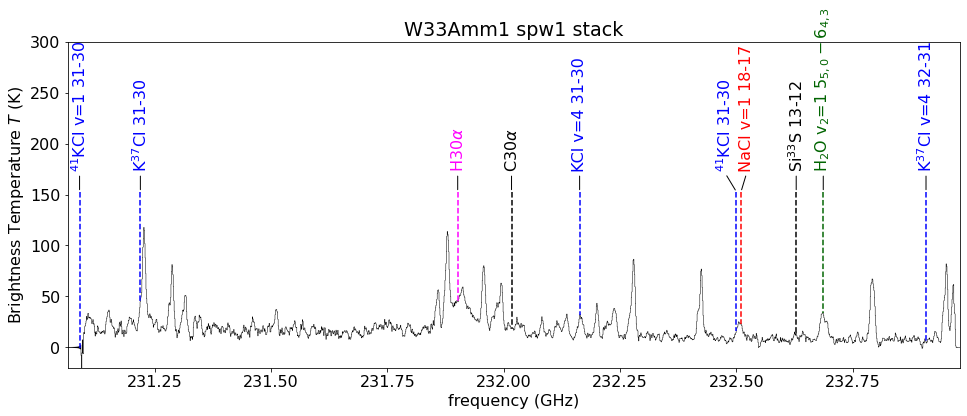}
    \includegraphics[width=0.7\textwidth]{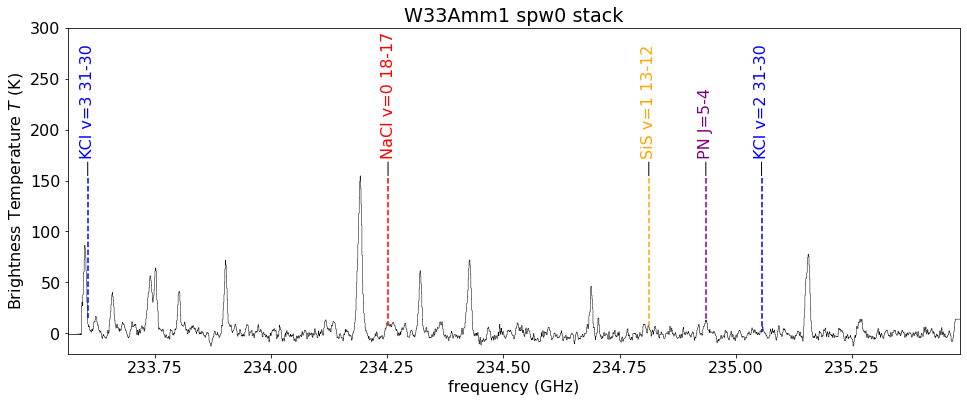}
    \caption{The W33A disk stacked spectrum. 
    See Figure \ref{fig:g351mm1} for additional description.
    }
    \label{fig:W33Aspectrum}
\end{figure*}

The labeled, stacked spectrum from W33A is shown in Figure \ref{fig:W33Aspectrum}.
{The four-panel moment map is not shown for this source because it is unresolved and shows no structure; it is consistent with a point source.}

\subsection{I16547A}
\label{appendix:i16547a}
\begin{figure*}
    \centering
    \includegraphics[width=0.7\textwidth]{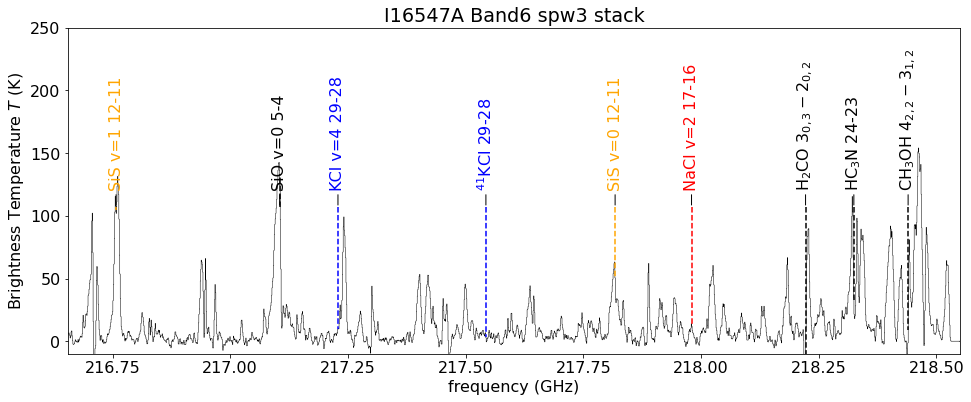}
    \includegraphics[width=0.7\textwidth]{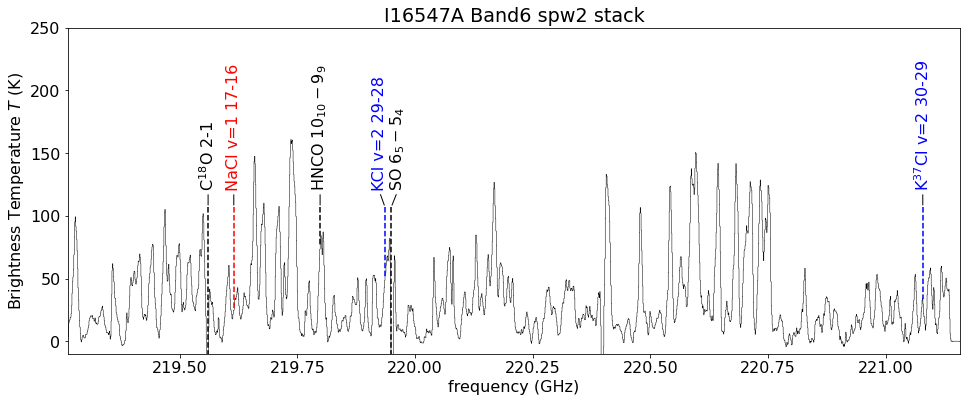}
    \includegraphics[width=0.7\textwidth]{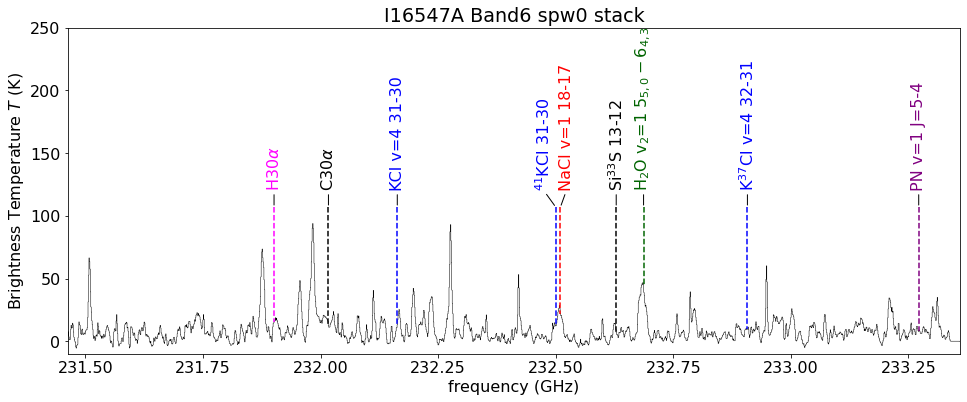}
    \includegraphics[width=0.7\textwidth]{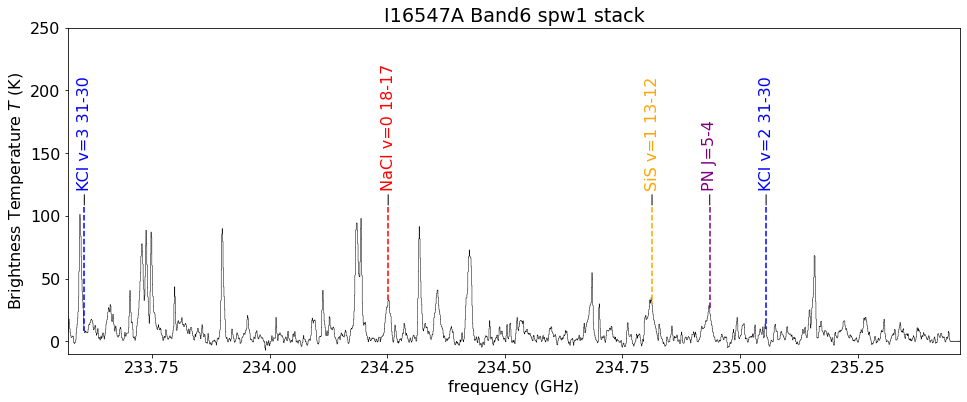}
    \caption{The I16547A disk stacked spectrum. 
    See Figure \ref{fig:g351mm1} for additional description.
    }
    \label{fig:I16547Aspectrum}
\end{figure*}

\begin{figure*}
    \centering
    \includegraphics[width=\textwidth]{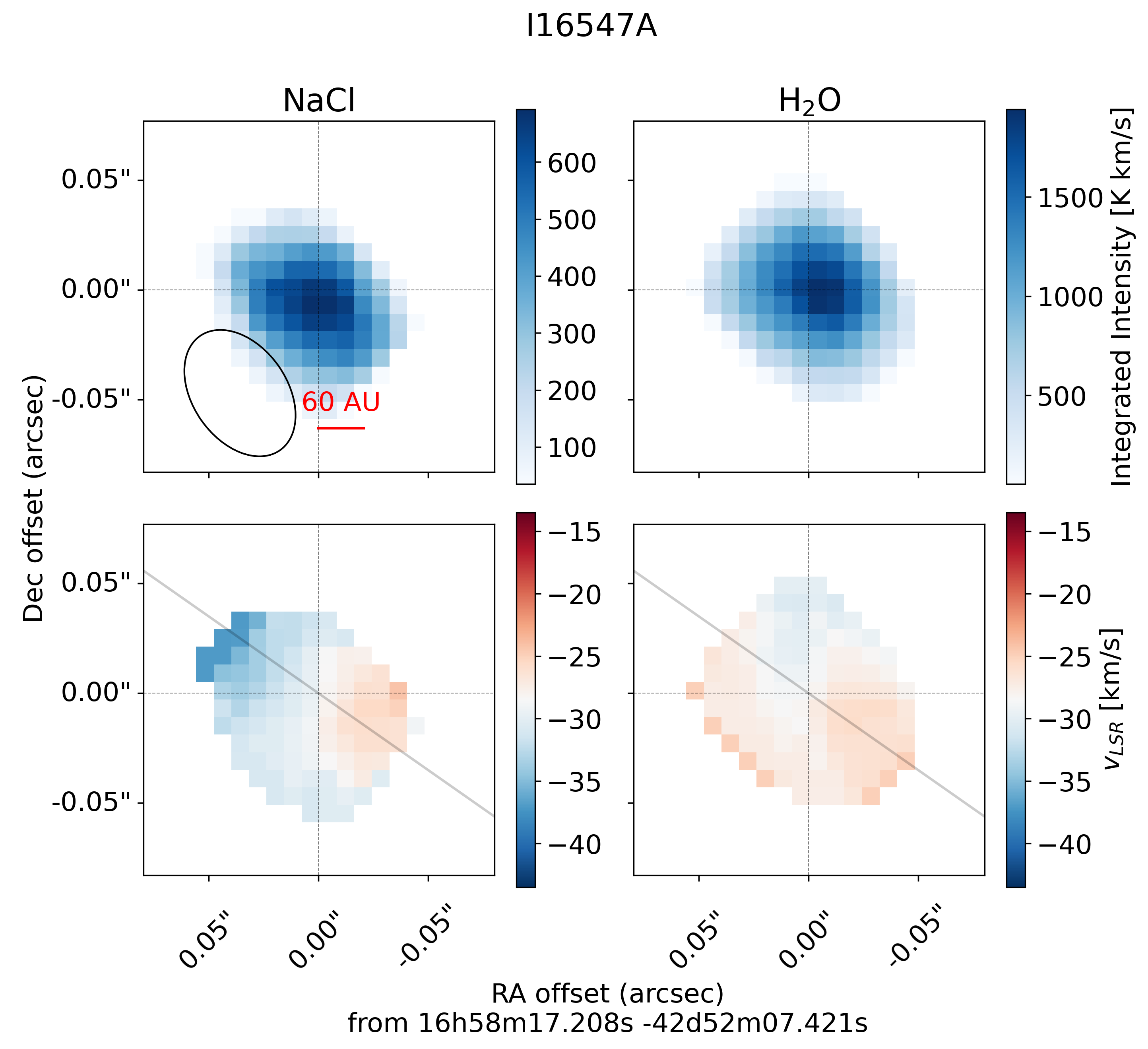}
    \caption{{Moment-0 and 1 images as in Figure \ref{fig:g351brinemoments},
    but for IRAS 16547A
    The solid gray line shows the orientation from which the position-velocity diagram
    (Figure \ref{fig:i16547pv}) is extracted, based on the angle determined in
    \citet{Tanaka2020}.
    The radio jet at PA=-16$^\circ$ is shown with arrows \citep{Tanaka2020}.
    }
    }
    \label{fig:i16547amoments}
\end{figure*}

The labeled, stacked spectrum from I16547A is shown in Figure \ref{fig:I16547Aspectrum}.
Moment-0 and moment-1 images are shown in Figure \ref{fig:i16547amoments}.
A position-velocity diagram, with overlaid Keplerian curves for an edge-on orbit with masses labeled,
is shown in Figure \ref{fig:i16547pv}.
We show these to provide order-of-magnitude mass estimates, but note that we have no constraint on the disk inclination and have not attempted to model the extent of the disk emission.
{No outflow is observed toward A \citep{Tanaka2020}.}

\subsection{I16547B}
\label{appendix:i16547b}
\begin{figure*}
    \centering
    \includegraphics[width=0.7\textwidth]{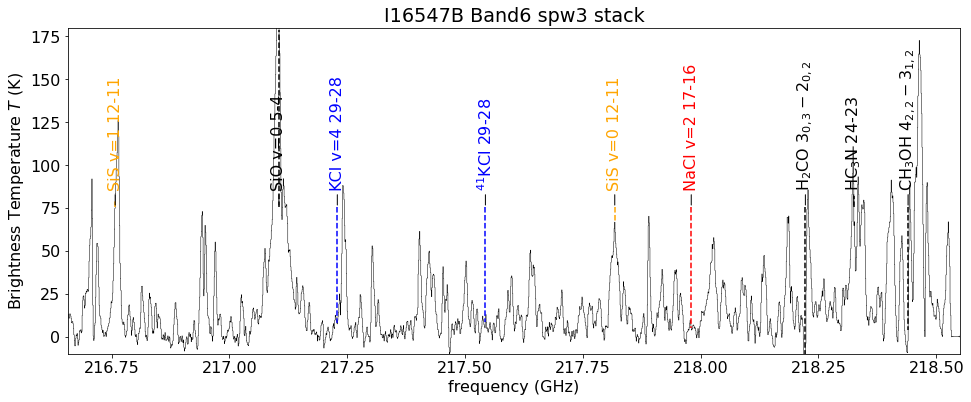}
    \includegraphics[width=0.7\textwidth]{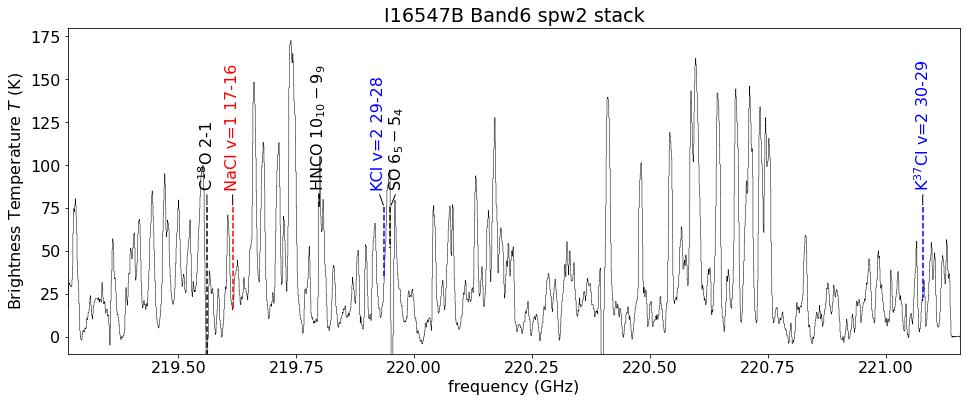}
    \includegraphics[width=0.7\textwidth]{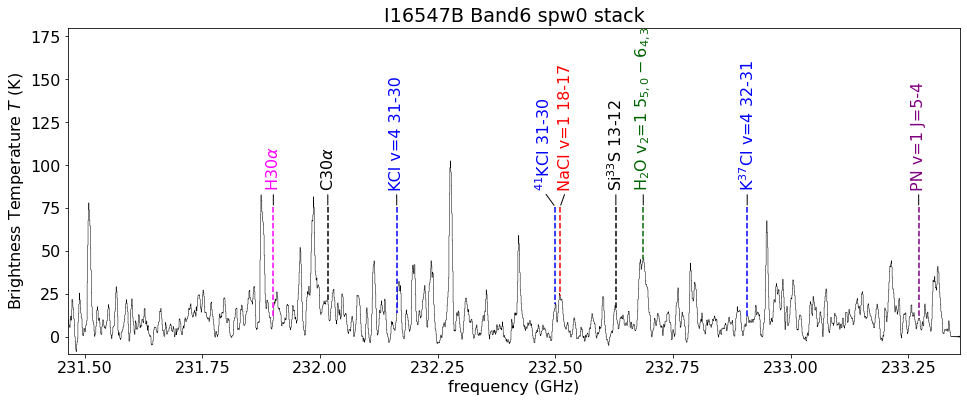}
    \includegraphics[width=0.7\textwidth]{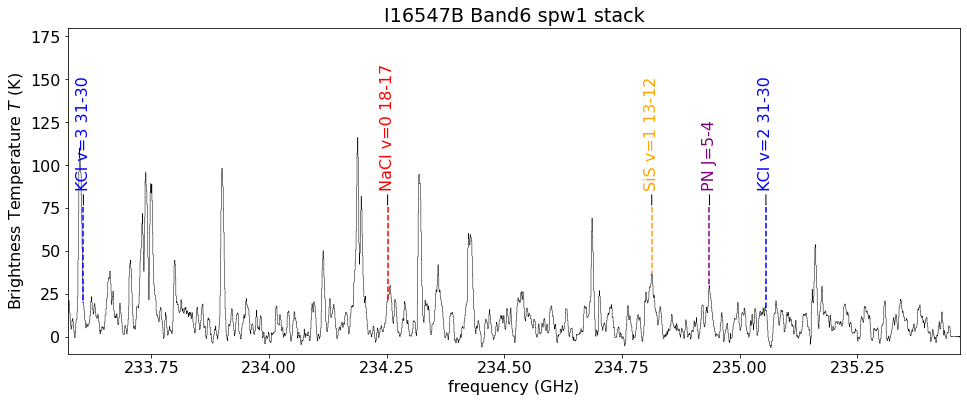}
    \caption{The I16547B disk stacked spectrum. 
    See Figure \ref{fig:g351mm1} for additional description.
    }
    \label{fig:I16547Bspectrum}
\end{figure*}

\begin{figure*}
    \centering
    \includegraphics[width=\textwidth]{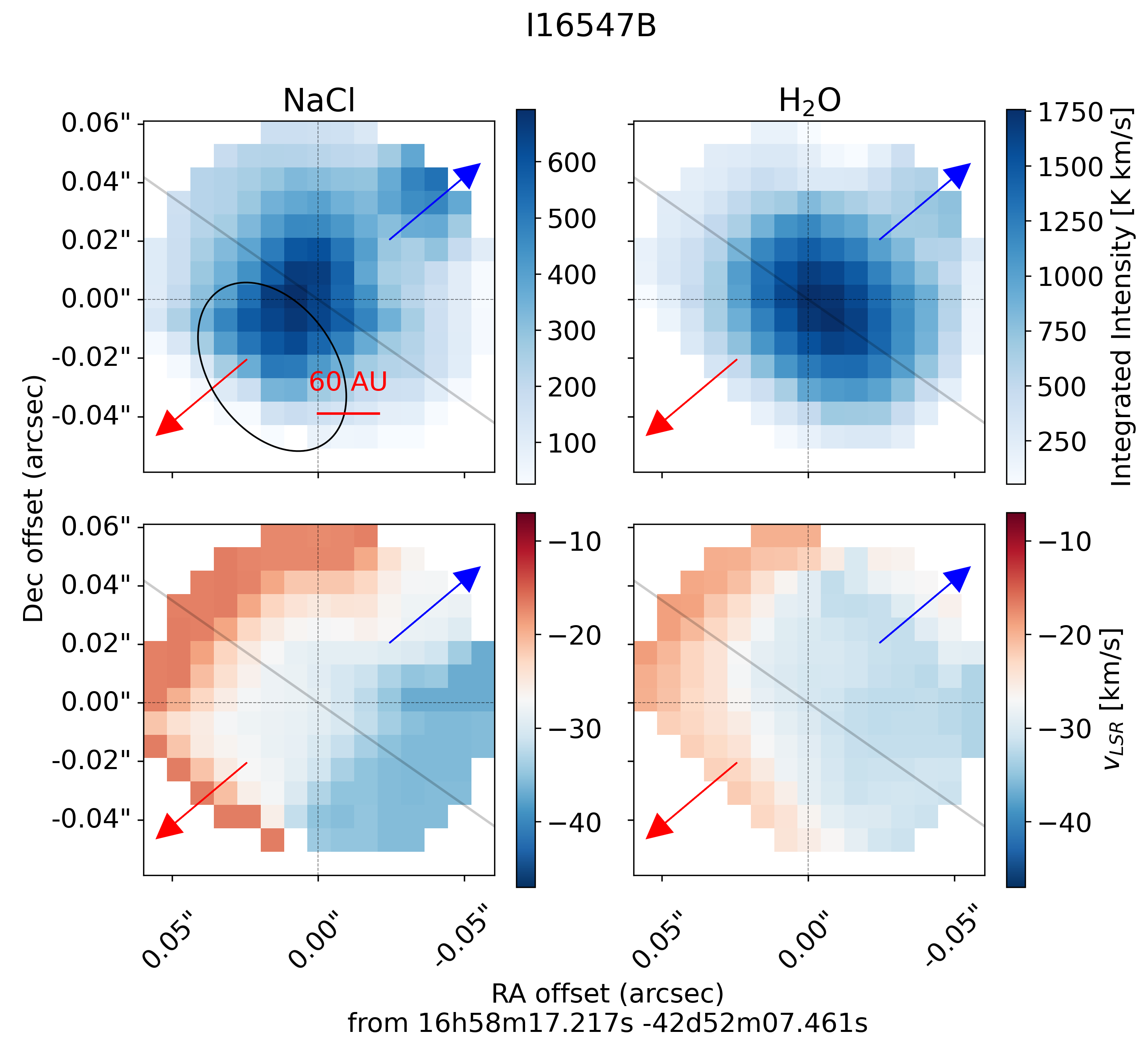}
    \caption{{Moment-0 and 1 images as in Figure \ref{fig:g351brinemoments},
    but for IRAS 16547B
    The solid gray line shows the orientation from which the position-velocity diagram
    (Figure \ref{fig:i16547pv}) is extracted, based on the angle determined in
    \citet{Tanaka2020}.
    The red and blue arrows indicate the direction of the SiO outflow noted in
    \citet{Tanaka2020}.
    }
    }
    \label{fig:i16547bmoments}
\end{figure*}

\begin{figure*}
    \centering
    \includegraphics[width=0.49\textwidth]{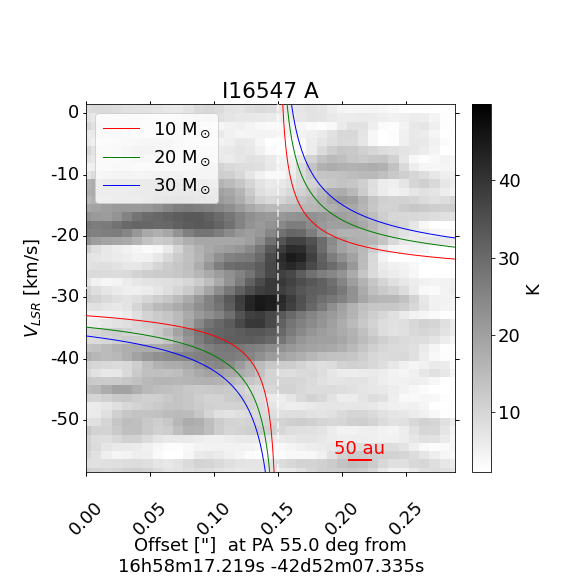}
    \includegraphics[width=0.49\textwidth]{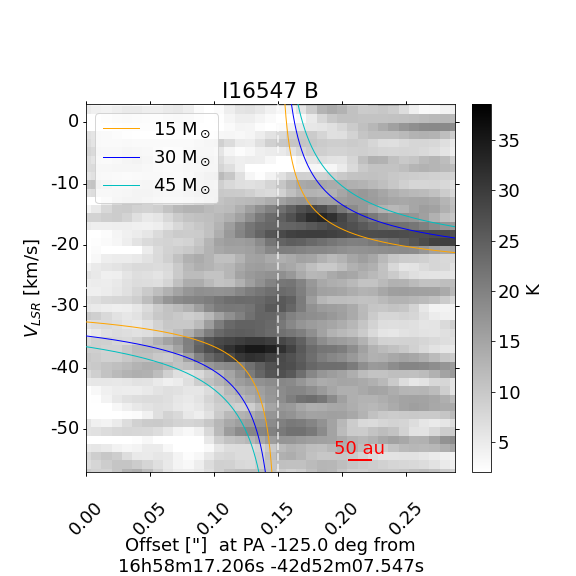}
    \caption{Position-velocity diagrams of the I16547 disks in the stacked NaCl lines.
    The overplotted curves show Keplerian rotation around central point sources with masses indicated in the caption,
    assuming an edge-on inclination.
    As noted in \citet{Tanaka2020}, the disks appear to be counter-rotating along similar position angles.
    }
    \label{fig:i16547pv}
\end{figure*}

The labeled, stacked spectrum from I16547B is shown in Figure \ref{fig:I16547Bspectrum}.
Moment-0 and moment-1 images are shown in Figure \ref{fig:i16547bmoments}.
A position-velocity diagram, with overlaid Keplerian curves for an edge-on orbit with masses labeled,
is shown in Figure \ref{fig:i16547pv}.
We show these to provide order-of-magnitude mass estimates, but note that we have no constraint on the disk inclination and have not attempted to model the extent of the disk emission.
{An SiO outflow is observed toward B, perpendicular to the gradient in the PV diagram \citep{Tanaka2020}.}

\subsection{Continuum}
\begin{figure*}
    \centering
    \includegraphics[width=0.49\textwidth]{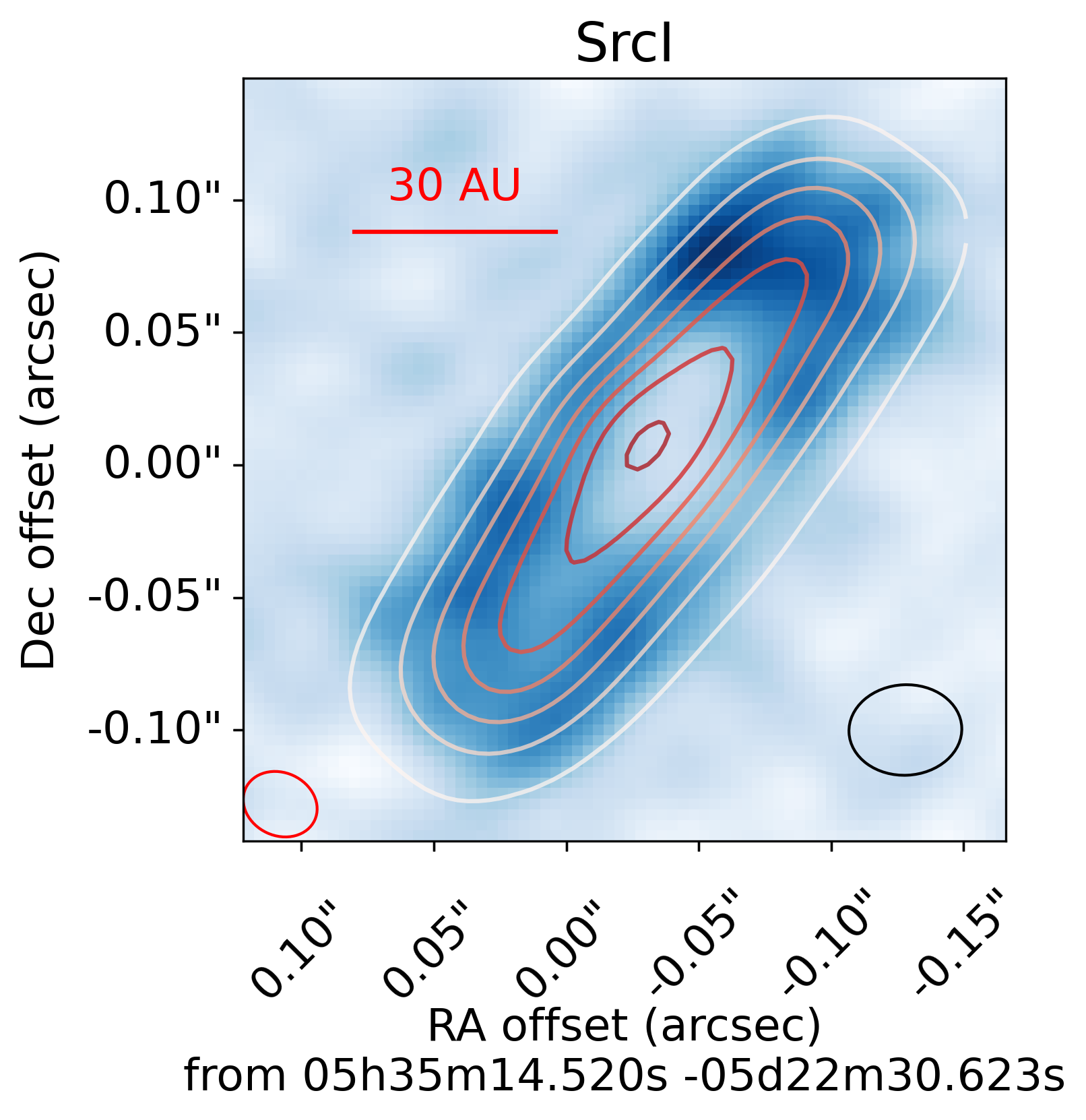}
    \includegraphics[width=0.49\textwidth]{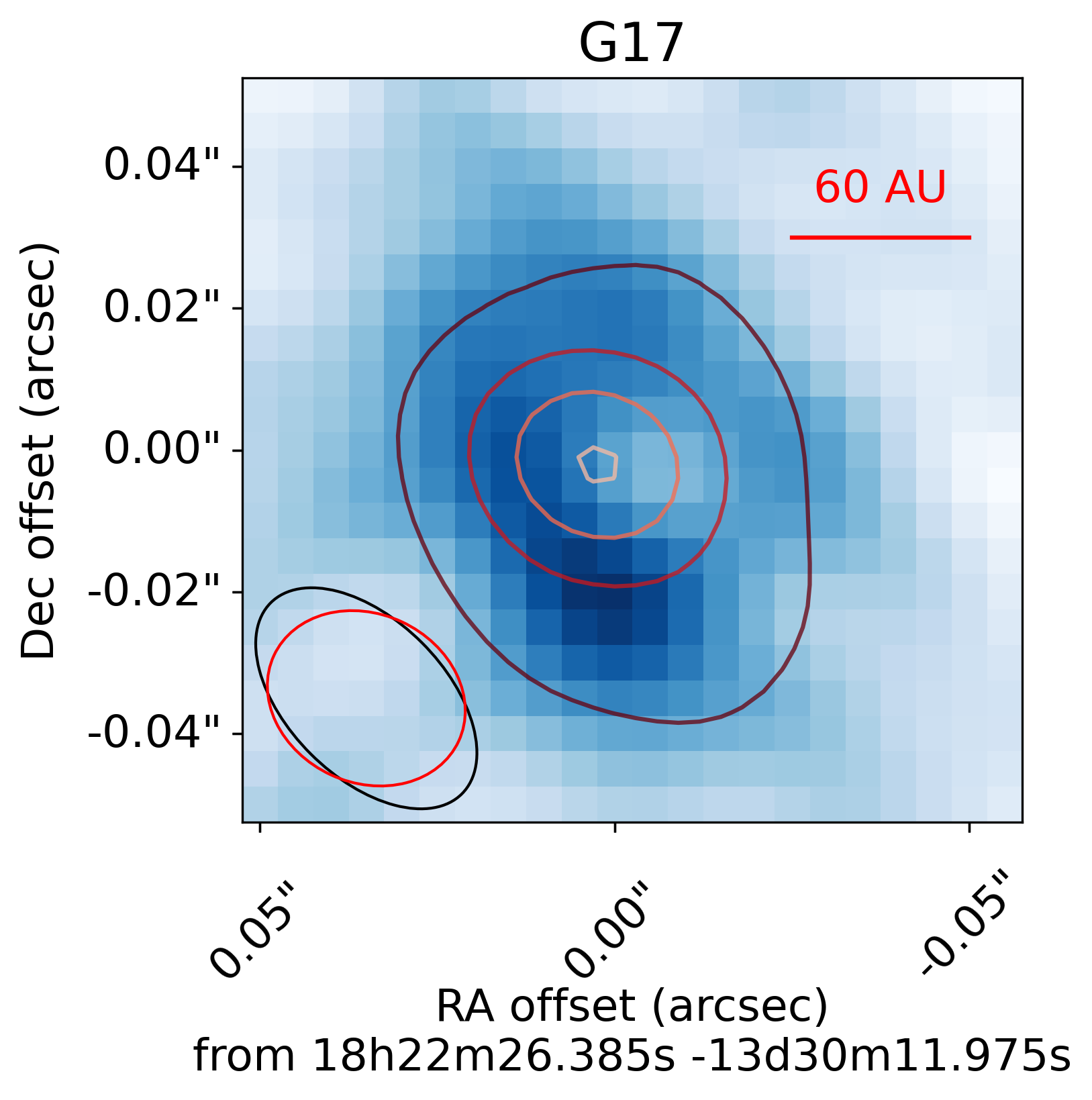}
    \includegraphics[width=0.49\textwidth]{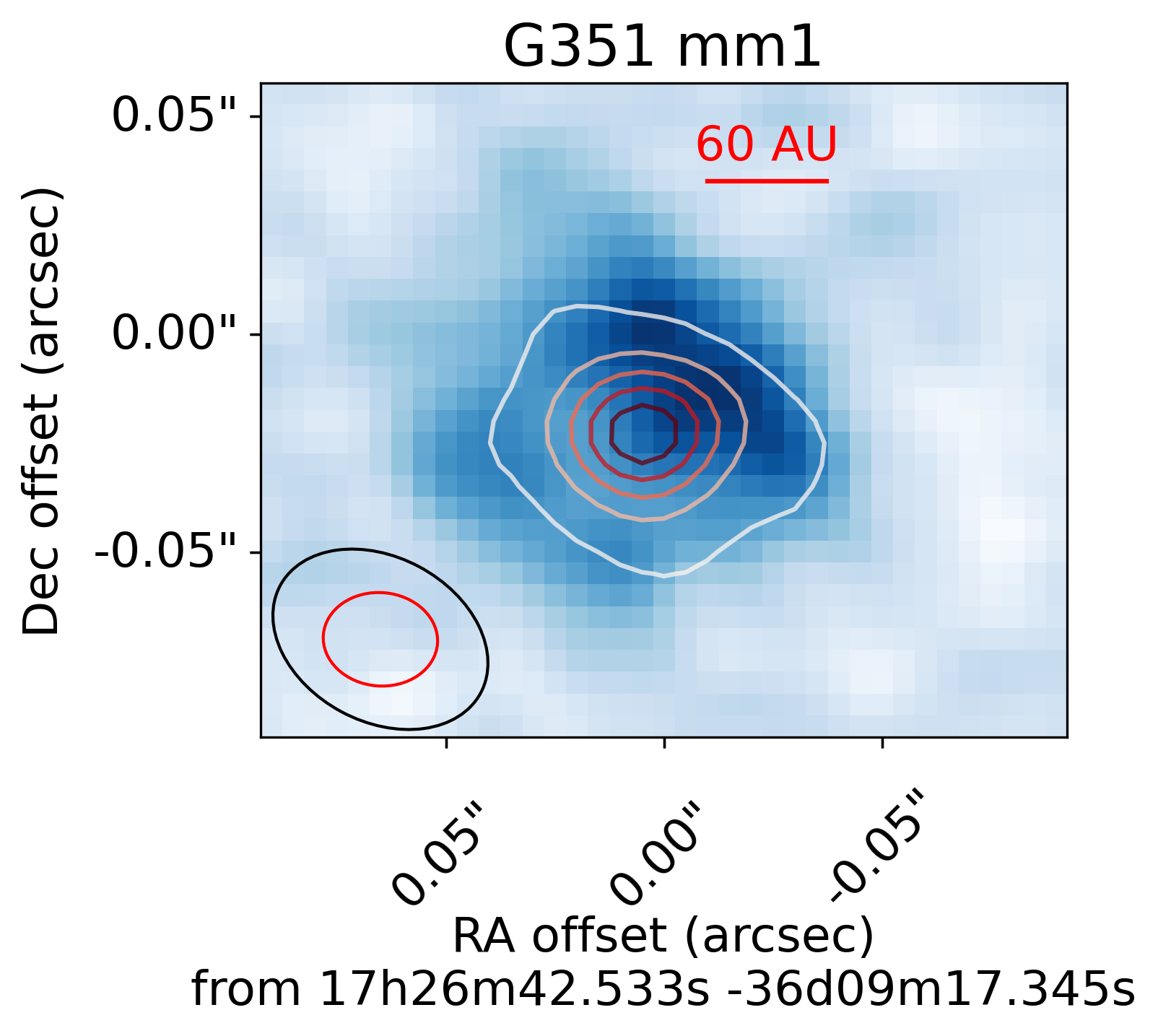}
    \includegraphics[width=0.49\textwidth]{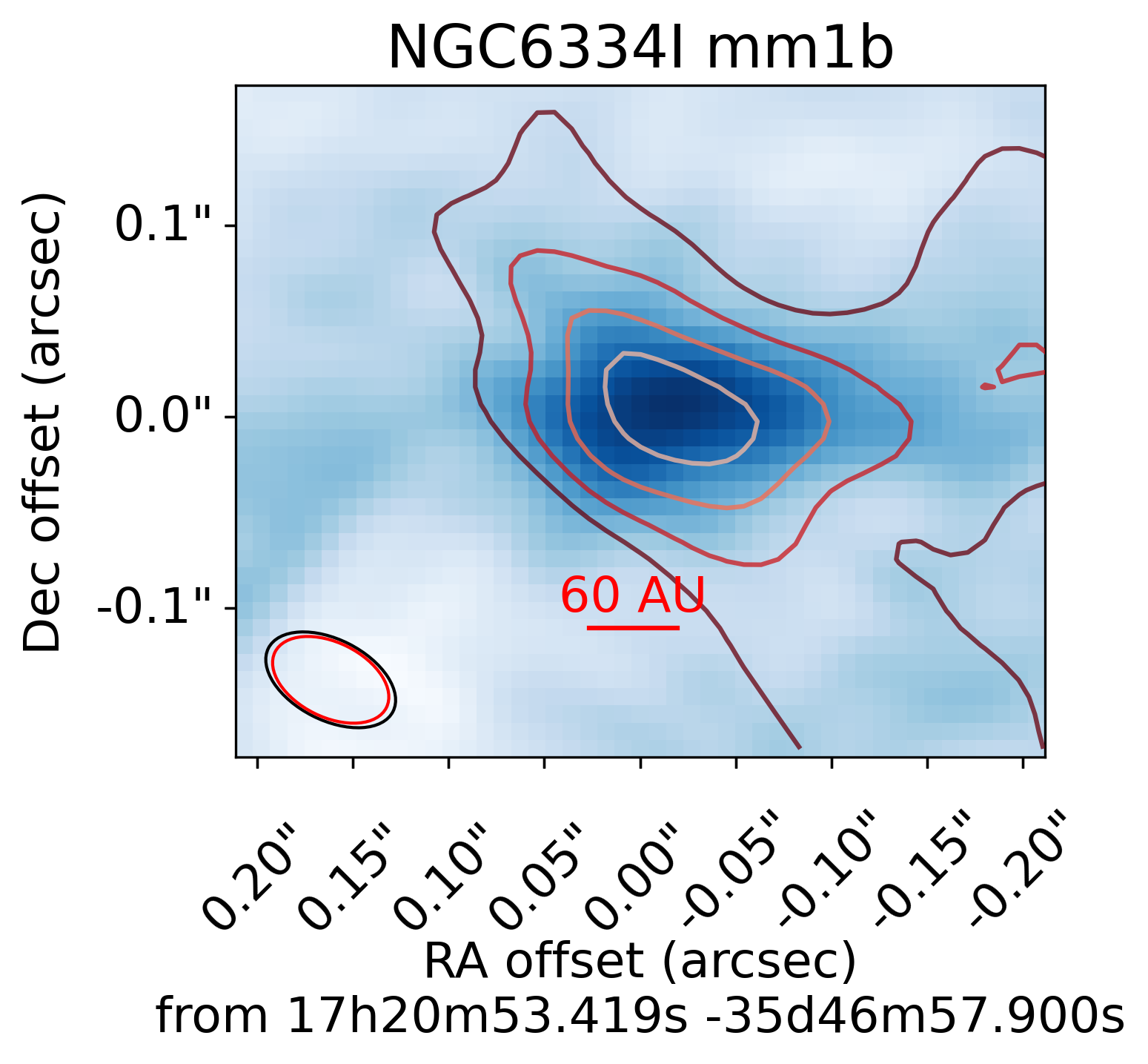}
    \caption{Moment-0 (integrated intensity) images of the resolved sources in NaCl lines
    as described in Figure \ref{fig:resolveddisks}.
     The red ellipse shows the continuum beam corresponding to the contours, while the black ellipse shows the line image beam.
     The contours are:
     SrcI $\sigma=0.21$ mJy/beam, contours at 25, 50, 75, 100, 125, 150 $\sigma$,
     G17 $\sigma=0.12$ mJy/beam, contours at 50, 100, 150, 200 $\sigma$,
     G351mm1 $\sigma=0.08$ mJy/beam, contours at 50, 75, 100, 125 $\sigma$,
     NGC6334Imm1b $\sigma=1$ mJy/beam, contours at 10, 15, 20, 25 $\sigma$.
     }
    \label{fig:momentswithcontours}
\end{figure*}

\begin{figure*}
    \centering
    \includegraphics[width=0.32\textwidth]{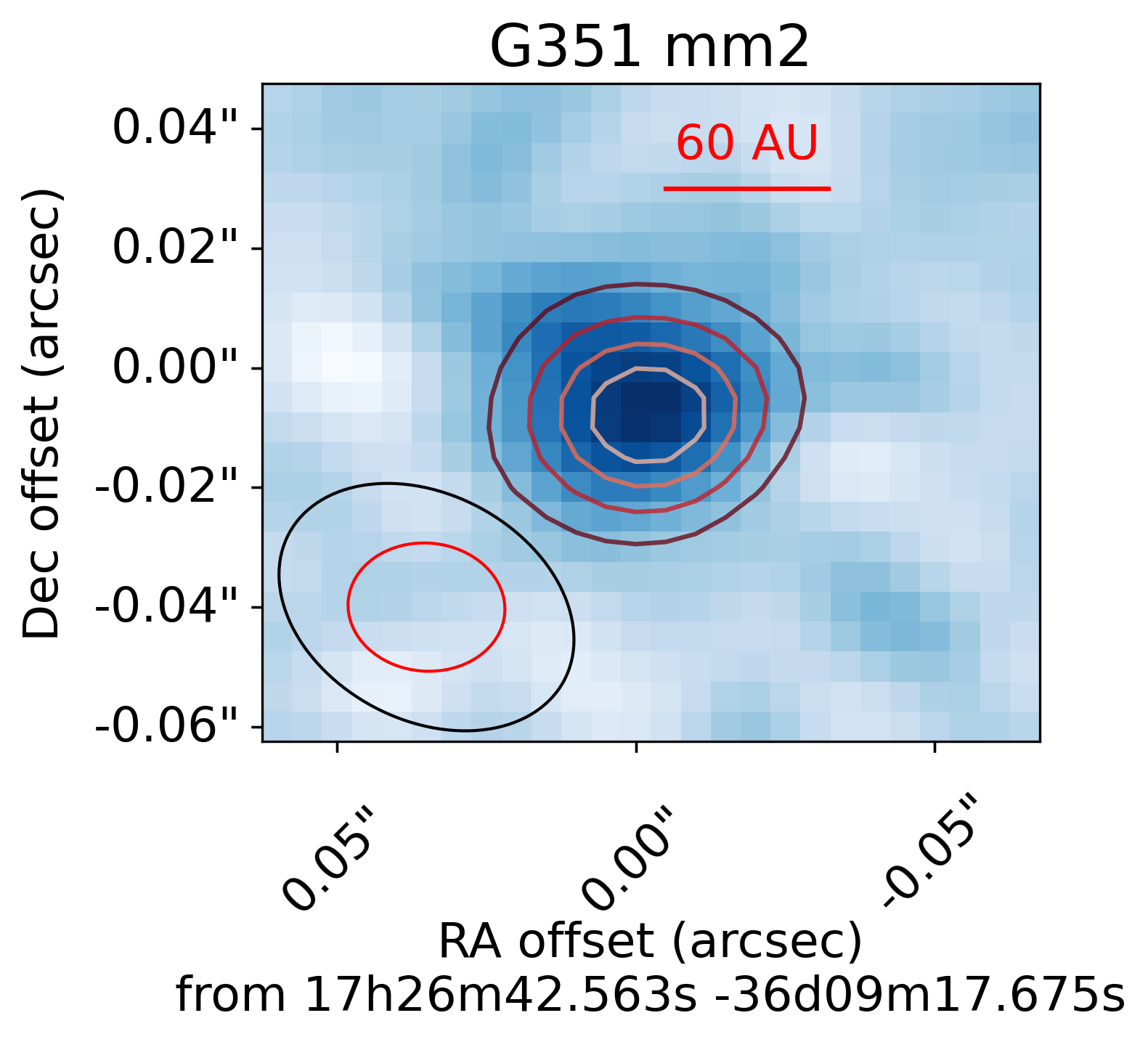}
    \includegraphics[width=0.32\textwidth]{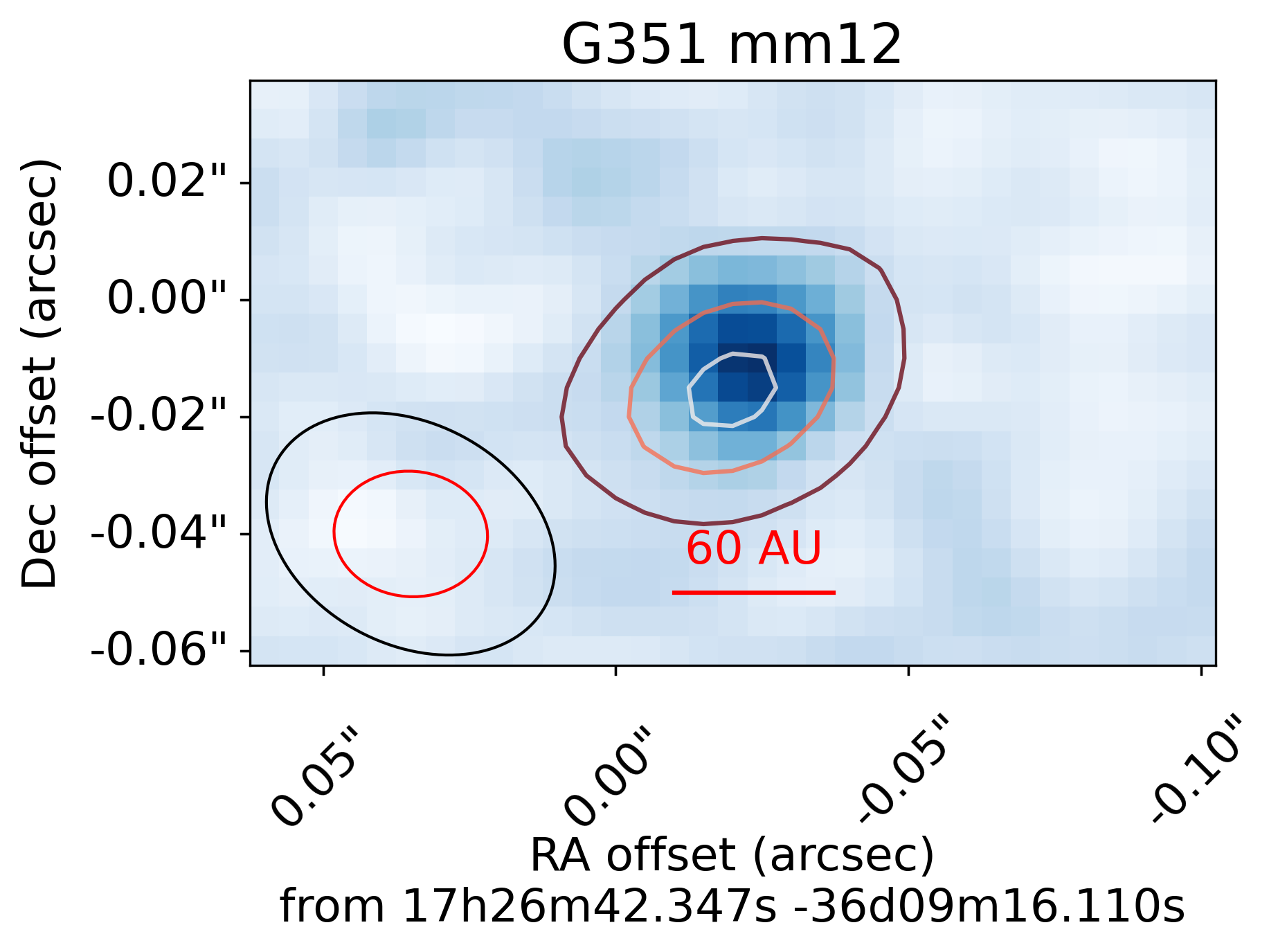}
    \includegraphics[width=0.32\textwidth]{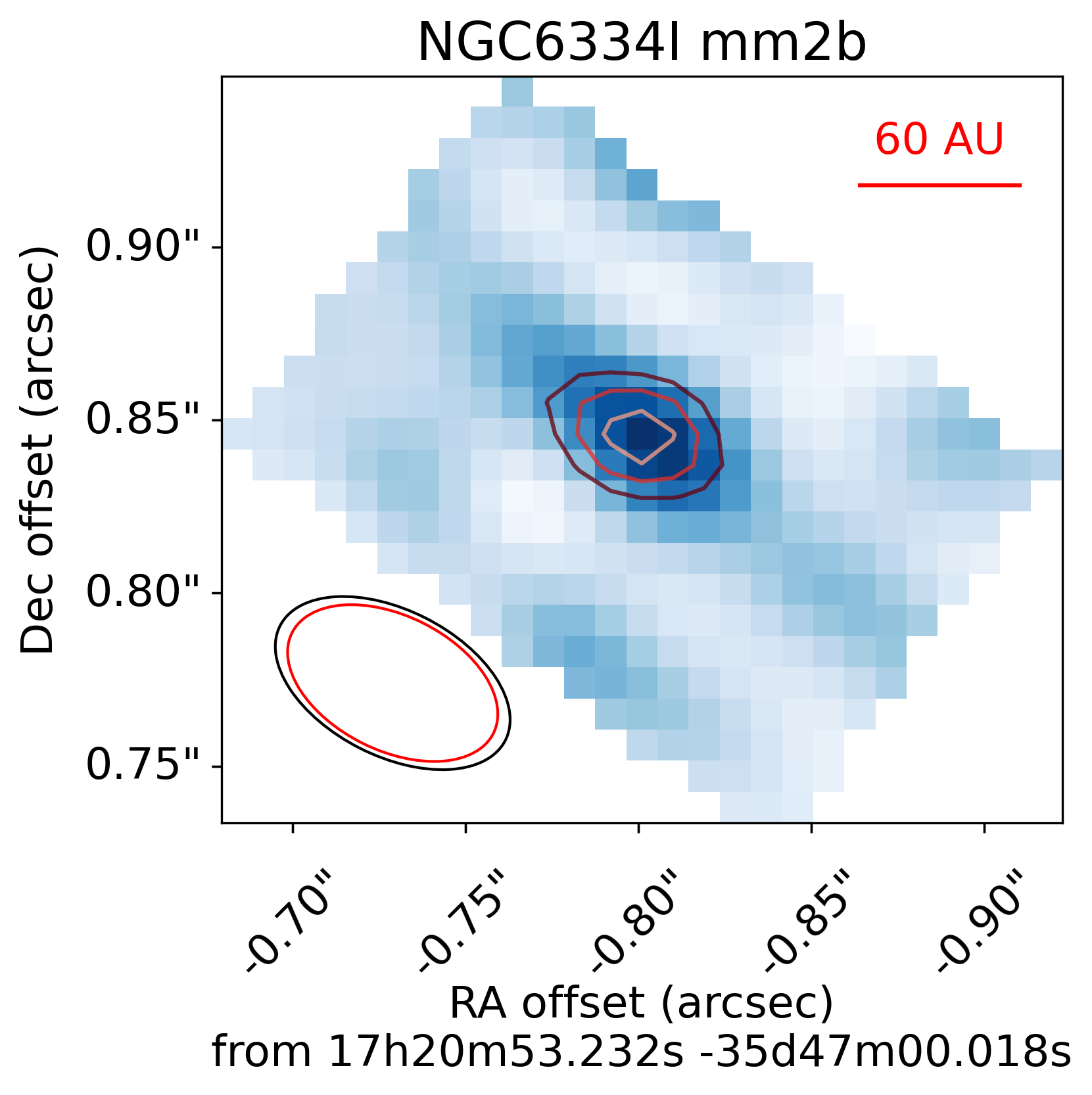}
    \includegraphics[width=0.32\textwidth]{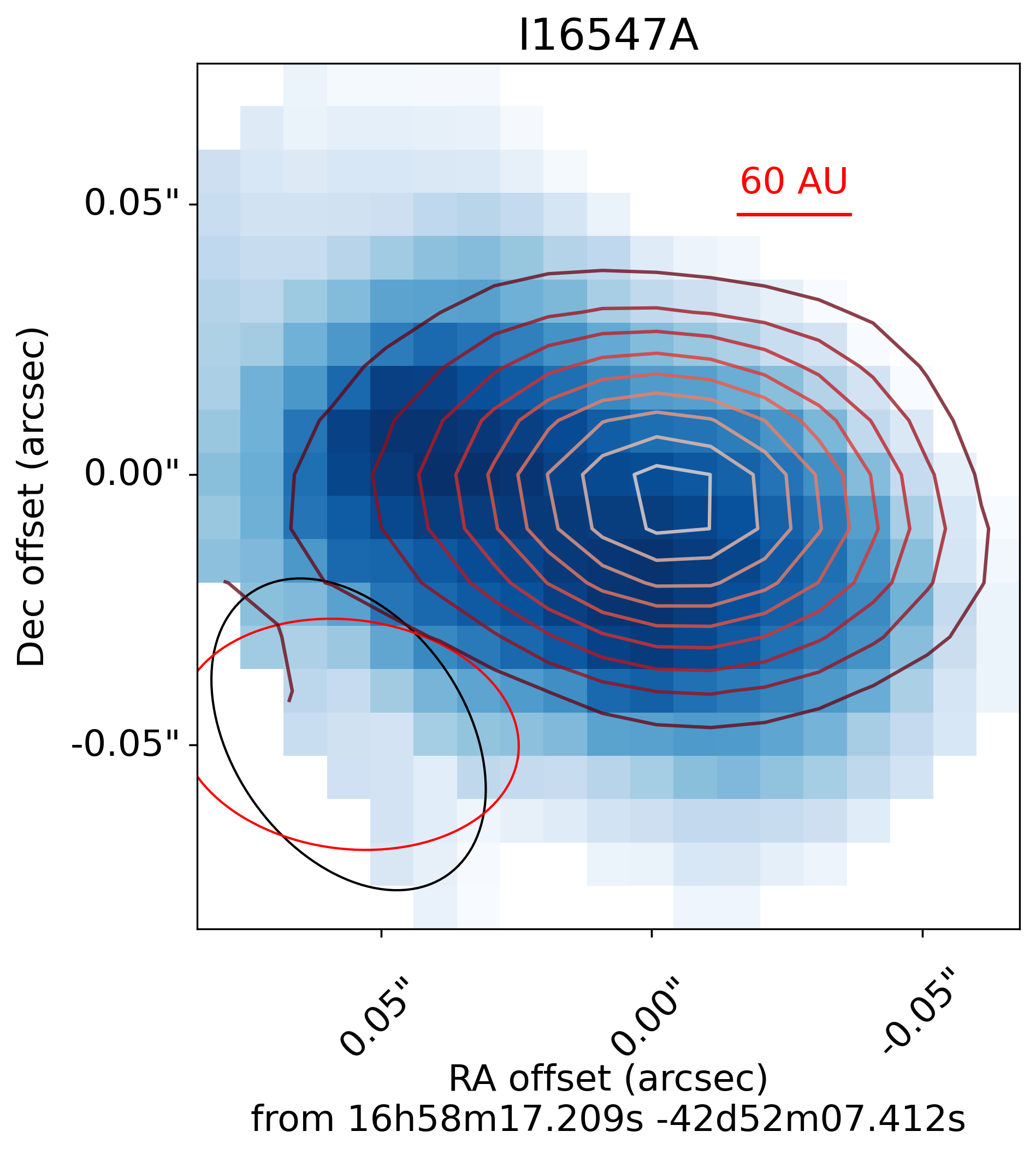}
    \includegraphics[width=0.32\textwidth]{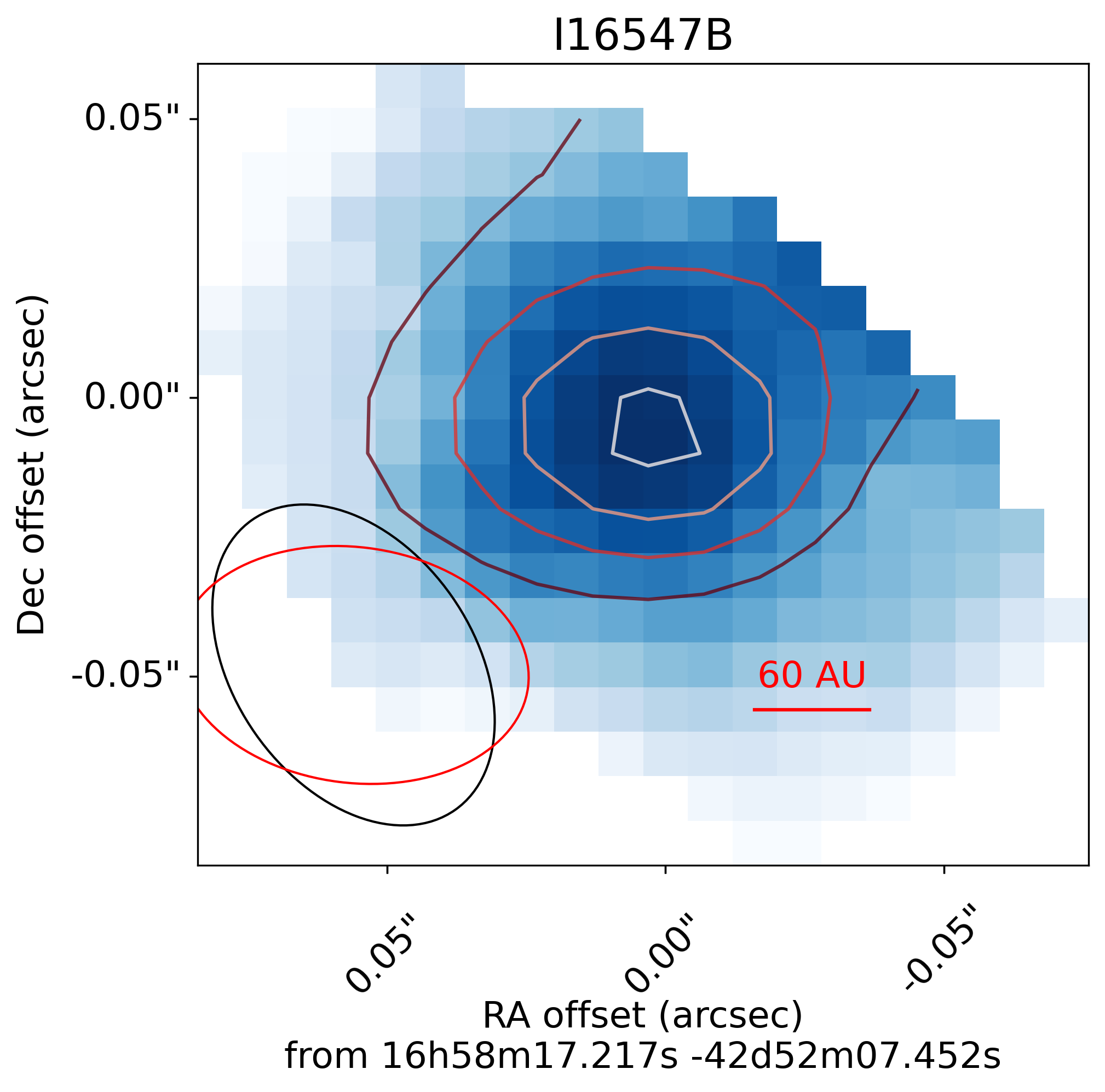}
    \includegraphics[width=0.32\textwidth]{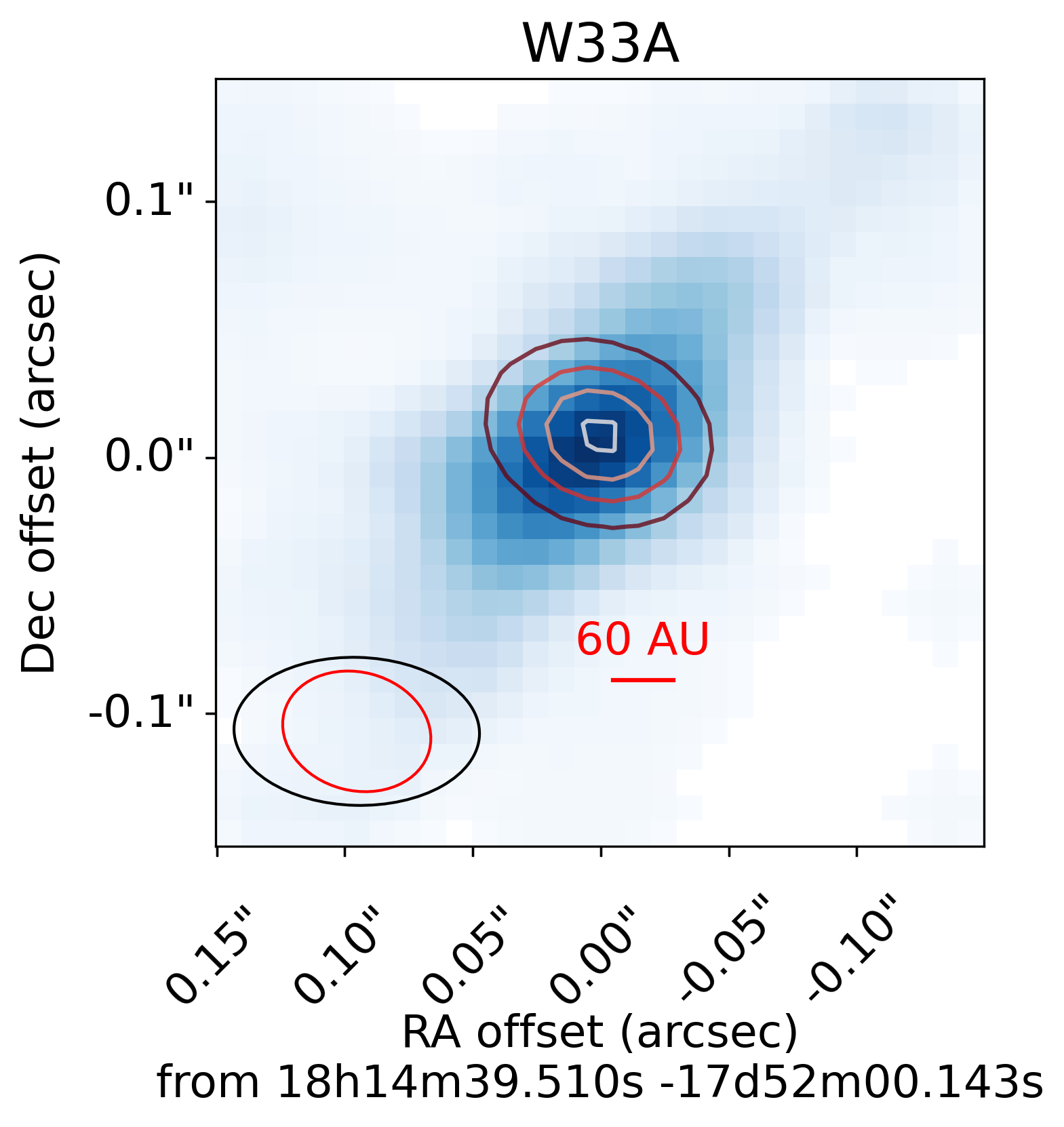}
    \caption{As in Figure \ref{fig:i16547}, but with continuum contours overlaid.
     The red ellipse shows the continuum beam corresponding to the contours, while the black ellipse shows the line image beam.
     The contours are:
     G351mm2 $\sigma=0.08$ mJy/beam, contours at 50, 75, 100, 125 $\sigma$,
     G351mm12 $\sigma=0.08$ mJy/beam, contours at 10, 20, 30 $\sigma$,
     NGC6334Imm2 $\sigma=1$ mJy/beam, contours at 2, 3, 4 $\sigma$,
     I16547A $\sigma=0.08$ mJy/beam, contours at 100, 150, 200, 250, 300 $\sigma$,
     I16547B $\sigma=0.08$ mJy/beam, contours at 75, 100, 125, 150 $\sigma$,
     W33A $\sigma=0.3$ mJy/beam, contours at 20,40,60,80 $\sigma$.
    }
    \label{fig:momentswithcontoursb}
\end{figure*}

We show Figures \ref{fig:resolveddisks} and \ref{fig:i16547} again, but this time with continuum contours overlaid, in Figures \ref{fig:momentswithcontours} and \ref{fig:momentswithcontoursb}.


\section{Unsalted Sources}
\label{appendix:unsalted}

\subsection{I16562}
IRAS16562-3959 is also known as G345.4938+01.4677.
This source shows SO in absorption against a continuum disk.
It exhibits RRL emission that shows a very tiny gradient;
\citet{Guzman2020} reported the detection of an ionized disk in this system based on RRL emission.
No other emission lines are associated with the RRL possible disk.
There is extensive CH$_3$CN emission around this source with kinematics more complicated than a simple Keplerian disk.

\subsection{G333}

We adopt naming from \citet{Stephens2015}, though their resolution was only $\sim2.5\arcsec$.

G333.23mm1 is the bright central region.  Morphologically, it looks like the hub at the center of several converging filaments.
There is no obvious sign of a disk in any of the lines we examined.
There is no hint of brinary lines.

G333.23mm2 shows some sign of a line gradient in \methanol, hinting that a disk is present.
However, no brinary lines are seen.

\subsection{G335}
\citet{Olguin2022} performed an extensive study of this system using the data presented here.
G335-ALMA1 is the bright central source that shows signs of rotation, but no clear disk.
Instead, it appears dominated by inflowing accretion filaments.
The spectrum is extremely rich, and we adopt the same \methanol line as in \citet{Olguin2022} as our velocity reference.
No brinary lines are detected.

\subsection{G5.89}
\label{sec:g5.89}
The G5.89 image is dominated by an extended HII region.
The only disklike source in the imaged field of view is mm15, so our cutout centered on that source.
It apparently exhibits \emph{no} line emission at all in the present data set.
We examined the other millimeter peaks in the region, but found no obvious signatures of disks or brinary lines.

\subsection{GGD27}
GGD27mm1 (otherwise referred to here as GGD27) is the driving source of HH80/81 \citep{Girart2017,Girart2018,Anez-Lopez2020}.
It is also known as IRAS 18162-2048.
It has a clear, well-defined, massive ($M\sim5\msun$) disk orbiting a $\sim20$ \msun star \citep{Anez-Lopez2020}.
There is no sign of NaCl, \water, KCl, or SiS in the spectrum of the disk in our data or previous observations \citep{Girart2017}.
It is bright in SO $6_5-5_4$, which we used to stack spectra to perform a deeper search.

This source is the archetype of non-salt-bearing disks.  Its well-defined Keplerian line profiles,
high central mass, and line-poor spectrum demonstrate that
not all HMYSO disks exhibit salt emission.
The system is very similar to SrcI in terms of its central stellar mass, disk size (though GGD27's is 2-4$\times$ larger in radius),
and luminosity, suggesting that none of these features are critical for releasing salt into the gas phase.
As the driver of the HH80/81 outflow, it is also clear that the simple presence of an outflow does not determine whether brinary lines are produced.
However, one notable difference is that the HH80/81 jet is highly collimated \citep{Rodriguez-Kamenetzky2017,Qiu2019}, while SrcI drives a broader disk wind \citep{Hirota2017,Tachibana2019}, hinting that the driving mechanism of the outflow may have a role in determining when salts are detectable.
The comparison of this source to SrcI and others in this
sample will be useful for future understanding of the origin of salts.

\subsection{IRAS18089}
\citet{Sanhueza2021} observed this source at high resolution, though they focused on the magnetic field.
Previously, \citet{Beuther2004a,Beuther2005} showed that this source was line-rich with SMA observations.
We used a \methanol as the stacking line because the kinematics were similar across all bright lines, including the SO $6_5-5_4$ line, but the SO line was affected by strong absorption toward the inner disk.
The line kinematics were reasonably disk-like, but not consistent with a single Keplerian disk; the velocity structure in this source requires more sophisticated modeling.
Nevertheless, there is no sign of brinary lines in the averaged or stacked spectra.

\subsection{G34.43mm1}
There is a structure toward the center of G34.43mm1 with a clear line gradient, but it does not trace disk-like kinematics and is quite extended. On the larger scale, G34.43mm1 drives a powerful outflow \citep{Sanhueza2010}. 
While this is region is quite nearby \citep[1.56 kpc][]{Xu2011}, it is not among the salt-bearing candidates.
Despite its relatively near distance, the observed spatial resolution is only $\sim400$ au, which is much larger than the detected disks.
The overall appearance of the inner region suggests that streamers are feeding in to a central region.
The chemically rich inflowing material results in a very spectrally dense spectrum \citep[e.g.,][]{Sanhueza2012,Liu2020G34}, which likely prevents detection of brinary features even if they are present.
This is a good candidate for followup at higher angular resolution.

\subsection{G29.96}
\citet{Beuther2007} and \citet{Beltran2011} studied this region at $\sim$arcsecond resolution.
The target region is centered on the brightest few sources at the center.
The distance to this object is unclear, with \citet{Beltran2011} reporting 3.5 kpc and \citet{Kalcheva2018} reporting 7.4 kpc.
There is no clear signature of any of the lines of interest, nor is there a clear signature of rotation.
There is a tentative detection of SiS v=0 12-11, but neither of the SiS v=1 lines (13-12 or 12-11) appear, so this detection is uncertain.

\subsection{S255IR}
Sh 2-255 IR SMA1 (S255IR) was the recent ($<10$ years) site of a major accretion outburst.
As such it is a strong candidate for being actively heated over its `normal' level.

No disk is obvious in the data cube from rotational signatures.  While several authors
\citep{Zinchenko2015,Zinchenko2020,Liu2020} have noted bulk rotation in the molecular gas around S255IR, these signatures are not evident on the smaller ($\sim70$ au) scales probed here.

There is, however, a hint of both \water and H30$\alpha$ emission along the innermost part of the outflow,
within about one resolution element of the central source.  There is also a hint of SiS here.
There is clearly SiO emission with a velocity gradient along the outflow axis, but curiously it is not very bright along the continuum jet.
This source warrants followup at higher resolution and sensitivity to try to identify the location of the actual disk, though it is not yet clear which lines to use for this search.

\subsection{NGC6334IN}
The NGC6334IN region contains several bright sources.  The brightest in our observations are SMA6 and SMA1 b/d \citep[names from][]{Sadaghiani2020}.
Neither source shows clear signs of rotation in any line.
SMA1b/d contains several sources that may exhibit some hint of \water emission, so we use that line as the basis for stacking
We do not detect any other lines.
Notably, both SO and SiO appear absent toward these sources.
SMA6 also shows a hint of a water line, but similarly shows no sign of any of the other targeted lines.

\subsection{G11.92mm1}
\citet{Ilee2018} reported on the disk G11.92mm1, showing a much larger ($r\sim800$ au) size than most of the sources in our sample.
We use SO $6_5-5_4$ as the guide line for stacking, since it shows clear disk kinematics.
No brinary lines are detected.

\end{document}